\documentclass[3p,authoryear]{elsarticle}
\usepackage[utf8]{inputenc}
\usepackage{caption}
\usepackage{subcaption}
\usepackage{graphicx}
\usepackage{soul, color}
\usepackage{amsfonts}
\usepackage{natbib}
\usepackage{calrsfs}
\usepackage{dsfont}
\DeclareMathAlphabet{\pazocal}{OMS}{zplm}{m}{n}
\usepackage{amsmath}
\usepackage{amssymb}
\usepackage{bbm}
\usepackage{tabularx}
\usepackage{booktabs}
\usepackage{longtable}

\usepackage{bm}
\usepackage{amsthm}
\usepackage{upgreek}
\usepackage{empheq}

\usepackage{enumitem}
\usepackage[colorlinks]{hyperref}
\hypersetup{
    linkcolor=blue,
    citecolor=blue,
    urlcolor=blue
}
\usepackage{pdflscape}
\usepackage[nameinlink]{cleveref}
\newtheorem{theorem}{Theorem}

\newtheorem{definition}{Definition}

\begin{document}
% Put here what will go to headers as author
\title{A Review of Max-Pressure Traffic Signal Control Algorithms: State of the Art and Research Opportunities}

% TODO: add macros for easier formatting of \author.
\author[1]{Abdullah Al Asif}
\ead{abdullah.asif@maine.edu}
\author[2]{Chaopeng Tan}
\ead{chaopeng.tan@tu-dresden.de}
\author[3]{Vikash V. Gayah}
\ead{gayah@engr.psu.edu}
\author[1]{Hao Liu\corref{cor1}}
\ead{hao.liu1@maine.edu}
\cortext[cor1]{Corresponding author}
\address[1]{Department of Civil and Environmental Engineering, University of Maine, Orono, ME, United States}
\address[2]{Chair of Traffic Process Automation, Technische Universit\"at Dresden,
Hettnerstra\ss e 3, 01069 Dresden, Germany}
\address[3]{The Pennsylvania State University, University Park, PA, United States}

\begin{abstract}
Max-Pressure (MP) has emerged as a widely studied decentralized traffic signal control framework, thanks to its desirable theoretical properties and strong control performance. Early research on MP focused on improving control performance by exploring different variables of the traffic state for the definition of pressure. Subsequent studies expanded MP to incorporate the complexity of real-world transportation systems and emerging technologies. This paper provides a comprehensive review of these developments. The review first introduces the theoretical foundations of MP and examines early developments in movement weight definition. It then synthesizes major extensions for multimodal and multiscale network control, followed by recent developments in connected and automated vehicle (CAV) environments and the integration of MP with artificial intelligence (AI) and data-driven methods. Beyond summarizing existing approaches, this review examines their underlying assumptions, data requirements, implementation considerations, and practical limitations. Based on this comprehensive review, key research gaps and future directions are identified, including remaining limitations in the theoretical foundations of MP and the underexplored integration of autonomous vehicle control with the MP framework. Overall, this review provides a structured perspective on the evolution and current state of MP control and highlights opportunities for developing more proactive, robust, and practically deployable MP-based traffic control methods.
\end{abstract}


\maketitle

\section{Introduction}\label{sec:intro}
Signalized intersections are the primary bottleneck for urban traffic operations in terms of both efficiency and safety, and thus are the subject of considerable focus in transportation research. Traffic signal control is the most investigated and deployed method to help improve operational efficiency and mitigate conflicts at intersections. Traffic signal control evolved from fixed-time plans toward adaptive systems that respond to real-time demands. Most of the early adaptive systems were centralized or hierarchical. Systems such as SCOOT \citep{hunt1982scoot} and SCATS \citep{lowrie1990scats} adjust coordinated signal timings using detector data gathered from across a network, then push updated settings back to local controllers, while rolling-horizon optimization methods such as OPAC \citep{gartner2001implementation} and RHODES \citep{mirchandani2001real} optimize signal timing using predicted arrivals, with OPAC combining intersection-level optimization with network-level synchronization and RHODES adopting a hierarchical, distributed architecture. These approaches have well-known drawbacks: communication and computational burden grows with network size, a central failure can disrupt adaptive coordination across large parts of the network, and the performance of prediction-based methods depends on forecasting future demand and turning movements, which becomes increasingly unreliable as the prediction horizon or network size grows. These limitations motivated a shift toward decentralized adaptive control, in which each intersection (or a small local cluster) sets its own signal timing using only locally observed, real-time information, without a central coordinator or demand forecasts. Early efforts in this direction included actuated control and rule-based self-organizing signals, later joined by distributed optimization and multi-agent learning approaches.

Max-pressure (MP), also known as backpressure (BP), originates in communication network theory and has become a popular decentralized real-time traffic signal control method. \citet{tassiulas1990stability} introduced the MP algorithm for scheduling and routing in multihop radio networks: each node selects which links to activate and how to route packets based on the differential backlog, defined as the queue-length gap between a transmitting and receiving node, favoring the link/route with the largest such "pressure." The study proved this policy is throughput-optimal: it keeps all queues bounded for any arrival rate within the network's capacity region, using only local queue information exchanged between neighboring nodes, with no need to know arrival statistics in advance. This decentralized structure, desirable performance, and the throughput-optimal property motivated its transfer to traffic signal control roughly two decades later. \citet{dai_maximum_2005} subsequently developed the MP formulation for a broader class of stochastic processing networks and explicitly introduced a family of \textit{maximum pressure} policies. In their framework, each feasible resource allocation is evaluated according to the current buffer levels (i.e., the number of jobs or materials waiting in each queue) and the network input-output relationships, and an allocation with maximum pressure is selected. Under their stated structural and service assumptions, they established throughput optimality using a fluid-model argument. \citet{wongpiromsarn2012distributed} and \citet{varaiya_max_2013} independently adapted the concept to networks of signalized intersections. The fundamental mechanism of an MP traffic signal algorithm is that, at each intersection, the controller computes the pressure of each candidate phase, which represents the need for service of the movements served by that phase, and then activates the phase with the highest pressure. This pressure is typically determined based on the difference in a traffic-state metric, such as queue length or travel time, between the upstream and downstream links associated with each movement. The upstream metric primarily reflects traffic demand for the corresponding movements, while the downstream metric accounts for the available downstream supply. This helps prevent excessive vehicle discharge toward links with limited receiving capacity, thereby reducing the risk of downstream spillback and avoiding ineffective signal timing when downstream supply is limited. \citet{varaiya_max_2013} showed that this policy maximizes throughput using a store-and-forward queuing model of the road network. Despite the shared mathematical backbone between the communication network and the transportation network, certain important features distinguish these two domains. 
\begin{enumerate}
    \item Wireless scheduling assumes discrete, synchronized time slots with negligible reconfiguration cost, while signals must have minimum green and clearance intervals, introducing switching losses. 
    \item Most traffic signal control methods follow a cyclic pattern, which is not required in wireless scheduling.
    \item Wireless queue capacity can generally be treated as unbounded, while road links have finite storage, so spillback can violate the point-queue assumptions underlying the store-and-forward model, a gap that later MP variants explicitly address. 
    \item Conflicting movements at an intersection follow fixed, geometry-constrained phase sets rather than the more flexible link-activation sets of radio scheduling, and turning movements face different physical and operational constraints from packet routing. Traffic networks must also accommodate pedestrians, whose service requirements have no direct analogue in packet networks.
\end{enumerate}
Because of these fundamental differences, the MP algorithm cannot be directly transferred from wireless communication to traffic signal control while retaining its optimal performance. Instead, substantial modifications and additional research are required to adapt the MP framework to the unique operational characteristics and constraints of traffic signal systems.

Over the past 15 years, numerous variants of the MP algorithm have been proposed to improve both its control performance and practical applicability in traffic signal control. These studies can be broadly classified into the following categories:
\begin{enumerate}
    \item Exploration of alternative traffic state variables for pressure definition;
    \item Enhancement through improved parameter calibration and control structures;
    \item Extension to multimodal and multiscale traffic operations; and
    \item Integration with emerging technologies, such as artificial intelligence (AI) and connected and automated vehicles (CAVs).
\end{enumerate}

The goal of this paper is to provide a comprehensive review of MP-based traffic signal control by synthesizing the major developments, identifying key gaps and limitations in the existing literature, and discussing promising directions for future research. Although a review of MP-based traffic signal control was published by \cite{levin_max-pressure_2023_review}, that study primarily focused on the methodological foundations, classical formulations, and early extensions of the MP framework through 2022. Compared with \citet{levin_max-pressure_2023_review}, the major contributions of this paper are threefold. First, we organize the MP literature using a multidimensional classification framework that captures both the fundamental design of MP algorithms and their broader practical evolution. Instead of summarizing individual studies or grouping them into broad topic areas, this review examines MP research through several interconnected dimensions, including theoretical formulations, implementation-oriented extensions such as multimodal applications, multiscale and system-level coordination, and integration with emerging transportation technologies. This organization moves beyond a simple enumeration of existing studies and provides a more comprehensive and hierarchical representation of the MP research landscape. Second, this review critically evaluates representative MP approaches in terms of their theoretical assumptions, sensing and data requirements, applicable scenarios, practical implications, and limitations, thereby highlighting the trade-offs among different methodological choices. Third, we incorporate recent developments across these research areas to provide an up-to-date understanding of the evolution and current state of MP traffic signal control, while identifying remaining research gaps and discussing promising directions for future research.

\begin{figure}[htbp]
	\centering
	\includegraphics[width=1\textwidth]{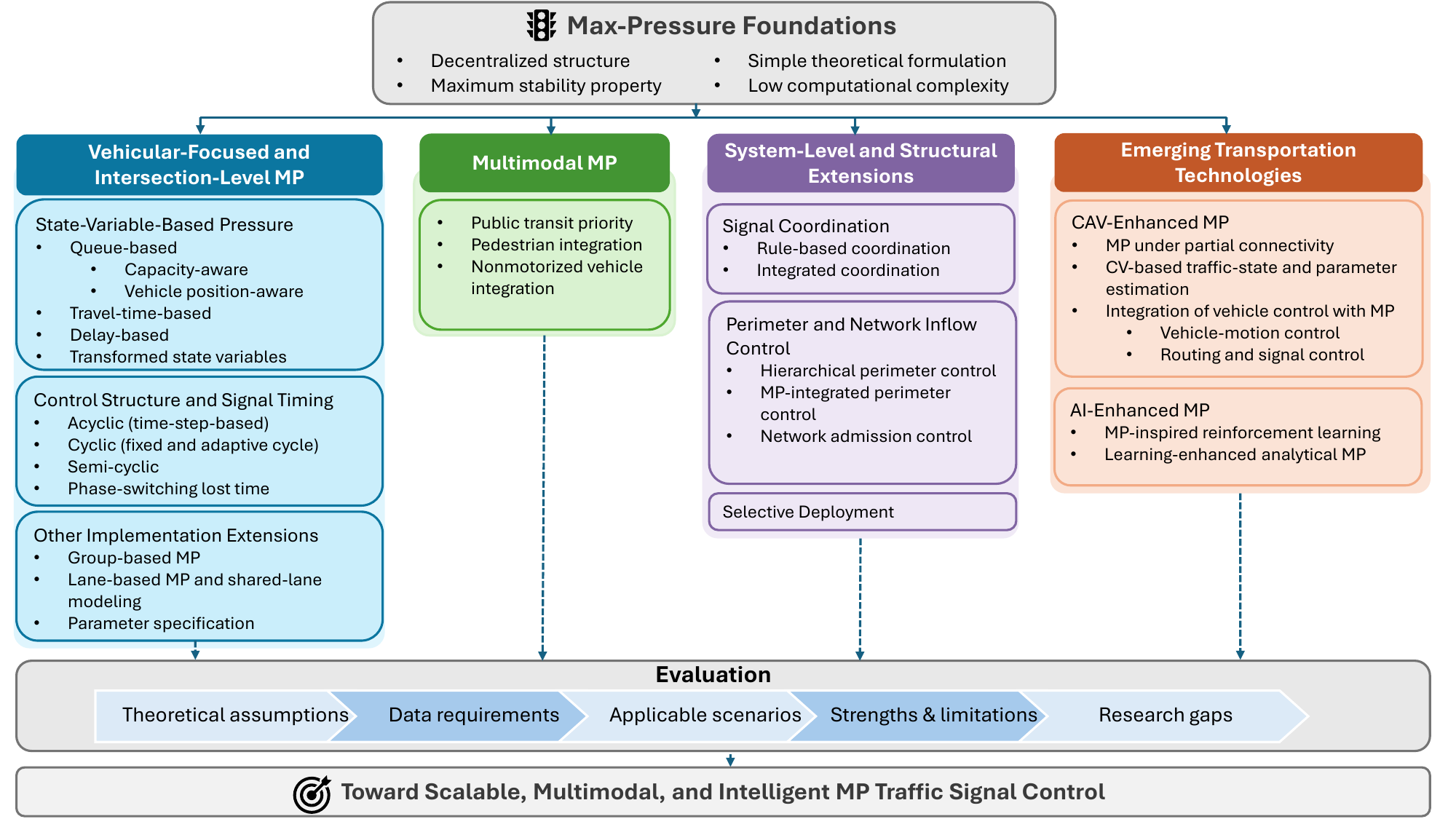}
	\caption{Taxonomy of MP signal control review}
	\label{fig:taxonomy}
\end{figure}

The remainder of this paper is organized as follows. \Cref{sec:review method} describes the literature search methodology, including the databases, keywords, and screening criteria used to identify and select relevant studies. \Cref{sec:theory} presents the theoretical foundations of MP traffic signal control. \Cref{sec: local} reviews recent advancements and variants of MP frameworks, with a focus on local intersection-level control. \Cref{sec: structure extension} discusses MP variants based on their control structure and compares representative MP variants under a common experimental setting. \Cref{sec: multi} discusses extensions of MP control to multimodal traffic, followed by \Cref{sec:system_and_structural}, which reviews the system-level and structural extension of MP works. \Cref{sec: emerging} examines the integration of MP with emerging transportation technologies. Finally, \Cref{sec: conclusion} summarizes the major findings, identifies remaining research gaps, and discusses promising directions for future research. To provide a structured overview of the extensive MP literature and guide the subsequent review, \autoref{fig:taxonomy} presents a taxonomy of the major research directions in MP traffic signal control.

\section{Literature search methodology}\label{sec:review method}
This paper adopts a critical narrative review approach to synthesize and evaluate developments in MP-based traffic signal control. The literature search described below supports this synthesis and provides transparency regarding study identification and selection; it is not intended to constitute a systematic review. To identify relevant studies on MP traffic signal control, we used a structured search and screening process. The search focused on English-language studies published between 2013 and 2026 and indexed in Scopus or Web of Science. The search combined terms related to max-pressure and traffic signal control: (\textit{"max pressure"} OR \textit{"max-pressure"} OR \textit{"backpressure control"}) AND (\textit{"signal control"} OR \textit{"traffic signal control"} OR \textit{"traffic signal"})

Peer-reviewed journal articles, conference papers, and review papers were included, whereas theses, editorials, and reports were excluded. The database search identified 237 records, including 104 from Scopus and 133 from Web of Science. After removing 89 duplicates, 148 papers were screened by title and abstract, of which 75 were retained for further assessment. Studies that merely used MP as a benchmark for comparison with other traffic signal control strategies, without introducing any new contribution to the MP framework, were excluded.  
Additional relevant studies, including arXiv preprints, were identified through prior knowledge, reference lists, and the requirements of specific review topics. These studies were assessed using the same relevance criteria but were not included in the reported database screening counts. Following the detailed assessment of the database-identified studies and the inclusion of these additional studies, a total of 90 studies were included in the final review.

\begin{figure}[htbp]
	\centering
	\includegraphics[width=0.85\textwidth]{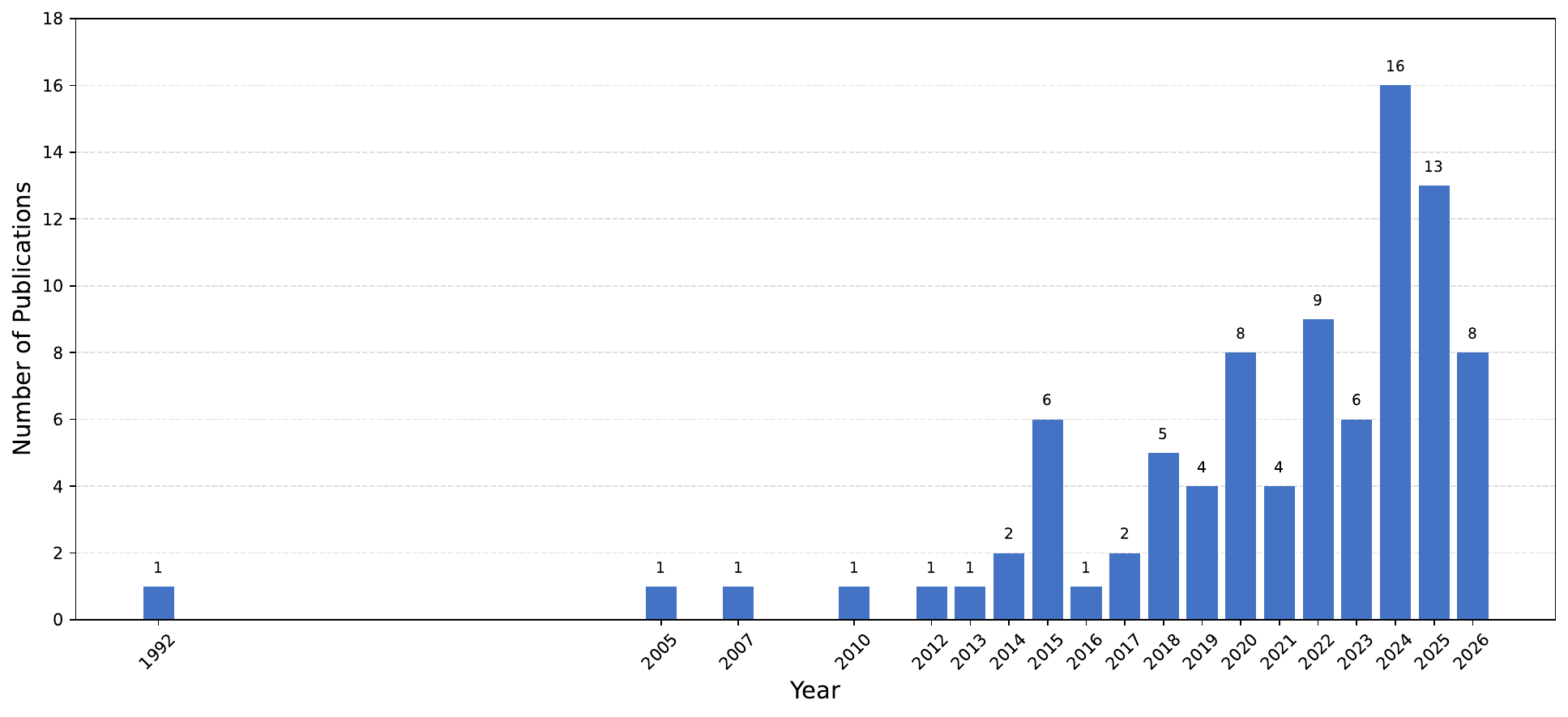}
	\caption{Number of MP-related publications included in this review by year }
	\label{fig:num_pubs}
\end{figure}

\autoref{fig:num_pubs} summarizes the number of MP-related publications included in this review by year. Note that early MP-related traffic signal control can be traced to \citet{wunderlich_stable_2007}, who proposed a queue-based maximal-weight matching algorithm for an isolated intersection. Building on this work, \citet{arel_reinforcement_2010} extended the framework to a multi-intersection setting by combining Longest-Queue-First (LQF)-controlled neighboring intersections with a Q-learning controller at the central intersection.  As shown in \autoref{fig:num_pubs}, research on MP control remained relatively limited before the mid-2010s. Since then, research interest has grown steadily, with a particularly notable increase after 2018. This growth has accelerated in recent years, as evidenced by the sharp rise in the number of publications since 2022 and the peak observed in 2024. Overall, this trend reflects the rapidly growing attention to MP control and highlights the increasing efforts to extend the original MP framework to address emerging challenges and application scenarios. Importantly, the research focus of MP control has evolved considerably over time. Early studies primarily focused on improving the control performance of MP itself, such as by exploring more appropriate metrics for pressure calculation beyond queue lengths. With the rapid growth of the literature, recent studies have substantially extended the original MP framework toward multiscale and multimodal transportation systems, connected and automated vehicles, and data-driven methods. These emerging research directions are reviewed in the following sections.

\section{Theoretical foundation of MP}\label{sec:theory}

Following the introduction of the MP model for traffic signal control by \citet{wongpiromsarn2012distributed} and \citet{varaiya_max_2013}, most MP algorithms model the queue evolution in a network using a store-and-forward process. Vehicle arrivals at entry links are assumed to be independent and identically distributed random processes; the average turning ratios at intersections are also assumed to be known. In addition, each turning movement has a separate queue, and these queues are assumed to be point queues, meaning the links have infinite queue capacity. The original MP formulation adopts a \textit{movement-based} structure, in which the basic unit of analysis is the vehicle movement, defined as an ordered pair of links $(l,m)$, where $l$ is an incoming link and $m$ is an outgoing link connected through the same intersection. In the MP framework, signal timing decisions are based entirely on phase pressures, while phase pressures are computed solely from the traffic states of vehicle movements. \autoref{table:notations} summarizes the notation used throughout this paper.

\begin{table*}[htbp]
\caption{Description of Notations}
\label{table:notations}
\centering
\begin{tabularx}{\linewidth}{lX}
\hline

\multicolumn{2}{l}{\textbf{Network characteristics and signal configuration}} \\

$\pazocal{L}_{\text{all}}$ & Set of all links (e.g., $l, m, k$) \\

$\pazocal{L}$ & Set of internal links \\
$\pazocal{L}_{\text{entry}}$ & Set of entry links \\

$\pazocal{L}_{\text{exit}}$ & Set of exit links \\

$\text{In}_l$ & Set of links input to the link $l \in \pazocal{L} \cup \pazocal{L}_{\text{exit}}$ \\

$\text{Out}_l$ & Set of links output from the link $l \in \pazocal{L} \cup \pazocal{L}_{\text{entry}}$ \\

$(l,m)$ &
Vehicle movement from incoming link $l$ to outgoing link $m$ \\

$S(l,m)(t)$ &
Signal indicator of movement $(l,m)$ at time $t$; 1 if green and 0 otherwise \\

$S(t)$ &
Network control matrix \\

$\pazocal{S}$ &
Set of feasible network control matrices \\

$\Phi_i$ & Set of all signal phases at intersection $i$ \\

$\pazocal{M}_i^\phi$ & Set of movements served by phase $\phi$ at intersection $i$ \\

$\pazocal{M}_i$ & Set of all valid movements at intersection $i$ \\

$\pazocal{M}$ & Set of all valid movements in the network \\

$\ell_m$ & Length of the link $m$ \\

$x_l^{max}$ & Maximum queue capacity of link $l$. \\

$x^{max}(l,m)$ &
Maximum queue capacity of the movement $(l,m)$ \\

$\lambda_{i}^\phi(n)$
& Proportion of the total allocable green time to phase $\phi$ at intersection $i$ \\

$e$
& Lost time associated with a phase switch   \\

$L_i$
& Total lost time of a cycle at intersection $i$     \\

$C_i^{cyc}$
& Cycle length at intersection $i$ \\

\hline

\multicolumn{2}{l}{\textbf{Traffic state and queue dynamics}} \\

$z(l,m)(t)$ &
Traffic-state measure associated with movement $(l,m)$ at the beginning of time period $t$ ,
such as queue length, travel time, or traffic delay \\

$x(l,m)(t)$ &
Number of vehicles associated with movement $(l,m)$ at the beginning of time period $t$ \\

$\Delta x^{\mathrm{out}}_{l,m}(t)$ &
Number of vehicles departing from movement $(l,m)$ during the interval $[t,t+1)$ \\

$\Delta x^{\mathrm{in}}_{l,m}(t)$ &
Number of vehicles entering movement $(l,m)$ during the interval $[t,t+1)$ \\

$x_l(t)$ &
Number of vehicles on the link $l$ at the beginning of time period $t$ \\

$x_{ln,l}(t)$ &
Number of vehicles on the lane $ln,l$ of link $l$ at the beginning of time period $t$ \\

% $H_l(t)$ &
% Spatial queue-length in the link $l$ at the beginning of time period $t$ \\

$X(t)$ &
Network state consisting of all movement queue lengths \\

$f_l$ &
Mean vehicle flow on link $l$, vehicles per period \\

$d_l$ &
Mean external demand entering entry link
$l\in\pazocal{L}_{\text{entry}}$, vehicles per period \\

$d(l,m)(t)$ &
Random demand for movement $(l,m)$ during time period $t$ \\

$r(l,m)$ &
Mean turning ratio (probability) that a vehicle leaving link $l$ enters link $m$ \\

$R(l,m)(t)$ &
Stochastic turning ratio from link $l$ to link $m$ \\

$c(l,m)$ &
Mean saturation flow of movement $(l,m)$, vehicles per period \\

$C(l,m)(t)$ &
Random saturation flow of movement $(l,m)$ during time period $t$ \\

\hline

\multicolumn{2}{l}{\textbf{Max-pressure algorithm}} \\

$w(l,m)(t)$ &
Weight of movement $(l,m)$ at time $t$ \\

$P_i^{\phi}(t)$              
& Pressure of phase $\phi$ at intersection $i$ at time $t$ \\
$\phi^{*}_i(t)$
& Selected phase in the Max-Pressure controller at time $t$  \\
\hline

\end{tabularx}
\end{table*}

\subsection{Assumptions in the Original MP}

MP algorithms are based on several assumptions regarding traffic dynamics and signal operations. These assumptions affect the theoretical modeling of the control algorithm, control performance, and analytical properties of MPs. Understanding these assumptions is important for interpreting the theoretical results and assessing the applicability of MP in real-world deployments. The assumptions adopted in MP algorithms are as follows: 

\begin{enumerate}[label=\roman*.]
     \item \textbf{Store-and-Forward model:} Traffic dynamics are modeled using a store-and-forward (SF) queueing model. Traffic is represented as queues on links, and vehicles are treated as ``packets'' moving from one queue to the next. Once a vehicle is discharged from an upstream queue during a time-step, it immediately joins the downstream queue, thereby neglecting link travel time.
    
    \item \textbf{Point queue model:} Within the SF model, queues are assumed to occupy no physical space and therefore have unlimited storage capacity. Thus, vehicles can always enter the downstream queue regardless of downstream congestion. This assumption is fundamental to proving the maximum stability property of MP. However, in real road networks, links have finite storage capacity, and downstream spillback may block upstream discharge. Therefore, although MP is theoretically stable under all feasible demand levels, the stability of MP algorithms is typically evaluated in microscopic traffic simulation using a limited number of demand scenarios.
    
    \item \textbf{Separate queue for each turning movement:} MP assumes an independent queue for every turning movement. This movement-based representation allows MP to calculate movement-specific pressure. However, in real networks, many intersection approaches have shared lanes, making movement-specific queue measurement or reliable estimation of vehicle turning intentions difficult.
    
    \item \textbf{Fixed routing probabilities:} Vehicle turning movements are assumed to follow fixed routing probabilities that are independent of time and signal control. These routing probabilities are required to estimate downstream pressure. However, routing proportions often vary with traffic conditions, incidents, and driver route choice. Consequently, inaccurate routing probabilities may reduce pressure estimation accuracy and controller performance.

\end{enumerate}

\subsection{Store-and-forward model}

Let $x(l,m)(t)$ denote the number of vehicles for the movement from link $l$ to link $m$ at time $t$. The evolution of the number of vehicles in an internal link is given by

\begin{equation}
\begin{split}
x(l,m)(t + 1) = {} & x(l,m)(t)
- \Delta x^{\mathrm{out}}_{l,m}(t)
+ \Delta x^{\mathrm{in}}_{l,m}(t),
\qquad
l \in \pazocal{L}, \;
m \in \mathrm{Out}_l .
\end{split}
\label{eq:queue_update_internal}
\end{equation}

Here,
\begin{equation} \label{eq:outflow}
\Delta x^{\mathrm{out}}_{l,m}(t) = \min\left(C(l,m)(t+1)S(l,m)(t), x(l,m)(t)\right)
\end{equation}
denotes the outflow from movement $(l,m)$, which is bounded by both the discharge capacity $C(l,m)(t+1)$ and the available queue length $x(l,m)(t)$. Here, $S(l,m)(t)$ is the signal control indicator, 1 if the movement $(l,m)$ is actuated and 0 otherwise. Similarly,

\begin{equation} \label{eq:inflow}
\Delta x^{\mathrm{in}}_{l,m}(t) = \sum_{k \in \mathrm{In}_l} [ \min( C(k,l)(t+1)S(k,l)(t), x(k,l)(t) ) ] R(l,m)(t+1)
\end{equation}

denotes the inflow to movement $(l,m)$, consisting of vehicles discharged from upstream movement $(k,l)$ and routed to movement $(l,m)$. Here, $R(l,m)(t+1)$ is the stochastic turning ratio from link $l$ to link $m$.

If $l$ is an entry link, the queue update equation becomes

\begin{equation}
\begin{split}
x(l,m)(t + 1) = {} &
x(l,m)(t)
-
\Delta x^{\mathrm{out}}_{l,m}(t)
+
d(l,m)(t + 1),
\qquad
l \in \pazocal{L}_{\mathrm{entry}}, \;
m \in \mathrm{Out}_l .
\end{split}
\label{eq:queue_update_entry}
\end{equation}
where $d(l,m)(t+1)$ is the demand for movement $(l,m)$ during time period $t+1$.

Note that the store-and-forward model does not directly affect signal timing determination, as will be shown in \Cref{subsec:mp_steps}, but it plays an important role in the stability analysis of MP algorithms.

\subsection{General MP control steps} \label{subsec:mp_steps}

According to the signal update mechanism, existing MP control methods can generally be categorized into \textit{time-step-based} (or \textit{acyclic}) and \textit{cyclic} approaches. Despite their differences in how and when signal decisions are updated, both approaches share the same fundamental principle: signal phases associated with larger pressure values are given higher priority for green time. Therefore, to illustrate the core mechanism of MP control, this section focuses primarily on the time-step-based formulation. The cyclic implementation of MP, along with its specific characteristics and extensions, will be discussed in the corresponding subsequent sections. 

Time-step-based MP algorithms, which adjust signal timing at a fixed frequency, generally follow a common three-step procedure. At each signal update instant, the controller performs the following steps:

\begin{enumerate}
    \item \textbf{Retrieve the traffic state and compute movement weights.}
    For each movement, the value of a selected traffic-state variable, such as queue length, travel time, or traffic delay, is obtained. The corresponding movement weight at time $t$, $w(l,m)(t)$, is computed as the difference between the traffic-state value of the upstream movement and the turning-ratio-weighted traffic-state values of its downstream movements:
    \begin{equation}
    w(l,m)(t)=z(l,m) (t)-\sum_{p\in\mathrm{Out}_m}r(m,p)\, z(m,p)(t)
    \label{eq:original_mp_weight}
    \end{equation}
    
    \noindent where $z(l,m)(t)$  and $z(m,p)(t)$ denote the traffic-state values of the upstream movement $(l,m)$ and downstream movement $(m,p)$ at time $t$, respectively. $r(m,p)$ is the average turning ratio from link $m$ to link $p$ and $\mathrm{Out}_{m}$ is the set of downstream links connected to link $m$. $r(m,p)=0$ if $m\in\pazocal{L}_{\mathrm{exit}}$. 

    \item \textbf{Compute pressure of each phase.} The pressure of phase $\phi$ at intersection $i$ at time $t$, $P_i^{\phi}(t)$, is obtained by summing the movement weights multiplied by their corresponding saturation flows, $c(l,m)$:
    \begin{equation}
    P_i^{\phi}(t)=\sum_{(l,m)\in \pazocal{M}_i^{\phi}}w(l,m)(t)\,c(l,m) \qquad \forall \phi\in\Phi_i
    \label{eq:original_mp_pressure}
    \end{equation}
    
    \noindent where $\pazocal{M}_i^{\phi}$ is the set of movements served by phase $\phi$ at intersection $i$, $c(l,m)$ is the saturation flow rate of movement $(l,m)$, and $\Phi_i$ is the set of all phases at intersection $i$.\\

    \item \textbf{Select the control action.} Finally, the phase with the highest pressure is selected and actuated for the next signal update interval:
    \begin{equation}
    \phi^{*}_i(t)=\arg\max_{\phi\in\Phi_i}P_i^{\phi}(t)
    \label{eq:original_mp_policy}
    \end{equation}
    
\end{enumerate}

\subsubsection{Choice of Signal Update Interval}

In the store-and-forward model, \autoref{eq:outflow} defines the discharge during one time step as the minimum of (i) saturation flow rate and (ii) the number of vehicles present on the link at the beginning of the step. Therefore, the time-step size or signal update interval should equal the free-flow travel time of the link. If the time-step exceeds the free-flow travel time, vehicles entering from the upstream link may join the queue and discharge within the same step, causing the discharge to be underestimated. On the other hand, if the time step is shorter than the free-flow travel time, vehicles near the upstream end of the link cannot reach the stop line within the current step, leading to an overestimation of the discharge \citep{liu_novel_2022}.

However, this assumption may not always be optimal for control performance in reality. Under low traffic demand, shorter green times are generally more desirable, whereas a time step equal to the free-flow travel time may be unnecessarily long and degrade performance.

\subsection{Maximum Stability Property of MP}

One of the most desirable properties of MP algorithms is \textit{maximum stability}. This property has also commonly been referred to as \textit{throughput optimality} in the literature \citep{levin_max-pressure_2019_AIM, levin_mp_2020_cyclic, wang_learning_2022, barman_throughput_2023}. The discussion in this subsection follows the stability-in-the-mean definition of \citet{varaiya_max_2013}, which forms the basis of the classical traffic-signal MP formulation. Under this definition, a queue-length process is stable in the mean when the time average of the expected total queue length is bounded uniformly in the horizon, and a control policy has the maximum stability property if it stabilizes the network for every demand that any admissible control can stabilize.

\begin{definition}[Feasible demand]
    \label{def:feasible_demand}
    \normalfont
Let $f=d \mathbf P$ denote the flow vector corresponding to the external demand vector $d$, where $\mathbf P$ depends only on the turning ratios; see \citet{varaiya_max_2013} for the derivation of $\mathbf P$ from flow conservation. Let 

\begin{equation}
co(\pazocal{S})=
\left\{
\sum_{S\in\pazocal{S}}\mu_S S
\;\middle|\;
\mu_S\ge0,\;
\sum_{S\in\pazocal{S}}\mu_S=1
\right\}
\end{equation}
denote the convex hull of the set of all feasible network control matrices. 
Define $D$ as the set of demand vectors $d$ for which there exists an average network control
matrix $\Sigma\in\operatorname{co}(\pazocal{S})$
satisfying $f_l r(l,m)\leq c(l,m)\Sigma(l,m)$
for all movements $(l,m)$. The interior of $D$,
denoted by $D^\circ$, consists of demand vectors
for which there exists
$\Sigma\in\operatorname{co}(\pazocal{S})$ such that

\begin{equation}
f_l r(l,m)<c(l,m)\Sigma(l,m),
\quad \forall(l,m).
\end{equation}

\noindent Following \citet{varaiya_max_2013},
$D^\circ$ is the set of feasible demands.
\end{definition}

\begin{definition}[Stability] \label{def:maximum_stability} \normalfont
The queue length process $X(t) = \{ x(l,m)(t) \}$ is stable in the mean, and the control policy is said to be stabilizing, if the long-run mean queue length is bounded, meaning it must be less than $K$, where $K < \infty$. This can be mathematically expressed as follows:
\begin{equation}
\limsup_{T\to\infty}
\frac{1}{T}
\sum_{t=1}^{T}\sum_{l,m}
\mathbb{E}[x(l,m)(t)]
< K,
\qquad K<\infty.
\label{eq:stability_def}
\end{equation}
\end{definition}

\begin{theorem}[Maximum Stability]  \label{theorem:maximum_stability} \normalfont
A signal control strategy has the maximum stability property if it can stabilize the network whenever $d = \{d_l\} \in D^\circ$. No control policy can stabilize the network when $d \notin D$.

\end{theorem}

The maximum stability property of MP control is typically proved using Lyapunov drift analysis. First, a Lyapunov function, commonly a quadratic function of network queue lengths, is defined to represent the overall level of congestion in the network. The drift of this function, which is the change of the function in one step, is then derived based on the queue dynamics. Next, it is shown that the MP policy minimizes an upper bound on the Lyapunov drift among all feasible signal control actions. Finally, for any feasible demand, the resulting expected drift can be shown to be bounded by a negative linear function of the number of vehicles in the network, and this negative-drift condition implies that the network queues remain stable. A simplified example of the maximum-stability proof, adapted from \citet{varaiya_max_2013}, is presented below. 

\begin{proof}

Let $S(t)$ be the network control matrix selected by
the MP policy. Following \citet[Appendix~A]{varaiya_max_2013}, for
$d\in D^{\circ}$ there exist constants $k_c<\infty$ and
$\varepsilon>0$ such that
\begin{equation}
\mathbb{E}\!\left[
|X(t+1)|^{2}-|X(t)|^{2}
\,\middle|\,X(t)
\right]
\le
k_c-\varepsilon |X(t)|,
\label{eq:drift}
\end{equation}
where $|X(t)|^{2}=\sum_{l,m}x^{2}(l,m)(t)$.

Taking unconditional expectations and summing over $t=1,\ldots,T$, the left-hand side telescopes to
\begin{equation}
\mathbb{E}|X(T+1)|^{2}
-
\mathbb{E}|X(1)|^{2}
\le
k_cT
-
\varepsilon
\sum_{t=1}^{T}\mathbb{E}|X(t)|.
\label{eq:telescoped}
\end{equation}

\noindent Rearranging \autoref{eq:telescoped} and dividing by
$\varepsilon T$ gives
\begin{equation}
\frac{1}{T}
\sum_{t=1}^{T}\mathbb{E}|X(t)|
\le
\frac{1}{\varepsilon}
\left[
k_c
+
\frac{1}{T}\mathbb{E}|X(1)|^{2}
-
\frac{1}{T}\mathbb{E}|X(T+1)|^{2}
\right]
\end{equation}

\noindent Here, $\mathbb{E}|X(T+1)|^{2}\ge0$. Therefore,

\begin{equation}
\frac{1}{T}\sum_{t=1}^{T}\mathbb{E}|X(t)|
\le
\frac{1}{\varepsilon}
\left[
k_c+\frac{1}{T}\mathbb{E}|X(1)|^{2}
\right],
\end{equation}

\noindent Since $X(1)$ is
bounded, $\mathbb{E}|X(1)|^{2}<\infty$, and thus
$\mathbb{E}|X(1)|^{2}/T\to0$ as $T\to\infty$. Therefore,
\begin{equation}
\limsup_{T\to\infty}
\frac{1}{T}\sum_{t=1}^{T}\mathbb{E}|X(t)|
\le
\frac{k_c}{\varepsilon}
<\infty.
\label{eq:X_bound}
\end{equation}

Since the network contains a finite number of queues,
\autoref{eq:X_bound} implies that the long-run time-average
expected total queue length is also bounded. Hence, there exists
a finite constant $K<\infty$ such that
\begin{equation}
\limsup_{T\to\infty}
\frac{1}{T}
\sum_{t=1}^{T}\sum_{l,m}
\mathbb{E}[x(l,m)(t)]
< K,
\qquad K<\infty.
\label{eq:stability_result}
\end{equation}

which is exactly \autoref{eq:stability_def}. Thus, the network is stable for every $d\in D^{\circ}$. Since MP
stabilizes the network for every demand vector in the feasible demand region, it satisfies the maximum-stability
property. See \citet{varaiya_max_2013} for the detailed proof.  

\end{proof}

\section{Vehicular-focused and intersection-level MP Control}\label{sec: local}

As shown in \Cref{subsec:mp_steps}, under the MP control framework, the choice of the traffic performance metric used in the weight definition plays a critical role in the effectiveness of MP control. This section reviews the evolution of MP algorithms, with a particular focus on advances in pressure formulation at the local intersection level and their implications for control performance.

\subsection{Number-of-vehicle-based MP}

Number-of-vehicle-based MP formulations are among the earliest and most widely studied MP approaches. However, a major limitation of the original MP \citep{varaiya_max_2013} is its reliance on point-queue models, which neglect the physical length of vehicles and finite link storage and therefore cannot explicitly prevent queue spillback. Two methods have been employed in the literature to address this issue.

\subsubsection{Capacity-aware MP}
One approach to addressing this limitation is to incorporate link queue capacity into the movement weight definition. This is typically done by replacing the actual queue length with a normalized queue length, where the corresponding queue capacity is used as the scaling factor. The core idea of this method is to assign higher priority to upstream links with smaller queue storage capacities and downstream links with larger storage capacities, as a given number of queued vehicles represents a higher level of congestion on a link with limited storage capacity than on one with a larger capacity.

\citet{kouvelas_maximum_2014} proposed a \textit{link-based} capacity-normalized MP formulation in which queue lengths are normalized by the corresponding link storage capacities. However, their method requires that all movements originating from the same link be served within the same phase, similar to a split-phasing configuration. Inspired by that idea, \citet{zoabi_enhanced_2026} proposed a movement-based capacity-normalized MP model.

This formulation replaces absolute queue lengths with normalized queue occupancies. Consequently, movements with fully occupied downstream links receive weights no more than zero. Thus, this model incorporates downstream storage availability into the MP decision and discourages discharge toward congested downstream movements.

As pointed out by \citet{gregoire_capacity-aware_2015}, when queue lengths are normalized by their corresponding queue capacities, an additional vehicle produces a larger normalized queue length in pressure on a lower-capacity road than on a higher-capacity road. This may introduce unfairness and lead to unnecessarily long green times being allocated to low-capacity roads, particularly under low-demand conditions. To address this issue, they proposed the following convex normalized pressure formulation:
\begin{equation}
P_l(t)
=
\min\left(
1,
\frac{
\dfrac{x_l(t)}{\Pi_\infty}
+
\left(2-\dfrac{x^{\text{max}}_l}{\Pi_\infty}\right)
\left(\dfrac{x_l(t)}{x^{\text{max}}_l}\right)^{\eta}
}{
1+
\left(\dfrac{x_l(t)}{x^{\text{\text{max}}}_l}\right)^{\eta-1}
}
\right)
\label{eq:convex_normalized_weight}
\end{equation}

\noindent where $P_l(t)$ is the pressure of link $l$ at time $t$, $x_l(t)$ is
the queue length on link $l$, $x^{\max}_l$ is its storage capacity,
and $\Pi_\infty$ and $\eta$ are parameters controlling the shape of the
pressure function. This formula is convex so that the marginal pressure generated by an additional vehicle increases as the road approaches its storage capacity, meaning the increase in pressure from an additional vehicle is higher when the link is more congested. Also, the pressure functions have the same slope at zero occupancy for all roads to ensure fairness under low traffic density. Based on \autoref{eq:convex_normalized_weight}, the weight associated with a movement
from the upstream link $l$ to the downstream link $m$ is:

\begin{equation}
w(l,m)(t)
=
\Upsilon_{lm}(t)
\max\left(
P_l(t)-P_m(t),0
\right),
\label{eq:gregoire_weight}
\end{equation}

\noindent where $\Upsilon_{lm}(t)$ is a binary variable indicating whether
vehicles are waiting at link $l$ to move toward link $m$. The authors also claimed that their method is \textit{work-conserving}, meaning that green time will not be allocated to a phase under which no vehicles can be served, resulting from either empty upstream links or completely jammed downstream links, as long as there is at least one vehicle that can be served by another phase.

\citet{hao_backpressure_2020} proposed an Extended Backpressure (EBP) algorithm that incorporates downstream link capacity into the queue-differential movement weight through a penalty term. A movement weight is defined as

\begin{equation}
w(l,m)(t)=\Delta q_{lm}(t)-V_{lm} \, h_b(l,m),
\label{eq:hao_weight}
\end{equation}

\noindent where the first term,
$\Delta q_{lm}(t)
=
\sum_{p\in\mathrm{Out}_m}
\left(x(l,m)(t)-x(m,p)(t)\right),$
represents the sum of queue-length differentials between the movement $(l,m)$ and all $(m,p)$. The second term incorporates the remaining capacity of the downstream link to reduce the risk of queue spillback. Specifically, $V_{lm}=\gamma_l (x^{max}_m-x_m)$, where $\gamma_l$ is a weighting coefficient and $x^{max}_m-x_m$ is the remaining storage capacity of downstream link $m$. The binary variable $h_b(l,m)\in\{-1,1\}$ is used to adjust the movement weight based on whether the remaining downstream capacity is sufficient to accommodate the upstream discharge. When the maximum transferable flow exceeds the remaining downstream capacity, $h_b(l,m)=1$, reducing the movement weight and thereby discouraging the controller from serving that movement. Otherwise, $h_b(l,m)=-1$. Using Lyapunov drift analysis, the authors showed that the proposed EBP maintains network stability, while simulation results demonstrated reduced queue spillback and improved operational performance.

\subsubsection{Vehicle position-aware MP} \label{sec:pwbp}
Another method that has been used to enhance queue-based MP control is to integrate vehicle positions along a link into pressure calculation. On an upstream link, a vehicle closer to the intersection represents a more immediate need to be served than a vehicle farther away. Similarly, on a downstream link, a vehicle closer to the upstream intersection has a greater influence on the receiving flows than one located farther downstream. Therefore, incorporating vehicle positions into the pressure definition can provide a more detailed representation of both upstream demand and downstream receiving conditions. \citet{li_position_2019} proposed the Position-Weighted Backpressure (PWBP) framework, which treats vehicle distribution as continuous in both temporal and spatial dimensions and explicitly incorporates the spatial distribution of vehicles by assigning greater importance to congestion located near the intersection. The movement weight is defined as:

\begin{equation}
w(l,m)(t) \equiv
\begin{cases}
\left|
\kappa_{l,m}\rho_l^m(t)
-
\displaystyle\int_{0}^{\ell_m}
\left|
\frac{\ell_m-y}{\ell_m}
\right|
\displaystyle\sum_{p\in\mathrm{Out}_m}
\kappa_{m,p}r(m,p)(t)\rho_m^p(y,t)\,dy
\right|,
& l\in\pazocal{L}_{\mathrm{src}},
\\[3ex]
\left|
\kappa_{l,m}
\displaystyle\int_{0}^{\ell_l}
\left|
\frac{y}{\ell_l}
\right|
\rho_l^m(y,t)\,dy
-
\displaystyle\int_{0}^{\ell_m}
\left|
\frac{\ell_m-y}{\ell_m}
\right|
\displaystyle\sum_{p\in\mathrm{Out}_m}
\kappa_{m,p}r(m,p)(t)\rho_m^p(y,t)\,dy
\right|,
& l\notin\pazocal{L}_{\mathrm{src}}.
\end{cases}
\label{eq:pwbp_weight}
\end{equation}

Here, $\rho_l^m(y,t)$ denotes the density at position $y$ on link $l$ 
of vehicles destined for downstream link $m$, $y$ is the distance from the upstream end of a link, $\ell_l$ and $\ell_m$ 
are the lengths of links $l$ and $m$, respectively, and $\pazocal{L}_{\mathrm{src}}$ denotes the set of fictitious source links. For source links, 
$\rho_l^m(t)$ represents the corresponding point-queue size. 
$\kappa_{l,m}$ is a nonnegative movement-specific weighting constant. 
For a physical upstream link, the factor $y/\ell_l$ assigns greater weight to vehicles located closer to the intersection. 
In contrast, the downstream factor $(\ell_m-y)/\ell_m$ assigns 
greater weight to vehicles near the entrance of the downstream link, where 
congestion is more likely to cause spillback. 
Another significant distinction of PWBP from previous MP models is that it considers finite queue capacities for internal links, while allowing unbounded queue accumulation on entry links. Under this setting, the authors proved that PWBP retains the maximum-stability property, i.e., the queue accumulation on entry links remains bounded.

Recognizing that PWBP requires continuous traffic density measurements along every link, which are impractical for real-world deployment, \citet{xu_approximate_2025} developed the Approximate Position-Weighted Backpressure (APWBP) controller. Instead of relying on continuous density observations, APWBP reconstructs traffic density using only upstream and downstream loop detector measurements together with a triangular flow--density relationship and kinematic wave theory. The authors proved that APWBP preserves the maximum-stability property while achieving nearly identical performance using substantially fewer sensors, although at the expense of a modest increase in computational complexity.

It should be noted that many other MP algorithms also define pressure based on the number of vehicles. However, their primary contributions lie in other aspects of MP control, such as extensions to multi-modal environments. Therefore, these studies are not included in this section and will instead be discussed in the corresponding sections that focus on their respective contributions.

\subsection{Travel Time-Based MP}

Although the number of vehicles provides a useful representation of the traffic state at a signal update instant, it essentially serves as a “snapshot” of the system and may not capture the full extent of traffic conditions experienced over time. For example, the number of vehicles alone does not reflect the travel time or delay experienced by vehicles on each movement, while such information is often important for traffic signal control. To capture a higher resolution of traffic conditions, MP algorithms based on other traffic state variables, such as travel time and travel delay, have also been developed. \citet{mercader_max-pressure_2020} introduced a link-based travel-time MP (TT-MP) that replaces queue-length-based weights with normalized travel-time weights within a cyclic-phase-control structure. The weight, $w_l(n)$, associated with incoming link $l$ is defined as:

\begin{equation}
w_l(n)
=
\frac{TT_l(n)}{TT_{\mathrm{ff},l}}
-
\sum_{m\in Out_l}
r(l,m)(n)
\frac{TT_m(n)}{TT_{\mathrm{ff},m}}
\label{eq:mercader_tt_weight}
\end{equation}

\noindent where $TT_l(n)$ is the average travel time on link $l$ during
cycle $n$, $TT_{\mathrm{ff},l}$ is the free-flow travel time of link $l$, $TT_m(n)$ is the average travel time on downstream link $m$
during cycle $n$, and $TT_{\mathrm{ff},m}$ is the free-flow travel time of
downstream link $m$. The controller maintains a fixed cycle length and a predefined cyclic phase sequence, while the available effective green time is distributed among the phases in proportion to their pressures, where each phase is guaranteed a minimum green time. Simulations show fewer unstable runs but longer travel times, while a field trial reports a 4\% travel-time reduction relative to optimized fixed-time control.

\citet{agarwal_fusing_2024} extended the practical applicability of TT-MP by integrating crowdsourced traffic data into a low-cost wireless adaptive traffic signal control architecture. While \citet{mercader_max-pressure_2020} estimated travel times using Bluetooth and loop-detector data, \citet{agarwal_fusing_2024} obtained travel-time information directly from crowdsourced routing services from Google Maps, HERE, and TomTom and ultimately selected the HERE Routing API for their proposed method. Beyond the MP implementation, the study's main contribution is a wireless control architecture using a Raspberry Pi and ESP32 controllers, together with fail-safe operation and crowdsourced-data-based emergency vehicle pre-emption. Simulation results showed that the proposed approach reduced total delay relative to both actuated control and queue-based MP.

\citet{maripini_queue_2025} proposed a queue dissipation time-based MP controller, in which phase pressure is defined as the difference between the queue dissipation times of the upstream phase and its corresponding downstream phase. The queue dissipation time is estimated from sparse vehicle re-identification travel-time data using the associated vehicle delays and downstream detection timestamps and represents the time required to discharge a queued phase. The proposed method adopts a cyclic MP structure with a predefined phase sequence, where the effective green time for the next cycle is allocated among phases in proportion to their queue-dissipation-time-based pressures, subject to minimum-green constraints.

\citet{tan_cv-mp_2026} proposed a non-cyclic connected-vehicle-based 
MP (CV-MP) controller that uses cumulative link travel time from 
connected vehicles (CVs) as the metric to define movement weight. For each CV $v$, the normalized real-time link travel time is calculated as

\begin{equation}
\tau_v(t)
=
\frac{t-t_v^{0}}{TT_{\mathrm{ff},(l,m)}}
=
\frac{1}{TT_{\mathrm{ff},(l,m)}}
\left(
\frac{y_v(t)}{v^{\mathrm{ff}}_v}
+
D_v(t)
\right)
\label{eq:tan_normalized_tt}
\end{equation}

\noindent where $t_v^{0}$ is the time at which CV $v$ entered the link, 
$TT_{\mathrm{ff},(l,m)}$ is the constant free-flow travel time of movement 
$(l,m)$ on link $l$, $y_v(t)$ is the distance traveled by the vehicle 
after entering the link, $v^{\mathrm{ff}}_v$ is its average free-flow speed, and 
$D_v(t)$ is its cumulative delay. Then, the sum of the normalized link travel time of all vehicles from the movement is used as the metric for weight calculation. The maximum stability property was proved under both fully and partially connected CV environments, although heterogeneous CV penetration can reduce the admissible demand region. Simulations showed that CV-MP generally reduced delay and spillover compared to other MP variants, while performance degraded at very low or highly heterogeneous CV penetration rates.

\subsection{Delay-Based MP}

Travel delay-based MP algorithms have also been developed. \citet{wu_delay-based_2018} first introduced a delay-based MP model for isolated intersections without considering downstream traffic. The proposed controller uses the sojourn time, $D_{l,m}^{\mathrm{HOL}}(t)$, of the Head-Of-Line (HOL) vehicle in movement $(l,m)$, which is measured from the time that vehicle entered the network, to define phase pressure:

\begin{equation}
P_i^{\phi}(t)=\sum_{(l,m)\in \pazocal{M}_i^{\phi}} \gamma_{l,m}  D_{l,m}^{\mathrm{HOL}}(t)\,c(l,m) \qquad \forall \phi\in\Phi_i
\label{eq:wu_delay_based}
\end{equation}

\noindent where $\gamma_{l,m}$ is a movement-specific weighting coefficient. The authors also proposed a weighted BP controller that combines the HOL vehicle sojourn time and movement queue length through separate weighting parameters, allowing the controller to balance delay fairness and queue-length performance.

\citet{dixit_simple_2020} proposed a crowdsourced, cycle-based delay control method inspired by MP control to reduce reliance on physical traffic detectors. The pressure is defined using the maximum delay-to-saturation-flow ratio among its served movements, and the available effective green time is allocated among the predefined phases in proportion to these pressures while preserving the original cycle length and phase sequence. Field implementation at seven intersections in India and Indonesia generally reduced approach delays relative to fixed-time control. The controller does not include downstream congestion in its pressure calculation and therefore lacks the upstream-downstream differential structure of conventional network MP.

Later studies progressively incorporated richer delay information. \citet{liu_novel_2022} proposed D-MP, which uses the delay incurred during the previous time step for computing movement weights. Since maximum stability is inherently characterized in terms of the number of vehicles in the network, this study established a connection between travel delay and the number of vehicles to prove the maximum-stability property of D-MP. Specifically, they assume that vehicles are either traveling at the free-flow speed or stopped with a speed of zero, and then the incurred delay at each time point is equivalent to the number of stopped vehicles. Based on this relationship, the maximum stability of D-MP was established. One limitation of D-MP is that low-demand movements may experience excessively large delays because delays accumulated before the previous time step are not reflected in the movement weight. To address this issue and improve delay equity, \citet{liu_total-delay-based_2023} extended D-MP by replacing the delay incurred during the previous time step in the upstream term with the cumulative delay experienced by all vehicles currently on the link since they entered the link. This modification was applied only to the upstream term, while the original time-step-based delay was retained for the downstream term. This is because the downstream term primarily represents the instantaneous congestion level and receiving condition at the signal update instant; therefore, incorporating historically accumulated delay into this term is neither necessary nor appropriate for representing the current downstream traffic state.

\citet{zoabi_max-pressure_2024} further extended delay-based MP control by proposing the Max-Pressure based on Sum of Travel Time Delays (MP-STTD) formulation. MP-STTD applies total movement delay to both the upstream and downstream terms. Rather than directly summing individual vehicle delays, the authors developed a new queuing-theory-based total movement delay model using cumulative arrival, actual departure, and virtual free-flow departure curves. The resulting controller therefore extends the cumulative-delay concept to the complete upstream-downstream pressure structure while explicitly targeting total delay minimization.

More recently, \citet{zoabi_experimental_2025} proposed the MP based on Travel Time Delays (MP-TTD) controller, which extends delay-based MP control by using average lane-level travel-time delay rather than the total or cumulative delay measures adopted in D-MP and TD-MP. For each upstream lane $l_n,l$ on link $l$, the weight is defined as:

\begin{equation}
w_{l_n,l}(t)
=
\sqrt[\beta]{
\frac{D_{l_n,l}(t)}
{TT_{\mathrm{ff},l_n,l}}
}
-
\sum_{m \in Out_l}
\sum_{l_n\in m}
r(l_n,l,l_n,m)(t)
\sqrt[\beta]{
\frac{D_{l_n,m}(t)}
{TT_{\mathrm{ff},l_n,m}}
}
\label{eq:zoabi_mpttd_weight}
\end{equation}

\noindent where the average travel-time delay is calculated as:

\begin{equation}
D_{l_n,l}(t)
=
\max
\left(
TT_{l_n,l}(t)
-
TT_{\mathrm{ff},l_n,l},
0
\right),
\label{eq:zoabi_mpttd_delay}
\end{equation}

\noindent where $TT_{l_n,l}(t)$ is the average observed travel time on lane $l_n,l$ of link $l$ during the corresponding control period, $TT_{\mathrm{ff},l_n,l}$ is the free-flow travel time of that lane, $r(l_n,l,l_n,m)(t)$ is the proportion of vehicles departing from lane $(l_n,l)$ of upstream link $l$ that enter lane $(l_n,m)$ of downstream link $m$, and $\beta$ is a fitting parameter.

A key improvement over the TT-MP formulation of 
\citet{mercader_max-pressure_2020} is the treatment of the relationship 
between travel time and queue occupancy. TT-MP directly substitutes 
normalized travel time, $TT/TT_{\mathrm{ff}}$, for normalized queue 
occupancy, effectively assuming a linear relationship between the two. 
In contrast, MP-TTD uses the nonlinear transformation 
$\sqrt[\beta]{D/TT_{\mathrm{ff}}}$, which the 
authors argue provides a more accurate representation under congested 
conditions. Moreover, whereas TT-MP is formulated at the link level, 
MP-TTD calculates pressures at the lane level, allowing traffic states 
associated with different lane and movement configurations to be 
represented more explicitly. The controller also adopts a cyclic fixed-phase-sequence structure and is developed for both fixed and time-varying cycle-length schemes.

\subsection{Transformed State Variables}
Moreover, several MP variants employ alternative traffic-state variables to better represent specific control objectives and traffic conditions. However, these state variables are typically derived from the number of vehicles through various transformations. 

\paragraph{Environmentally weighted MP}
To address the environmental impact of traffic signal control, \citet{riehl_green-pressure_2025} proposed Green-Pressure, a sustainability-oriented extension of queue-based MP control that incorporates vehicle-category information into the pressure calculation. Instead of treating every queued vehicle equally, the controller defines link pressure as a weighted queue length, where passenger cars, motorcycles, transporters, trucks, and buses are assigned different weights according to their vehicle class. The vehicle-class weights were optimized offline using a stochastic hill-climbing algorithm to minimize a weighted combination of CO$_2$, CO, HC, NO$_x$, and particulate-matter emissions, resulting in substantially larger weights for trucks and buses than for passenger cars. Simulation results showed that Green-Pressure reduced total Air Quality Index (AQI)-weighted emissions and total travel time relative to the corresponding unweighted MP controller, primarily by reducing the delay and waiting time of heavier and more polluting vehicles. The authors noted that systematically prioritizing high-emission vehicles could create undesirable long-term incentives by effectively providing better service to more-polluting vehicle types.

\paragraph{Arrival-Based MP}
The original MP algorithm determines signal phases based on queue differences and does not explicitly account for real-time arrival flows, which may limit its responsiveness under rapidly changing demand. To address this limitation, \citet{gao_arrival-based_2022} proposed an Arrival-Based Backpressure (ABP) controller that explicitly incorporates real-time movement-level arrival flows into the phase evaluation. The proposed method uses the differential queue weight, $w(l,m)(t) = x_l(t) - x_m(t)$ for a movement from upstream link $l$ to downstream link $m$, and modifies the movement pressure by combining the saturation flow, $c(l,m)$, with the observed arrival flow, $A(l,m)(t)$. Accordingly, the pressure of a feasible phase $\phi$ is calculated as

\begin{equation}
P_i^{\phi}(t)=\sum_{(l,m)\in \pazocal{M}_i^{\phi}}w(l,m)(t)
\left[c(l,m)+A(l,m)(t)\right].
\end{equation}

This formulation combines the conventional queue-based feedback term with an arrival-based feed-forward term, allowing the controller to respond more effectively to fluctuating upstream demand. The authors showed through Lyapunov analysis that ABP preserves the stability and maximum-throughput properties of BP control, and simulations demonstrated improved throughput, queue length, and delay, particularly under highly fluctuating arrival patterns.

\subsection{Note on Asymmetric Upstream and Downstream Traffic-State Formulations}

Although conventional MP formulations generally use the same form of traffic-state measure for the upstream and downstream components of movement weight, this symmetry is not required, and several extensions formulate the two components differently to reflect their distinct roles. The upstream term generally represents the demand or urgency for service, whereas the downstream term represents the receiving condition or available supply. For example, TD-MP \citep{liu_total-delay-based_2023} uses the cumulative delay experienced by vehicles currently on the upstream movement since entering the link, while retaining the delay incurred during the previous time step for the downstream term. This allows accumulated delay to influence service priority while using a more instantaneous measure to represent downstream congestion. PWBP \citep{li_position_2019} introduces a different form of asymmetry through spatial weighting: vehicles near the downstream end of the upstream link receive greater weight, whereas vehicles near the upstream end of the downstream link receive greater weight, emphasizing traffic that is most relevant to immediate service demand and downstream receiving conditions. Similarly, Transit-MP \citep{tan_transit-mp_2027} uses passenger-occupancy-weighted normalized link travel time for the upstream traffic state but normalized link travel time without passenger-occupancy weighting for the downstream state, because occupancy is relevant to transit service priority but not to the physical vehicle-storage capacity of the receiving link. These studies demonstrate that the upstream and downstream components of MP weight can be designed differently to better represent their respective demand and supply functions.

\section{Control Structure and Implementation Extensions}\label{sec: structure extension}
In addition to extensions in metric selection and transformation for weight and pressure calculation, considerable efforts have also been made to improve the control structure and practical implementation of MP algorithms. This section reviews extensions related to phase selection and signal timing, including cyclic MP frameworks and methods that account for phase-switching lost time. In addition, traditional MP algorithms typically rely on average values for important input parameters, such as saturation flow rates and turning ratios, which may not adequately capture their temporal variations under real-world traffic conditions. Moreover, the requirement for movement-level traffic state information can be particularly restrictive at intersections with shared lanes. Efforts aimed at addressing these issues are also reviewed in this section.

\subsection{Cyclic MP}

In time-step-based MP algorithms, no predefined phase sequence is imposed, and low-demand phases are not guaranteed a minimum service time within a given time period, which may create operational and safety concerns, particularly during green-to-red transitions where unexpected signal changes may require abrupt braking. When the time step is small, frequent phase changes can also increase switching losses and reduce intersection efficiency. To address these limitations and better align MP control with conventional signal operation, researchers have developed cyclic MP formulations in which phases follow a predefined sequence while green-time allocation remains pressure-responsive. These approaches can be broadly classified into two categories: fixed-cycle MP and adaptive-cycle MP.

\subsubsection{Fixed-cycle MP}

\citet{kouvelas_maximum_2014} proposed a fixed-cycle MP controller in which the green split of each phase is updated according to its pressure at the beginning of every cycle. After assigning the minimum green time to each phase, the remaining effective green is then distributed among the phases in proportion to their nonnegative pressures. This cyclic MP formulation updates the green splits once per cycle while preserving a minimum green time for every phase. \citet{mercader_max-pressure_2020} also adopted the same proportional green time allocation method of  \citet{kouvelas_maximum_2014} in their TT-MP. Instead of allocating green time proportionally to phase pressures, \citet{pumir_stability_2015} simply assign the minimum green time to all phases except the highest-pressure phase and allocate the remaining green time to the latter. The control performance of this algorithm was extensively evaluated through numerical simulations \citep{anderson_stability_2018}.

\citet{le_decentralized_2015} used a different approach to allocate service time within each fixed cycle. Instead of allocating green time directly in proportion to the phase pressure, they used an exponential softmax function of the phase weights to determine the green-time proportions as follows:

\begin{equation}
\lambda_i^{\phi}(n)
=
\frac{
\exp\left(\theta P_i^{\phi}(n)\right)
}{
\displaystyle\sum_{\phi' \in \Phi_i}
\exp\left(\theta P_i^{\phi'}(n)\right)
}
\label{eq:le_green_split}
\end{equation}

where  $\theta>0$ is a tuning parameter, regulating how strongly the algorithm favors high-pressure phases. As $\theta \rightarrow 0$, the allocation approaches an equal split among the phases, whereas as $\theta \rightarrow \infty$, nearly all available service time is assigned to the highest-pressure phase or phases, approaching the conventional MP decision. The authors proved the maximum stability property for their cyclic MP structure. However, unlike the previous cyclic MP models, \citet{le_decentralized_2015} did not impose an explicit minimum green time for each phase.

\subsubsection{Adaptive-Cycle MP}
Although earlier cyclic MP frameworks improved practical applicability by enforcing a prescribed phase sequence and minimum green time, their fixed cycle lengths limit responsiveness to traffic fluctuations occurring within a cycle. In particular, when phase durations are determined at the beginning of the cycle, newly arriving vehicles cannot immediately influence the signal timing, while unnecessarily long fixed cycles can increase waiting times when demand is relatively low. To address these limitations, \citet{levin_mp_2020_cyclic} proposed a cyclic MP controller with an adaptive cycle length formulated as a rolling-horizon model predictive control (MPC) problem. This method has a fixed phase sequence and minimum green time. Rather than imposing a fixed cycle duration, they imposed an upper bound for the cycle length. The signal decisions are obtained by maximizing the pressure-weighted service over a prediction horizon $H$:

\begin{equation}
\max
\frac{1}{H}
\sum_{h=0}^{H-1}
\sum_{(l,m)\in \pazocal{M}}
w(l,m)(t) \, c(l,m)\, S(l,m)(t+h)
\left[1-L_{l,m}(t+h)\right]
\label{eq:levin_mpc}
\end{equation}

\noindent where $h$ is the look-ahead time-step index, $H$ is the prediction horizon, and $L_{l,m}(t+h)$ represents the lost-time fraction associated with phase switching. The optimization is subject to the cyclic phase-order, one-phase-at-a-time, minimum-phase-service, lost-time, and maximum-cycle-length constraints. The authors show that a prediction horizon equal to one maximum cycle length is sufficient. At time $t$, the controller optimizes the phase sequence over the future horizon, implements only the first control action, and discards the remaining planned actions. The authors proved that the proposed optimization algorithm has a decentralized solution: the optimal solution at one intersection is independent of the solutions at other intersections.

\citet{barman_performance_2022} extended the work of \citet{levin_mp_2020_cyclic} to improve control performance by reducing wasted green time in two ways. First, if no vehicles remained to be served by the current phase, the controller advanced to the next phase before the regular MP decision instant. Second, if no vehicles were waiting to be served by the next phase, that phase was skipped, and the controller continued through the cyclic sequence until a phase with waiting demand was identified. Simulation results comparing acyclic MP, cyclic MP without phase skipping (CMP-NS), and the proposed cyclic MP with phase skipping (CMP-S) showed that acyclic MP generally achieved the lowest average delay, primarily because it generated the most flexible control without a predefined cyclic phase sequence. Nevertheless, CMP-S generally outperformed CMP-NS by avoiding unnecessary service to empty phases. CMP-S was also shown to reduce the worst-lane delay but with more frequent phase changes. The study further demonstrated that MP performance was sensitive to the decision interval and maximum cycle length, with no single combination of these parameters consistently performing best across all intersections and demand levels.

To maintain the control flexibility of acyclic MP while avoiding excessive delays for low-demand phases, \citet{robbennolt_microsimulation_2022} introduced a "semi-cyclic" MP controller by imposing a maximum interval between successive phase activations on a time step-based MP. Under normal conditions, the proposed controller operates similarly to conventional acyclic MP, selecting the phase with the highest pressure at each decision instant. However, a maximum allowable number of decision steps between successive activations is imposed for each phase, determined based on a global multiplier and the number of phases at the intersection. Even though the phase sequence is not fixed, this mechanism prevents low-pressure phases from remaining unserved for excessively long periods. Simulation results in SUMO showed that the proposed method reduced travel delay compared with cyclic MP, while mitigating the excessive delays experienced by low-demand phases under acyclic MP.

Overall, the existing studies generally suggest that, although cyclic MP preserves the conventional cyclic operation of traffic signals and is therefore more consistent with existing signal control practices, which can mitigate safety concerns, time-step-based acyclic MP generally achieves higher operational efficiency due to its greater flexibility in phase selection \citep{levin_mp_2020_cyclic, sun_simulation_2018}. 

\subsection{Consideration of Phase Switching Lost Time}

Phase switching loss is the unavoidable time interval during which no vehicle can effectively pass through an intersection when the signal transitions from one phase to another. In principle, switching loss should affect signal timing decisions because the currently active phase can utilize the entire time step if it continues to be served, whereas a newly selected phase can utilize only the remaining portion of the time step after accounting for the phase-transition time. However, early MP algorithms do not explicitly account for this difference in effective green time in their control formulation.

Cyclic MP algorithms with fixed cycle length \citep{kouvelas_maximum_2014, mercader_max-pressure_2020, pumir_stability_2015} account for lost time by subtracting the total lost time per cycle from the cycle length to determine the total time available for green allocation.

For time-step-based MP control structure, \citet{hsieh_throughput-optimal_2017} proposed a Biased MP (B-MP) controller to account explicitly for phase-switch loss. Let $e$ be the lost time associated with a phase switch, during which all movements at the intersection are prohibited, and the throughput is zero. To reduce the total loss caused by frequent switching, B-MP introduces a bias in favor of the currently active phase. Let $P_i^{\mathrm{cur}}(t)$ denote the pressure of the phase currently active at intersection $i$ at time $t$, and let $P_i^{\max}(t)$ denote the largest pressure among all admissible phases at that intersection at the same time. The controller changes phase only when
\begin{equation}\label{eq:bias}
\left(1+B_i(t_i^{sw})\right)
\left[P_i^{\mathrm{cur}}(t)\right]^+
<
\left[P_i^{\max}(t)\right]^+,
\end{equation}
where $t_i^{sw}$ is the time of the most recent phase switch event at that intersection, and $B_i(t_i^{sw})$ is the switching-bias factor evaluated at $t_i^{sw}$. The bias is defined as
\begin{equation}
B_i(t_i^{sw})
=
\zeta \, e
\min
\left\{
1,
\left[
\left(
\sum_{(l,m)\in M_i} w(l,m)(t_i^{sw})
\right)^+
\right]^{-\alpha_{b}}
\right\},
\end{equation}
where $\zeta>0$ is a tuning parameter, $M_i$ is the set of traffic movements at intersection $i$, and $\alpha_b\in(0,1)$ is a parameter controlling how the bias varies with pressure. Because $B_i(t_i^{sw})$ is proportional to $e$, a larger phase-switch loss requires a larger pressure advantage before the controller leaves the current phase. Therefore, B-MP does not shorten the clearance time associated with an individual switch; rather, it reduces the frequency of phase changes and hence the accumulated switch-over loss.  The authors prove that it remains throughput-optimal in the presence of switch-over delay. \citet{sun_max-pressure_2025} proposed a similar method to offer priority to the currently activated phase over idle phases, in which the bias parameter in \autoref{eq:bias} is a pre-defined constant.

\citet{liu_novel_2022} and \citet{liu_total-delay-based_2023} used a similar method to explicitly incorporate phase-switch lost time by reducing the effective saturation flow of an idle phase. Let $\phi$ denote the candidate phase being considered and $\phi_i^{\mathrm{cur}}(t)$ denote the currently active phase at intersection $i$. The effective saturation flow of movement $(l,m)$ can be expressed as

\begin{equation}
c_{\mathrm{eff}}(l,m)
=
\begin{cases}
c(l,m), &\phi =\phi_i^{\mathrm{cur}}(t), \\[4pt]
c(l,m)\dfrac{\Delta t-e}{\Delta t}, &\phi \neq \phi_i^{\mathrm{cur}}(t),
\end{cases}
\label{eq:switch_loss_saturation}
\end{equation}

\noindent where $c_{\mathrm{eff}}(l,m)$ is the effective saturation flow used when evaluating movement $(l,m)$, $c(l,m)$ is the actual saturation flow of movement $(l,m)$, and $\Delta t$ is the signal-update time-step length. Consequently, the pressure of a phase requiring a switch is reduced relative to that of the currently active phase, thereby accounting for the capacity lost during the phase transition and discouraging unnecessary phase switching. Note that switch lost time is considered only in their numerical analysis rather than in their theoretical modeling.

Subsequent research shifted from simply reducing switching frequency to developing adaptive switching mechanisms that respond to real-time traffic conditions. \citet{wang_learning_2022} extended MP control by explicitly modeling switching loss, dividing each movement’s operation into discharge intervals, during which vehicles can pass, and switching intervals, during which no vehicles are served. They proposed the Switching-Curve-based Max Pressure (SCMP) controller, which maintains the current phase until the pressure gain from switching exceeds a threshold defined by a switching curve that is a monotonically increasing sublinear function of the total queue length. SCMP was proven to be throughput-optimal under switching loss. To better capture realistic traffic dynamics, the authors further developed Extended-SCMP (ESCMP), which incorporates distributed switching and position-weighted pressure, and Learned-ESCMP (LESCMP), in which the weight curve and switching curve are optimized using Proximal Policy Optimization. SUMO simulations showed that, compared with PWBP, ESCMP reduced total delay by approximately 10.3\%, while LESCMP reduced total delay by approximately 44.1\% and maintained strong performance across different traffic demand levels.

\subsection{Other MP Extensions}

\subsubsection{Group-Based MP}
The cyclic and acyclic MP controllers reviewed above generally rely on predefined phase configurations, in which the sets of movements served simultaneously are fixed. Departing from this framework, \citet{ghosh_pascal_2026} proposed PASCAL, a group-based MP controller that dynamically selects compatible movement groups without relying on predefined phases. At the end of each dynamic time slot, the highest-pressure eligible movement is selected as the critical movement from a critical movement selection pool, followed by the selection of a compatible complementary movement set that maximizes the combined pressure among predefined candidate sets. The selected group is activated for a dynamically determined green duration, and all served movements are removed from the pool for subsequent critical-movement selection. The pool is reset when empty or when no remaining movement has positive pressure. Simulation results showed reduced delays and queue lengths compared with selected cyclic, acyclic, and capacity-aware MP benchmarks.

\subsubsection{Lane-Based MP and Shared-Lane Modeling}

Most MP algorithms are movement-based and treat each movement independently, without explicitly accounting for interactions between movements from shared lanes. Also, multiple lanes serving the same movement may have different vehicle counts and discharge rates, but traditional MP algorithms typically assume that saturation flow is proportional to the number of lanes, without accounting for uneven traffic conditions across lanes. To address these limitations, \citet{cheng_lane-based_2024} proposed a lane-based MP formulation in which the queue state, discharge rate, and downstream relationships are represented separately for each lane. Specifically, the weight of lane $l_n,l$ of link $l$ is defined as
\begin{equation}
w_{l_n,l}(t) = x_{l_n,l}(t) - \sum_{{l_n,m}\in \mathrm{Out_{l_n,l}}}
x_{l_n,m}(t) r({l_n,l}, {l_n,m})(t)
\end{equation}
where $w_{l_n,l}(t)$ is the weight of lane $l_n,l$ at time $t$, $ x_{l_n,l}(t)$ is the number of vehicles in this lane.  Thus, unlike movement-based MP, the proposed controller accounts for lane-specific vehicle counts, discharge rates, and turning relationships when selecting the maximum-pressure signal state. The authors further derived a Lyapunov-drift-based stability result for the lane-based formulation and showed numerically that its benefits become more pronounced when lanes have heterogeneous discharge rates or turning patterns, particularly under oversaturated conditions.

\citet{zoabi_enhanced_2026} explicitly considered shared lanes, for which the conventional movement-based representation may inaccurately represent different turning movements using the same physical lane as separate movement queues. They proposed a lane-structured MP formulation in which the traffic state and saturation flow are represented separately for each physical lane rather than for each movement. A shared lane containing multiple movements is therefore represented by a single lane queue, while lane-to-lane turning ratios describe the proportions of vehicles traveling from that lane to the possible downstream lanes. Consequently, the weight of an upstream lane is calculated using its queue length and the turning-ratio-weighted queue lengths of the relevant downstream lanes. This formulation allows multiple movements sharing the same lane to be represented through their common physical queue while retaining their different downstream destinations. The authors also established a stability guarantee for the lane-structured formulation.

\subsubsection{Parameter Specification}

\begin{figure}[htbp]
	\centering
	\includegraphics[width=0.45\textwidth]{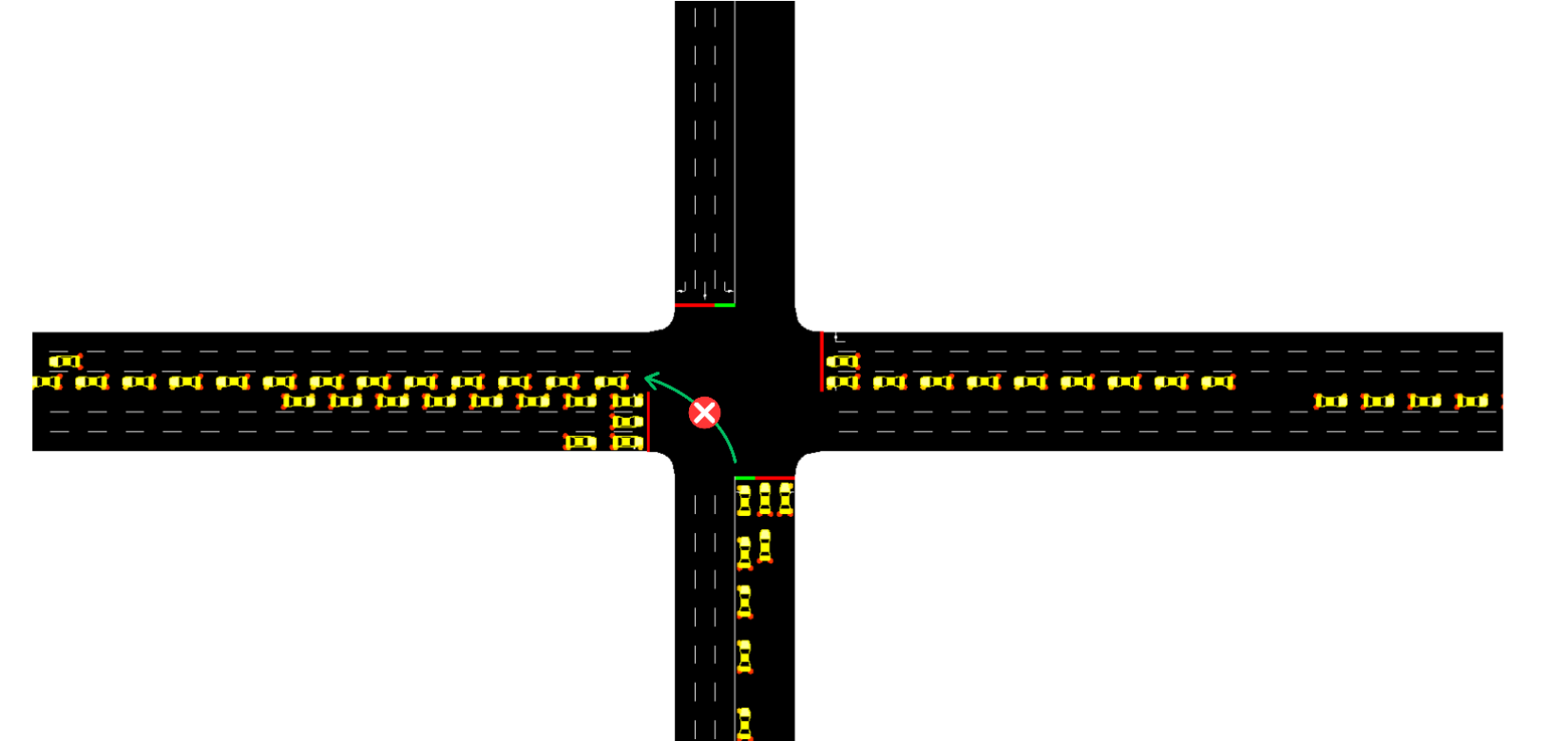}
	\caption{ Upstream blockage from a jammed downstream lane despite available space in other lanes.}
	\label{fig:spillback}
\end{figure}

Average turning ratios and average saturation flow rates were used in weight (\autoref{eq:original_mp_weight}) and pressure calculation (\autoref{eq:original_mp_pressure}) for many MP algorithms despite their variation under real-time traffic conditions. \citet{varaiya_max_2013} introduced an adaptive MP, in which estimated turning ratios and saturation flow rates replace their nominal values, and established stability when these estimates converge to the true parameters. \citet{gregoire_back-pressure_2014} proposed a BP controller that does not require turning ratios or movement-specific queue lengths, relying instead on aggregated queue lengths and stop-line detectors. More recently, \citet{lin_unveiling_2025a,lin_unveiling_2025b} investigated the influence of imminent saturation flow rates on MP control and showed that more accurate information about near-term service capacity can enlarge the network capacity region. In addition to saturation flow, \citet{zoabi_enhanced_2026} showed that conventional turning-ratio calculations may inaccurately represent downstream phase demand because they neglect the upstream origins of vehicles and proposed a revised turning-ratio update mechanism. This impact can be amplified under congested conditions, particularly when vehicle distributions are highly uneven across downstream lanes. An illustrative example from our simulation is shown in \autoref{fig:spillback}, where the leading vehicle of the northbound left-turn movement intends to enter the leftmost lane of the downstream link. However, because this lane is congested, the vehicle cannot proceed and consequently blocks all following vehicles. When average turning ratios are used to characterize the downstream traffic state, the severe congestion in the leftmost lane may be masked by the relatively uncongested conditions in the other downstream lanes. As a result, the phase may still be assigned a high pressure, as illustrated in the figure, even though its discharge is effectively blocked. This issue can be mitigated by using real-time turning ratios, which place greater emphasis on the traffic state of the leftmost downstream lane in this example. A distinct form of blocking occurs when vehicles belonging to one movement occupy lanes serving other movements, obstructing their discharge. Such lane occupancy can also complicate movement-specific queue estimation.

Both \Cref{sec: local} and \Cref{sec: structure extension} focus on vehicular-focused and intersection-level MP algorithms. To facilitate comparison across the studies reviewed in these two sections, \autoref{tab:mp_studies_comparison} summarizes the key features of selected MP algorithms, including their weight metrics, signal update structures (time-step-based or cyclic), maximum stability properties, and data requirements.

\begin{landscape}
\begingroup
\setlength{\tabcolsep}{1pt}
\renewcommand{\arraystretch}{1.15}
\fontsize{10}{12}\selectfont
\begin{longtable}{@{} >{\raggedright\arraybackslash}p{0.10\linewidth} >{\raggedright\arraybackslash}p{0.09\linewidth} >{\raggedright\arraybackslash}p{0.14\linewidth} >{\raggedright\arraybackslash}p{0.13\linewidth} >{\raggedright\arraybackslash}p{0.10\linewidth} >{\raggedright\arraybackslash}p{0.13\linewidth} >{\raggedright\arraybackslash}p{0.09\linewidth} >{\raggedright\arraybackslash}p{0.17\linewidth} @{}}

\caption{Comparison of Vehicular-Focused and Intersection-Level MP Studies}
\label{tab:mp_studies_comparison}\\
\toprule
\textbf{Paper} & \textbf{Category} & \textbf{Weight Metric} & \textbf{Time-step-based or Cyclic} & \textbf{Theoretical Stability Guarantee} & \textbf{Treatment of Negative Weights} & \textbf{Supporting Evidence} & \textbf{Data Requirements} \\
\midrule
\endfirsthead

\multicolumn{8}{l}{\textit{Continued from previous page}} \\
\toprule
\textbf{Paper} & \textbf{Category} & \textbf{Weight Metric} & \textbf{Time-step-based or Cyclic} & \textbf{Theoretical Stability Guarantee} & \textbf{Treatment of Negative Weights} & \textbf{Supporting Evidence} & \textbf{Data Requirements} \\
\midrule
\endhead

\midrule
\multicolumn{8}{r}{\textit{Continued on next page}} \\
\endfoot

\bottomrule
\endlastfoot

\citet{mercader_max-pressure_2020}
& Travel time-based
& Normalized travel time
& Fixed-cycle
& Not proven
& Link weights may be negative, but phase pressure is always non-negative.
& Simulation and field test
&Link travel times, free-flow travel times, turning ratios, and saturation flows\\

\citet{wu_delay-based_2018}
& Delay-based
& Sojourn time of the Head-of-Line (HOL) vehicle
& Time-step-based
& Yes
& Not applicable; isolated intersection
& Theoretical proof and simulation
&Movement-level HOL vehicle sojourn times, saturation flows\\

\citet{liu_novel_2022}
& Delay-based
& Total traffic delay, measured over the previous time step
& Time-step-based
& Yes
& Weight and pressure can both be negative
& Theoretical proof and simulation
& Real-time movement-level delay; turning ratios; saturation flows\\

\citet{liu_total-delay-based_2023}
& Delay-based
& Upstream cumulative delay of current vehicles since entering the link; downstream delay from previous time step
& Time-step-based
& Not proven
& Weight and pressure both can be negative
& Simulation only
&Per-vehicle link entry/exit times; turning ratios; saturation flows\\

\citet{zoabi_max-pressure_2024}
& Delay-based
& Total movement delays, accumulated until vehicles leave the links
& Time-step-based
& Not proven
& Weight and pressure both can be negative
& Simulation only
& Movement-level cumulative arrivals and departures; turning ratios; saturation flows; free-flow travel times\\

\citet{zoabi_experimental_2025}
& Delay-based
& Nonlinear normalized lane-level travel-time-delay difference between upstream and downstream
& Cyclic. Includes fixed-cycle and adaptive bounded/\allowbreak unbounded cycle length
& Not proven
& Lane weights may be negative; negative phase pressure is set to zero
& Simulation and field test
& Real-time lane-level travel times; turning ratios; saturation flows; free-flow travel times\\

\citet{zoabi_enhanced_2026}
& Queue-based
& Lane-level queue difference. The capacity-aware version uses normalized lane queues.
& Time-step-based
& Yes
& Weight and pressure both can be negative
& Theoretical proof and simulation
& Lane-level queues, lane-to-lane turning ratios, lane saturation flows, and lane capacities\\

\citet{gregoire_capacity-aware_2015}
& Capacity-aware
& Difference between convex, capacity-normalized upstream and downstream queue pressures
& Time-step-based
& Not established under finite capacities
& Negative movement weight is set to zero
& Analytical proof of work conservation and simulation
& Real-time aggregate queue lengths and movement-presence indicators; link capacities and saturation flows\\

\citet{hao_backpressure_2020}
& Capacity-aware
& Movement queue difference, adjusted by a penalty based on the downstream link's remaining capacity
& Time-step-based
& Yes
& Weight and pressure both can be negative
& Theoretical proof and simulation
& Movement queues; downstream link capacities; saturation flows\\

\citet{li_position_2019}
& Position-aware
& Position-weighted vehicle density
& Time-step-based
& Yes
& Absolute value of the weight
& Theoretical proof and simulation
& Spatial traffic densities, turning ratios, link lengths, and fundamental-diagram parameters\\

\citet{xu_approximate_2025}
& Position-aware
& Position-weighted estimated densities
& Time-step-based
& Yes
& Absolute value of the weight
& Theoretical proof and simulation
& Cumulative vehicle counts (2 detectors per link), turning ratios, and fundamental diagram parameters\\

\citet{maripini_queue_2025}
& Travel time
& Queue dissipation time
& Cyclic MP: both fixed-cycle and adaptive-cycle
& Stability argued; no rigorous maximum-stability proof
& Pressure can be negative
& Brief analytical stability argument and simulation
& Sample vehicle travel times and detection timestamps, red duration, historical delay-polygon slope\\

\citet{riehl_green-pressure_2025}
& Transformed state variables
& Vehicle-category-weighted queue length; upstream information only
& Time-step-based
& Not proven
& Not applicable: Only upstream consideration
& Simulation only
& Vehicle counts by vehicle category (loop-detector data)\\

\citet{gao_arrival-based_2022}
& Transformed state variables
& Queue length. Arrival flow is added only in the phase-pressure calculation.
& Time-step-based
& Yes
& Weight and pressure both can be negative
& Theoretical proof and simulation
& Queue lengths, movement arrival flows, saturation flows\\

\citet{le_decentralized_2015}
& Cyclic
& Queue length
& Fixed-cycle cyclic MP
& Yes
& Weight and pressure can both be negative
& Theoretical proof and simulation
& Queue lengths, estimated turning ratios, saturation flows\\

\citet{anderson_stability_2018}
& Cyclic
& Queue length
& Fixed-cycle cyclic MP
& Stability is retained for a reduced set of feasible demands
& Weight and pressure can both be negative
& Theoretical proof and simulation
& Movement-level queue lengths, turning ratios, saturation flows\\

\citet{levin_mp_2020_cyclic}
& Cyclic
& Queue length
& Adaptive-cycle cyclic MP
& Yes, within the reduced stability region
& Weight and pressure can both be negative
& Theoretical proof and simulation
& Queue lengths, turning ratios, saturation flows\\

\citet{barman_performance_2022}
& Cyclic
& Queue length
& Adaptive-cycle MP with phase skipping
& Not proven
& Weight and pressure can both be negative
& Simulation only
& Movement queue lengths, turning ratios, saturation flows\\

\citet{robbennolt_microsimulation_2022}
& Cyclic
& Queue length
& Semi-cyclic: normally acyclic, but each phase must be served within a maximum number of time steps
& Not proven
& Weight and pressure both can be negative
& Simulation only
& Queue lengths, saturation flow rates, turning ratios, link storage capacity\\

\citet{hsieh_throughput-optimal_2017}
& Phase switching lost time
& Queue length
& Time-step-based
& Yes
& Movement and phase pressures can be negative; negative pressure set to zero for phase-switching decision.
& Theoretical proof and simulation
& Movement queue lengths, turning ratios, saturation flows
\\

\citet{wang_learning_2022}
& Phase switching lost time
& Queue-length weight (SCMP); position-weighted vehicle counts (ESCMP/LESCMP)
& Time-step-based
& SCMP: Yes. ESCMP/\allowbreak LESCMP: Not proven
& Weight and pressure both can be negative
& Theoretical proof and simulation
& Cell-level vehicle counts, signal state, turning ratios, saturation flows\\

\citet{sun_max-pressure_2025}
& Phase switching lost time
& Queue length
& Time-step-based
& Not proven
& Weight and pressure can both be negative
& Simulation only
& Vehicle counts, turning ratios, saturation flows, downstream link length\\

\citet{cheng_lane-based_2024}
& Lane-based MP
& Queue length
& Time-step-based
& Yes
& Weight and pressure can both be negative
& Theoretical proof and simulation
& Lane queue lengths, lane-to-lane turning ratios, lane discharge rates\\

\citet{gregoire_back-pressure_2014}
& Unknown routing rate
& Aggregated queue length
& Time-step-based
& Not proven
& Negative movement weights are set to zero
& Theoretical analysis and simulation
& Aggregated queue lengths; movement-specific loop detector measurements, saturation flows\\

\end{longtable}
\endgroup
\end{landscape}

\subsection{Fundamental Research Gaps in MP Control}
Collectively, these developments reviewed in the previous two sections have significantly improved the flexibility and performance of MP control and facilitated its application to more realistic traffic conditions. Nevertheless, several fundamental issues remain insufficiently addressed. 

First, interactions among different vehicular movements remain insufficiently investigated in existing MP formulations. Although several studies have considered shared-lane operations or multiple lanes serving the same movement, these extensions do not fully capture the complex interactions and conflicts among vehicles from different movements. For example, under permitted left-turn operations, the service of left-turning vehicles depends directly on the availability of acceptable gaps in the conflicting through traffic, making the effective service rate of the left-turn movement dependent on the state of another movement. Similar interactions arise for right turns on red (RTOR), particularly when right-turning and through vehicles share a lane, where queued through vehicles may block right-turning vehicles. Moreover, for shared lanes serving multiple movements, it remains unclear how the traffic states and service opportunities associated with different movements should be represented and incorporated into their respective movement weights when they compete for the same lane space and discharge capacity. Therefore, a more general framework that explicitly captures interactions among movements, lanes, and conflicting traffic streams is still needed.

Second, there still exists a gap between the theoretically elegant formulation of MP control and its practical implementation at signalized intersections. The original MP framework is built upon a relatively simple control logic, whereas practical implementation requires numerous additional design decisions. For example, lost time associated with phase transitions needs to be properly incorporated; negative movement weights or phase pressures need to be handled appropriately; and additional constraints or control rules may be required to ensure desirable properties such as work conservation. Other implementation choices, including minimum green times for cyclic MPs, decision intervals, and phase-transition rules, can also affect control performance significantly. Existing studies have adopted different treatments for these issues, while some have simplified or completely ignored them. Consequently, despite the extensive literature on MP control, there remains no unified implementation framework or set of evaluation criteria for determining which formulation is appropriate under specific operational conditions. To provide a more consistent assessment of these fundamental design choices, we therefore employ a unified simulation framework to compare the performance of several representative MP formulations under common network, demand, and evaluation settings.

The following six models have been selected for the comparison:
\begin{enumerate}
    \item Queue-based MP (Q-MP): the original MP formulation proposed by \citet{varaiya_max_2013}.
    
    \item Delay MP (D-MP): the delay-based MP formulation proposed by \citet{liu_novel_2022}.
    
    \item Total-Delay MP (TD-MP): the total-delay-based MP formulation proposed by \citet{liu_total-delay-based_2023}.
    
    \item Travel-Time MP (TT-MP): a modified version of the travel-time-based MP formulation proposed by \citet{mercader_max-pressure_2020}. The original formulation is link-based, whereas in this study, we adopt a movement-based formulation to maintain consistency with the other MP controllers considered in the comparison.
    
    \item Capacity-Aware MP (CA-MP): the capacity-aware MP formulation proposed by \citet{zoabi_enhanced_2026}.
    
    \item Position-Weighted Backpressure (PWBP): the position-weighted backpressure formulation proposed by \citet{li_position_2019}.
\end{enumerate}

In addition to the model-specific formulations, three common modifications, shown below, are applied to all six algorithms to ensure methodological consistency and enable a fair and direct comparison.

\begin{enumerate}
    \item Non-negative movement weight: The movement weight is constrained to be non-negative. Specifically, the weight is calculated as:
    \begin{equation}
        w = \max\left(0,\; \text{upstream metric} - \text{downstream metric}\right).
    \end{equation}
    The rationale behind this treatment is that adding a movement to a phase should not reduce the phase's need for green time, even if the added movement has a relatively large downstream metric value.
    \item Work-conservation and spillback-prevention rule: A movement is assigned zero weight if no vehicles are present in its upstream movement or if any downstream link associated with the movement has reached its storage capacity. The latter is to prevent further discharge toward the saturated downstream link.
    
    \item Phase-switching loss: The effective discharge capacity associated with a phase switch is adjusted using the phase-switch loss factor defined in \autoref{eq:switch_loss_saturation}. This modification accounts for the lost time associated with signal transitions and ensures that switching losses are treated consistently across all controllers.
\end{enumerate}

\begin{figure}[htbp]
	\centering
	\includegraphics[width=0.6\textwidth]{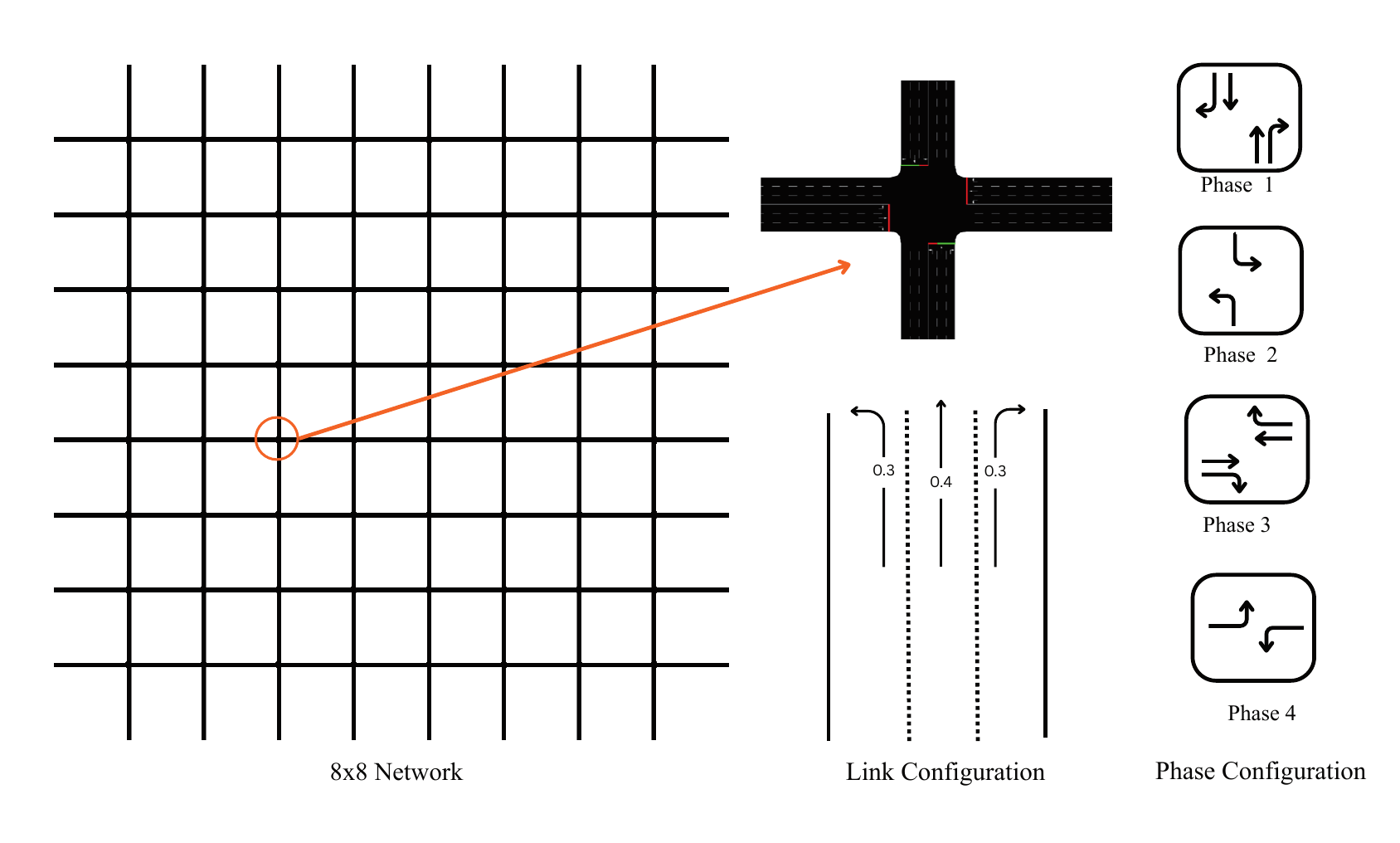}
	\caption{ Network configuration for simulation}
	\label{fig:Simulation_network_review_paper}
\end{figure}

We conducted our simulation on an $8\times8$ grid network consisting of 64 signalized intersections, as shown in \autoref{fig:Simulation_network_review_paper}. Traffic microsimulation software, Simulation of Urban MObility (SUMO) \citep{sumo2018}, has been used to test the MP controllers. Each road segment has bidirectional traffic flow. Each link has three dedicated lanes for left, through, and right movements, with saturation flows of 1750 veh/h, 1764 veh/h, and 1542 veh/h, respectively. These values have been calibrated from SUMO. For simplicity, the turning ratios are assumed to be the same at all the intersections: 0.3 (left), 0.4 (through), and 0.3 (right).   Vehicle length is 5 m. The Intelligent Driver Model (IDM) car-following model was used with SUMO’s default parameter values. The distance between intersections is 300 m, and the free-flow speed in each lane is 20 m/s.   The signal consists of four phases as shown in \autoref{fig:Simulation_network_review_paper}. Yellow time is 3 seconds, and all-red time is 2 seconds during signal phase changes. The demand is 700 veh/h at each entry link, corresponding to a degree of saturation of 0.56 at each intersection.

We tested all models using seven signal update time steps: 6, 9, 12, 15, 18, 21, and 24 s. For each model and time step, five independent simulation runs were conducted using five different random seeds to account for stochastic variability.

\begin{figure}[htbp]
	\centering
	\includegraphics[width=0.5\textwidth]{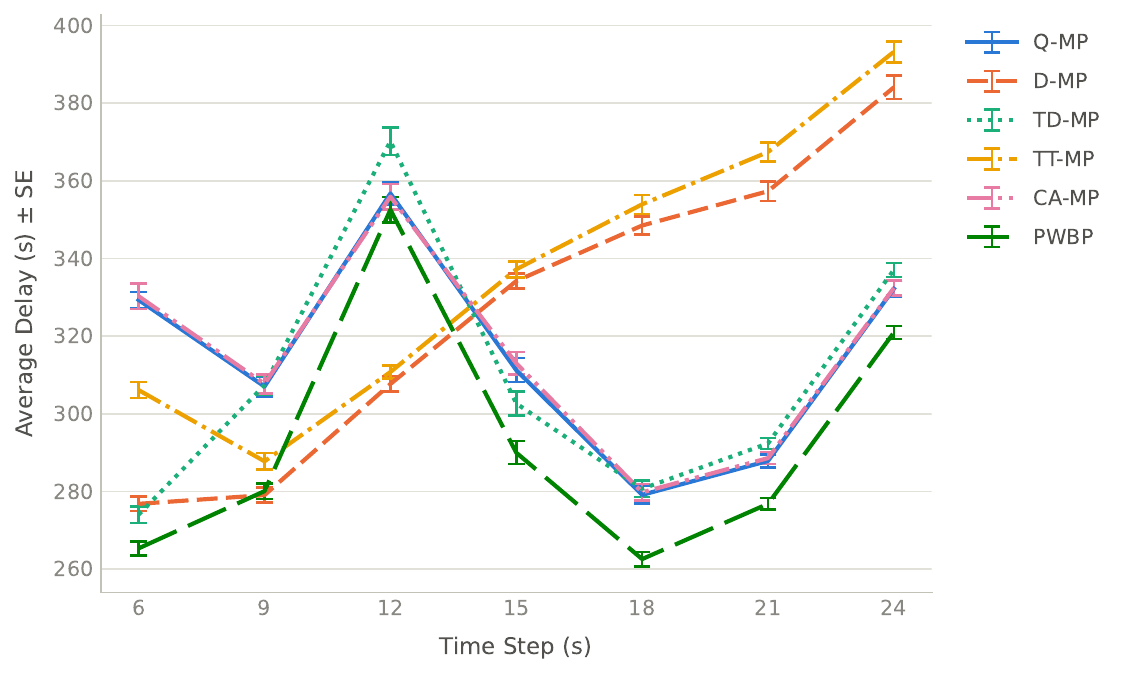}
	\caption{Delay vs. time step for MP extensions}
	\label{fig:delay_vs_time_step}
\end{figure}

\autoref{fig:delay_vs_time_step} shows the average delay per vehicle for the six MP extensions at different time steps. Q-MP and CA-MP exhibit nearly identical trends across all tested time steps, while TD-MP follows a similar pattern with some variation. For all tested algorithms, the time step size has a significant impact on the control performance. D-MP and TT-MP show a generally increasing delay as the time step increases, although D-MP performs slightly better than TT-MP. PWBP achieves the lowest average delay in most cases. As discussed in \Cref{sec:pwbp}, PWBP incorporates vehicle-position information into its pressure calculation, allowing it to distinguish between vehicles near the stop line and those farther upstream. This spatial information may help the PWBP controller make better signal-update decisions.

\begin{figure}[htbp]
     \centering
     \begin{subfigure}[h]{0.47\textwidth}
         \centering
         \includegraphics[width=\textwidth]{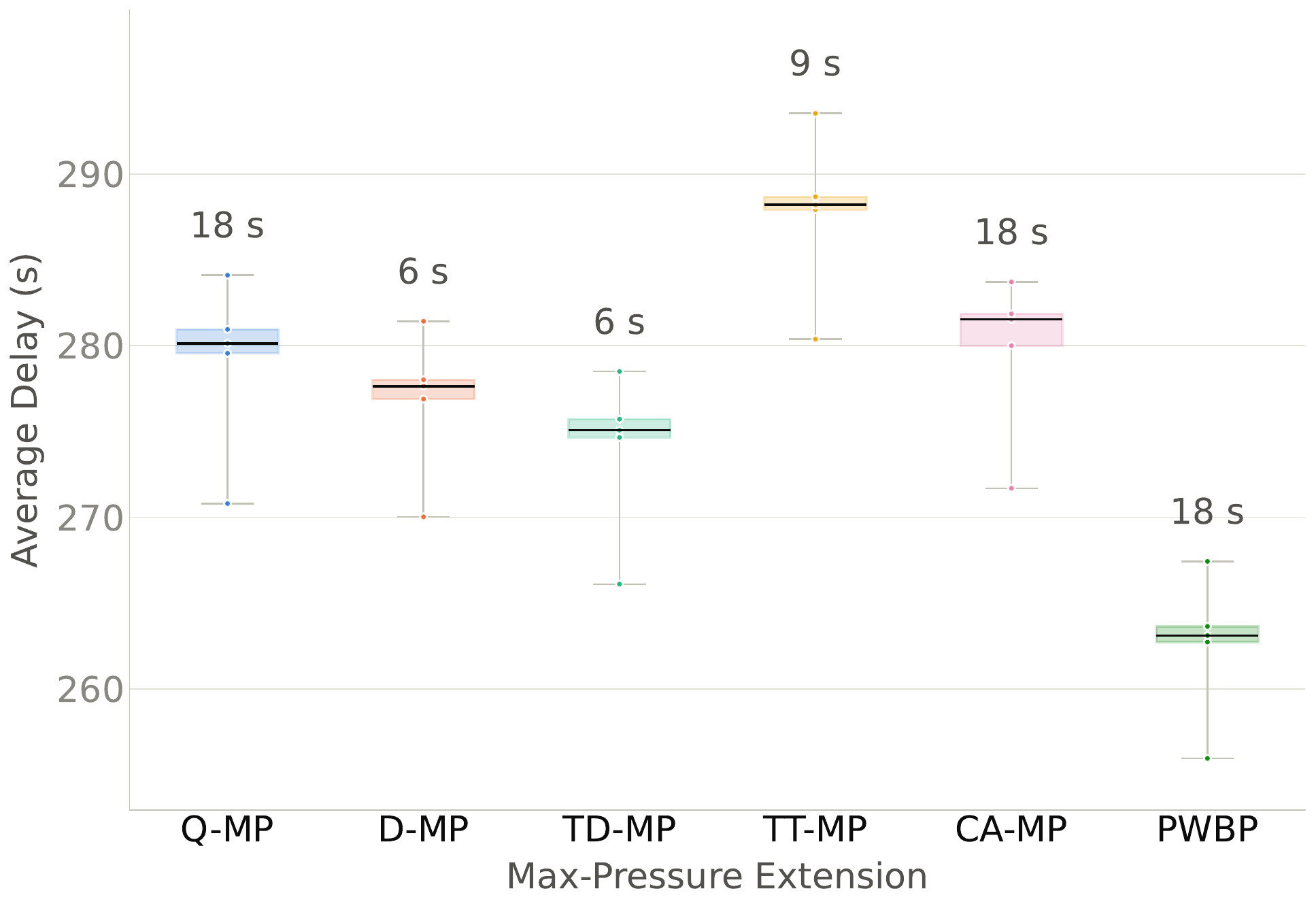}
         \caption{Average delay}
         \label{fig:delay_at_best_time_step}
     \end{subfigure}
     \hfill
     \begin{subfigure}[h]{0.47\textwidth}
         \centering
         \includegraphics[width=\textwidth]{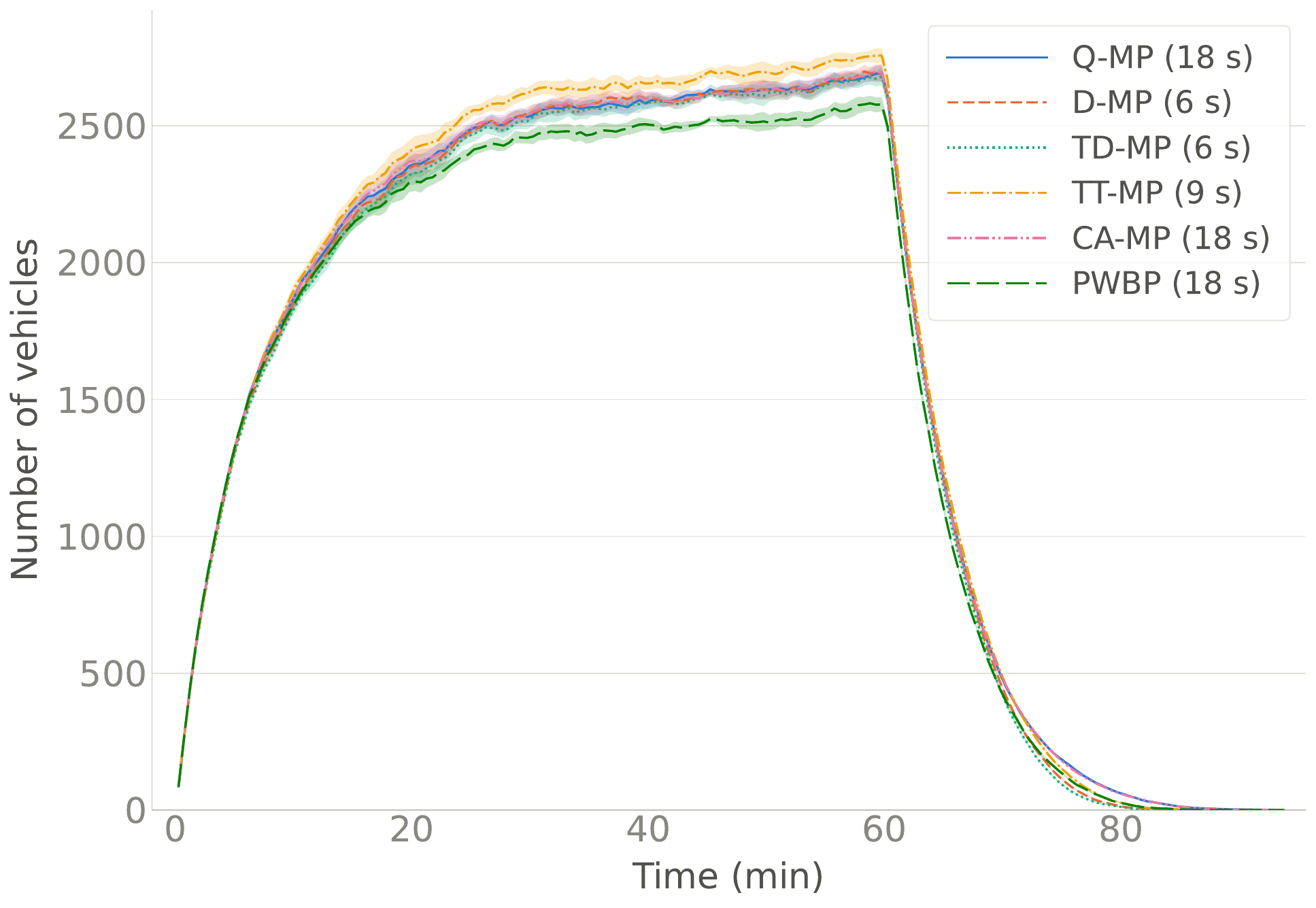}
         \caption{Evolution of accumulation}
         \label{fig:vehicles_network_vs_time}
     \end{subfigure}
     \caption{Performance of MP variants at their best time step}
     \label{fig:actual_predicted_safety}
\end{figure}

\autoref{fig:actual_predicted_safety} shows the average delay and the evolution of network accumulation for the six tested MP algorithms at their respective best-performing time steps. First, the time steps that lead to the best performance differ substantially across the six algorithms, emphasizing the importance of time-step calibration when implementing and comparing MP algorithms. Second, PWBP achieves the lowest average delay and vehicle accumulation in the network, at the cost of the highest data requirements among the tested algorithms. A potential improvement for D-MP, TD-MP, and TT-MP is to replace their time-aggregated downstream metrics with a ``snapshot'' metric, such as the number of vehicles, which may provide a more robust and direct representation of downstream congestion. Meanwhile, an adjustment factor may be needed to ensure compatibility between the upstream and downstream terms in the movement weight formulation.

We emphasize that the purpose of this comparison is solely to demonstrate the need for careful and comprehensive experimental design when comparing different MP algorithms, rather than to identify the best- or worst-performing algorithm. The results are specific to the tested network and parameter settings and therefore should not be interpreted as general conclusions regarding the relative performance of these algorithms. In addition, to ensure consistency across the comparison, we adopt a unified implementation setting for all tested algorithms, which may differ from the settings used in their original studies. For example, the original TT-MP was formulated at the link level and designed for cyclic control, whereas it is implemented at the movement level with time-step-based control in this comparison. Such implementation differences may affect the performance of individual algorithms and should therefore be considered when interpreting the results.

Third, the results shown in \autoref{fig:delay_vs_time_step} indicate that the signal update frequency substantially affects the control performance of all tested time-step-based models. However, this effect may depend on other factors, including demand level, network layout, and lane configuration. Therefore, the third research gap concerns how to select effective adaptive signal update intervals across different MP formulations and operating conditions. Most existing methods make control decisions at a predefined and fixed time step, regardless of how rapidly traffic conditions evolve. A fixed temporal resolution may limit control responsiveness under rapidly changing conditions, while unnecessarily frequent decision-making may provide little additional benefit when traffic conditions are relatively stable. Although \citet{Zoabi_dynamic_time_step} advanced this direction by proposing traffic-dependent activation durations that account for switching losses and demand and supply constraints, their formulation does not explicitly optimize these durations against a defined network performance objective, and its effectiveness across different MP formulations and heterogeneous networks remains to be established. Future research could therefore investigate how to jointly select phases and activation durations to improve network performance, and determine how suitable interval-selection rules vary across MP formulations and traffic conditions.

Fourth, although considering downstream traffic conditions when determining signal priorities is one of the key features of MP, there is debate over how these conditions should be incorporated. For example, \citet{smith_backpressure_2024} questioned whether incorporating downstream pressure always improves network performance, particularly when drivers can adjust their route choices in response to travel times. Using a simple analytical network with two signalized intersections, two origin--destination pairs, and four routes, they compared delay-based pressure controllers and a continuum-modified version of queue-based MP with and without downstream pressure under Wardrop user-equilibrium routing. Their results showed that incorporating downstream pressure increased equilibrium delays at one intersection without reducing delays at the other. Similarly, \citet{hoang_optimal_2024}  proposed a platoon-based Max Weight controller for connected-vehicle networks, in which movement weights are formulated using upstream platoon delay, vehicle position, and speed-related characteristics, without explicitly accounting for downstream pressure.  However, these findings were obtained for a specific network configuration under restrictive assumptions and therefore cannot be generalized to arbitrary networks or traffic conditions. Furthermore, under heavily congested conditions, ignoring downstream traffic conditions may lead to inefficient signal decisions and increase the risk of queue spillback. Therefore, further research is needed to determine when and how downstream traffic conditions should be incorporated into pressure calculations to improve control performance, particularly considering the impact of downstream congestion, queue spillback, and drivers’ route-choice behavior.

\section{MP in Multi-Modal Transportation Environments}\label{sec: multi}

Early MP traffic signal control research primarily focused on vehicular traffic and on improving the definitions of movement weights and pressures. However, urban intersections serve multiple transportation modes with different operational characteristics and service priorities. Consequently, recent studies have extended the MP framework to explicitly incorporate public transit, pedestrians, and non-motorized vehicles. Existing multimodal MP approaches can generally be classified as either \textit{rule-based} or \textit{integrated} methods. Rule-based approaches retain the conventional vehicular MP formulation and introduce additional constraints or intervention mechanisms to provide priority to another mode when specified conditions are satisfied. In contrast, integrated approaches incorporate the traffic state or priority of the additional mode directly into the MP weight or pressure formulation, allowing different modal demands to be considered within the pressure-based decision process. The following subsections review these approaches for public transit and vulnerable road users.

\subsection{Transit and Bus Priority}

\citet{xu_integrating_2022} proposed a rule-based MP framework incorporating transit signal priority (TSP) for bus rapid transit (BRT) operating on exclusive bus lanes. The proposed method maintains separate movement-level queues for buses and private vehicles. At each time step, the controller detects whether any bus is waiting at the intersection. If no bus is detected, the original signal-control policy is applied. If one or more buses are waiting, however, at least one movement containing a waiting bus must be activated. This requirement is expressed as

\begin{equation}
\sum_{(l,m)\in\pazocal{M}_i}
x^{B}(l,m)(t) \, S(l,m)(t)>0
\quad \text{if} \quad
\sum_{(l,m)\in\pazocal{M}_i}
x^{B}(l,m)(t)>0,
\label{eq:xu_tsp}
\end{equation}

$x^{B}(l,m)(t)$ is the number of buses waiting to make movement $(l,m)$ at time step $t$. Thus, whenever at least one bus is waiting, the feasible signal control is restricted so that at least one waiting bus movement receives green. If the detected bus movements do not conflict, the transit-priority signal is activated directly. If conflicting bus-priority movements exist, the MP controller uses the optimal solution of its pressure-maximization problem to determine which bus moves first. When no bus is present, the signal decision is again governed by the conventional MP objective based on private-vehicle traffic states. The proposed controller achieves maximum stability within the feasible region constrained by transit priority. Simulation results showed lower average bus and private-vehicle travel times than the tested fixed-time and adaptive controllers with transit signal priority. However, unconditional bus priority can reduce the private-vehicle stability region relative to standard MP without transit priority and increase delays for conflicting private-vehicle movements. Moreover, the approach is limited to dedicated bus lanes, whereas most urban networks involve shared lanes for both transit vehicles and private automobiles.

To address these issues, \citet{ahmed_occ-mp_2024} proposed an occupancy-based MP framework (OCC-MP) that incorporates passenger occupancy into the movement-weight calculation to implicitly prioritize transit and other high-occupancy vehicles (HOVs). Rather than replacing vehicle queues with passenger counts, OCC-MP provides a more flexible control by scaling the conventional queue-based movement weight by the average occupancy of vehicles queued on the upstream movement. Specifically, the weight of movement $(l,m)$ is defined as
\begin{equation}
w(l,m) = o(l,m) \, max[0, w_q(l,m)]
\end{equation}
where $o(l,m)$ is the average occupancy of all vehicles in the upstream movement $(l,m)$ and $w_q(l,m)$ is the corresponding weight under the original queue-based MP. Importantly, only passenger occupancy of the upstream movement is considered. The authors argued that the downstream queue represents the available supply or storage on the receiving link, which depends on the number of vehicles occupying that link rather than on the number of passengers carried by those vehicles. Thus, a movement containing buses or other HOVs receives greater weight when its upstream average occupancy is high, while downstream congestion continues to constrain its priority.
Unlike the rule-based method, which always gives priority to transit vehicles, OCC-MP can still favor a heavily congested conflicting private-vehicle movement when appropriate, reducing the adverse effect of transit priority on general traffic. Simulation results show that OCC-MP significantly reduces bus travel times and improves overall passenger travel time in most scenarios, with only marginal increases in private vehicle delay. 

\citet{tan_transit-mp_2027} recently developed another MP algorithm, called Transit-MP, which extends their CV-MP framework \citep{tan_cv-mp_2026} to prioritize public transit vehicles in a partially connected environment. They assume all public transit vehicles are connected so that their occupancies and positions are always available. The proposed algorithm explicitly accounts for the effects of transit stops and dwell times in the pressure calculation. Transit-MP activates signal phases according to transit-adjusted movement pressures, thereby accounting for both general traffic conditions and the additional priority assigned to transit vehicles. The weight in Transit-MP at time $t$:

\begin{equation}
w(l,m)(t) = 
\sum_{v\in V^{cv}_{l,m}}\beta_v o_v\tau_v
-
\sum_{p \in Out_m}
r(m,p)
\sum_{v\in V^{cv}_{m,p}}\beta_v\tau_v
\end{equation}

\noindent where $V^{cv}_{l,m}$ and $V^{cv}_{m,p}$ are the sets of connected vehicles on the upstream movement $(l,m)$ and downstream movement $(m,p)$, respectively, $o_v$ is the occupancy of vehicle $v$, $\tau_v$  is the normalized link-travel time of CV $v$, $\beta_v$ is set to 1 for connected personal vehicles and buses that have already passed the bus stop, and to 0 for buses that have not yet reached or are dwelling at the stop. This treatment prevents green time from being unnecessarily allocated to a bus that cannot pass through the intersection during the next decision step. In addition, occupancy is applied only to the upstream traffic state while the downstream state is not occupancy-weighted because the downstream term represents the ability of the downstream link to accommodate vehicles rather than passengers. Consequently, Transit-MP prioritizes high-occupancy vehicles based on their accumulated link travel time while avoiding unnecessary transit priority during station dwell. This model maintains theoretical rigor by proving network stability even under partially connected vehicle environments, despite using different upstream and downstream pressure formulations. Simulation results demonstrate that the method improves both transit and general traffic performance while reducing spillover. Furthermore, it preserves the stability region under uniform conditions and enhances practical stability through a modified version (mTransit-MP), which addresses data sparsity in low CV penetration scenarios by incorporating historical traffic data.

\subsection{Vulnerable Road Users} Pedestrians and users of non-motorized vehicles are generally considered vulnerable road users (VRUs) because they lack the physical protection provided by the structure of a motor vehicle and are therefore more exposed to injury in the event of a collision. Their safety and service needs also differ substantially from those of motorized traffic, creating additional considerations for traffic signal control. Several extensions of the MP framework have incorporated pedestrians and non-motorized traffic into the control formulation to account for these needs. 

\subsubsection{Pedestrians}
\citet{chen_stability-based_2020} incorporated pedestrians into MP control by jointly considering vehicle and pedestrian movements. Because pedestrian queues were difficult to measure directly, they were estimated using an inflow--outflow model based on pedestrian waiting time and mean arrival rates, and these estimates were used to calculate pedestrian weights. The controller maximizes the combined vehicle--pedestrian pressure while preventing vehicles from moving when they conflict with an activated crosswalk. For the conflict-region model, the authors proved throughput optimality for demand within the interior of the stability region. The framework assumes fully autonomous, connected vehicles rather than conventional human-driven traffic.

\citet{xu_ped-mp_2024} proposed a rule-based pedestrian-friendly MP signal control (Ped-MP) for phase-based signalized intersections. Ped-MP retains the original vehicle-based MP objective but adds constraints to guarantee pedestrian access. Rather than using pedestrian queue length in the pressure calculation, the controller tracks how long each pedestrian crossing has remained unavailable. Ped-MP forces the crosswalk to be activated once the predefined waiting-time threshold is reached, while simultaneously serving the highest-pressure vehicular phase that does not conflict with the activated pedestrian phase. The authors prove maximum stability for vehicles within the pedestrian-constrained stability region, which can be smaller than that of the original MP because pedestrian service reduces the feasible service available to vehicles. Simulations show that shorter pedestrian tolerance times substantially reduce pedestrian delay but also reduce vehicle throughput and increase vehicle delay. Moreover, because conflicting vehicle and pedestrian movements are treated as mutually exclusive, the formulation does not represent permissive movements such as right-turning vehicles yielding to pedestrians during an active pedestrian phase.

To address these limitations, \citet{liu_max_2024} proposed a pedestrian-queue-based MP algorithm (PQ-MP) that incorporates pedestrian states directly into the pressure calculation while retaining the decentralized, time-step-based MP control structure. PQ-MP defines weights for both vehicular and pedestrian movements using the number of vehicles and pedestrians as the metric for weight definition, in a manner similar to Q-MP \citep{varaiya_max_2013}. For a pedestrian movement waiting to cross the intersection, its weight is defined as the number of pedestrians waiting for that movement minus the number of pedestrians at the downstream corner, weighted by the proportion expected to continue crossing the intersection in the perpendicular direction. Pedestrians traveling on sidewalks are modeled in the pedestrian dynamics but are not included in the pressure calculation because they are assumed to walk at free-flow speed without incurring delay. The pressure of an admissible phase $\phi$ is then calculated jointly from vehicle and pedestrian movements as

\begin{equation}
P_i^{\phi}(t)
=
\sum_{(l,m)\in\pazocal{M}^{vi}}
w^{v}(l,m)(t) \, \tilde{C}^{v}(l,m)(t) \, S^{v,i}(l,m)(t)
+
\psi \sum_{(i_{\alpha},i_{\beta})\in\pazocal{M}^{pi}}
w^{p}_{i_{\alpha},i_{\beta}}(t) \, C^{p}_{i_{\alpha},i_{\beta}}(t) \, S^{p,i}_{i_{\alpha},i_{\beta}}(t)
\end{equation}

where $\pazocal{M}^{vi}$ and $\pazocal{M}^{pi}$ are the sets of vehicle and pedestrian crosswalk movements at intersection $i$, respectively; $\tilde{C}^{v}(l,m)(t)$ is the effective saturation flow of vehicle movement $(l,m)$ after accounting for pedestrian conflicts; $C^{p}_{i_{\alpha},i_\beta}(t)$ is the pedestrian saturation flow; $S^{v,i}(l,m)(t)$ and $S^{p,i}_{i_{\alpha},i_\beta}(t)$ indicate whether the corresponding vehicle and pedestrian movements are served by the phase; and $0<\psi<1$ scales the pedestrian contribution to avoid excessively favoring pedestrian movements because of their comparatively high saturation flows. Unlike  \citet{xu_ped-mp_2024}, PQ-MP explicitly accounts for right-turn vehicles yielding to conflicting pedestrians by reducing their effective saturation flow according to the fraction of the time step required to serve the conflicting pedestrian queue. Thus, pedestrian service is determined jointly with vehicle service through the MP optimization rather than being triggered by a predefined pedestrian waiting-time threshold. The authors further prove maximum stability of the combined vehicle and pedestrian queue process for demand within the interior of the feasible region, subject to the assumption that the distribution of the pedestrian-adjusted saturation flow of right-turn movements remains fixed. Microscopic simulations show that PQ-MP produces lower total person delay than both the original queue-based MP and the rule-based Ped-MP across the tested vehicle-demand levels, although the rule-based approach can achieve a larger vehicle stability region when a sufficiently large pedestrian waiting-time threshold is used.

\subsubsection{Non-Motorized Vehicles}
\citet{li_max_2025} extended MP control to jointly account for motor and nonmotor vehicles (NMVs) on roads with segregated facilities. For motor vehicles, the pressure is based on queue length normalized by their corresponding road-section lengths. For NMVs, the pressure is based on the actual queue length. The combined phase pressures for both motor vehicles and NMVs are calculated from their respective pressures and then combined as follows:
\begin{equation}
P_i^{\phi} (t)
=
\delta_\text{MV} P_i^{\phi,\text{MV}}(t)
+
\alpha \, \delta_{\text{NV}} P_i^{\phi,\text{NV}}(t)
\end{equation}
where $P_i^{\phi,\text{MV}}(t)$ is the motor-vehicle phase pressure, $P_i^{\phi,\text{NV}}(t)$ is the NMV phase pressure,  $\delta_\text{MV}$ and $\delta_\text{NV}$ are the percentages of motor and nonmotor vehicles on the road, respectively, and $\alpha$ is the bicycle conversion factor, taken as $0.22$ based on \citet{wang_research_2008}. To reduce the risk of motor-vehicle queue overflow, when the motor-vehicle queuing ratio exceeds $0.85$, the integrated pressure is calculated using only the motor-vehicle component, and the NMV contribution is removed under severe motor-vehicle congestion. Simulation results show that, compared with MP considering only motor vehicles, the proposed joint controller reduced both motor- and nonmotor-vehicle queue lengths and lowered average arterial delay, while slightly increasing the throughput of both traffic types. 

\subsection{Research Gaps and Future Directions for Multi-Modal MP Control}
Despite recent advances in extending MP control to pedestrians, cyclists, and public transit, several transportation modes have received limited or no explicit attention. Emergency vehicles represent a particularly important research gap, as their integration requires not only signal priority or preemption but also consideration of how such interventions affect network-wide traffic states and how the MP controller recovers after the preemption ends. Rail-based transit has also received limited attention within the MP framework, despite its distinct priority requirements and operational constraints at signalized intersections. Freight traffic has primarily been considered indirectly, for example through environmental or emission-related weighting \citep{riehl_green-pressure_2025}, while freight-specific mobility and priority needs remain largely unexplored. Similarly, powered two-wheelers and emerging micromobility modes, such as shared electric scooters, have rarely been explicitly incorporated into MP formulations despite their growing presence in urban transportation systems. These gaps highlight opportunities to further extend MP from a predominantly road-traffic control framework toward a more comprehensive multimodal traffic management approach.

Current studies extend MP by considering only one additional user class, whereas urban intersections simultaneously serve multiple modes. Developing a unified multi-modal MP framework is therefore a promising research direction. However, combining multiple priority schemes may further reduce the stability region, while pressure-based approaches require appropriate weighting among heterogeneous traffic measures. Moreover, vehicle queues, pedestrian demand, and bicycle queues are measured in different units, making their integration into a single pressure definition challenging. A person-based pressure measure may provide a more consistent framework. Whether a unified multi-modal MP controller can preserve network stability remains an open research question.

\section{System-Level and Structural Extensions of MP Control} \label{sec:system_and_structural}

Traditional MP control is fundamentally decentralized, with signal decisions determined primarily from local upstream and downstream traffic states. Although this structure allows neighboring traffic conditions to influence local decisions indirectly, it does not explicitly coordinate signal actions across intersections, regulate traffic at a regional level, or address where MP should be deployed when network-wide implementation is impractical. Research on system-level and structural extensions of MP has therefore developed along three main directions: explicit signal coordination, integration with regional perimeter control, and selective deployment of MP at a subset of intersections. These developments are reviewed below.

\subsection{Signal Coordination with MP}

Signal coordination aims to provide smooth vehicular progression through closely spaced signalized intersections by synchronizing signal timings to create a ``green wave,'' thereby reducing stops, travel time, delay, fuel consumption, and rear-end crash risk. These benefits are particularly important along arterial corridors with strong directional traffic demand. Although the local upstream--downstream pressure structure of classical MP implicitly captures some interactions between neighboring intersections, signal decisions are still made independently without explicit inter-intersection coordination. This may limit network-wide efficiency when coordinated traffic progression is desirable. To address this limitation, several coordination extensions of MP have been developed. Similar to the classification of multimodal MP algorithms in \Cref{sec: multi}, these approaches can be broadly categorized as \textit{rule-based} or \textit{integrated}. Rule-based approaches introduce coordination through an external mechanism that modifies or influences the signal decisions generated by the MP controller, whereas integrated approaches incorporate coordination considerations directly into the MP formulation and decision-making process.

\citet{xu_smoothing-mp_2024} proposed a rule-based MP, called Smoothing-MP, to encourage platoon progression along selected corridors. When the upstream coordinated movement $(l,m)$ receives a green signal at time $t$, the corresponding downstream movement $(m,p)$ receives a coordination incentive in the following decision interval. This incentive is incorporated into the pressure calculation as shown below:

\begin{equation}
P^\phi_i (t)
=
\sum_{(l,m)\in\pazocal{M}_i^\phi}
[ w(l,m)(t) \, c(l,m)
+
\xi_{l,m} e_{l,m}^r(t)]
\end{equation}
where $\xi_{l,m}\geq0$ is the coordination weight and $e_{l,m}^r(t)$ is the signal coordination indicator for the turning movement $(l,m)$ at time $t$. Thus, the rule-based coordination term biases the signal decision toward maintaining progression along the designated corridor, while the final phase selection is still determined by maximizing the augmented MP objective. The authors proved that Smoothing-MP preserves the maximum-stability property and the stable region of the original MP controller. Simulation results showed smoother traffic progression, higher speeds, and lower delays along coordinated corridors, although stronger coordination generally deteriorated performance in conflicting directions.

Rather than a rule-based policy, \citet{tutsch_distributed_2024} formulated coordination as a distributed optimization problem through the Coordinated Maximum Pressure-plus-Penalty (CMPP) policy. CMPP expands the local control problem of traditional MPs to include the central intersection and its neighboring intersections. Specifically, intersection $i$ determines the neighborhood signal control as

\begin{equation}
\boldsymbol{\sigma}_i^{\mathrm{CMPP}}(t)
=
\arg\max_{\boldsymbol{\sigma}_i\in\pazocal{F}_i}
\left[
\sum_{j\in\pazocal{N}_i\cup\{i\}}
\sum_{\phi\in\Phi_j}
P_j^\phi(t)
-
V E_i(t)
\right].
\end{equation}

\noindent where $\boldsymbol{\sigma}_i$ is a neighborhood signal-control vector containing the phase-selection variables for intersection $i$ and its neighbors, $\pazocal{F}_i$ is the set of feasible neighborhood signal-control vectors, and $\pazocal{N}_i$ is the set of neighboring intersections of intersection $i$. The parameter $V>0$ weights the penalty $E_i(t)$ associated with the central intersection. Both the pressure contributions and the penalty depend implicitly on the neighborhood signal-control vector $\boldsymbol{\sigma}_i$.
The penalty discourages control actions for which estimated upstream or downstream movement queues exceed a prescribed threshold. It also penalizes repeated phase activation over a backward-looking horizon to discourage prolonged green times. Thus, each intersection proposes both its own signal action and those of its neighbors by considering the total neighborhood pressure and the penalty. Since adjacent neighborhoods overlap, the same intersection may receive different proposed signal actions from different local optimization problems. Coordination is achieved by imposing consensus constraints that require these overlapping decisions to agree on the signal action for each shared intersection. The resulting distributed optimization problem is addressed using either an Alternating Direction Method of Multipliers (ADMM)- based algorithm or a greedy heuristic with majority voting. Under the stated assumptions, the authors proved stability of the queuing network under CMPP. In the simulation experiments, CMPP achieved lower average travel time, average waiting time, and network congestion than classical MP and capacity-aware BP.

Taking a different approach, \citet{ahmed_c-mp_2025} proposed Coordinated Max Pressure (C-MP), which achieves coordination by incorporating vehicle space-mean speed (SMS) into the pressure calculation. Instead of relying on neighboring signal information or predefined offsets, C-MP identifies moving platoons directly from local traffic conditions and modifies both the upstream and downstream components of the conventional queue-based MP weight. Specifically, the C-MP movement weight is calculated as
\begin{equation}
w(l,m)(t)
=
x(l,m)(t)
\left(
1+\omega_1 \frac{v_{(l,m)}(t)}{v^\mathrm{ff}}
\right)
-
\sum_{p\in {Out}_m}
x(m,p)(t) \, r(m,p)(t)
\left(
1-\omega_2\frac{v_{(m,p)}(t)}{v^\mathrm{ff}}
\right),
\end{equation}

where $v_{(l,m)}(t)$ and $v_{(m,p)}(t)$ denote the instantaneous SMS of vehicles associated with movements $(l,m)$ and $(m,p)$, $v^\mathrm{ff}$ is the free-flow speed, and $\omega_1$ and $\omega_2$ are tuning parameters controlling the strength of the upstream and downstream speed-based adjustments, respectively. The upstream multiplier $1+\omega_1 v_{(l,m)}(t)/v^\mathrm{ff}$ increases the weight of movements containing faster-moving vehicles, thereby prioritizing approaching platoons before they are forced to stop. In contrast, the downstream multiplier $1-\omega_2 v_{(m,p)}(t)/v^\mathrm{ff}$ reduces the contribution of moving downstream vehicles, since they are more likely to continue progressing and vacate downstream space, while stopped downstream queues continue to exert their full restricting effect. The authors analytically proved that these speed-dependent modifications preserve the maximum-stability property without reducing the feasible demand region. In microsimulation, C-MP achieved the lowest network and corridor delay and fuel consumption among the tested Q-MP, PWBP, TT-MP, Smoothing-MP, and SCATS strategies, while its time--space diagrams demonstrated coordinated progression in both arterial directions. C-MP also remained effective when traffic-state information was obtained from only a subset of connected vehicles, particularly maintaining lower fuel consumption and travel time than the other MP-based policies under partial CV penetration.

\subsection{Integration of Network Perimeter Control}

\begin{figure}[htb]
	\centering
	\includegraphics[width=0.6\textwidth]{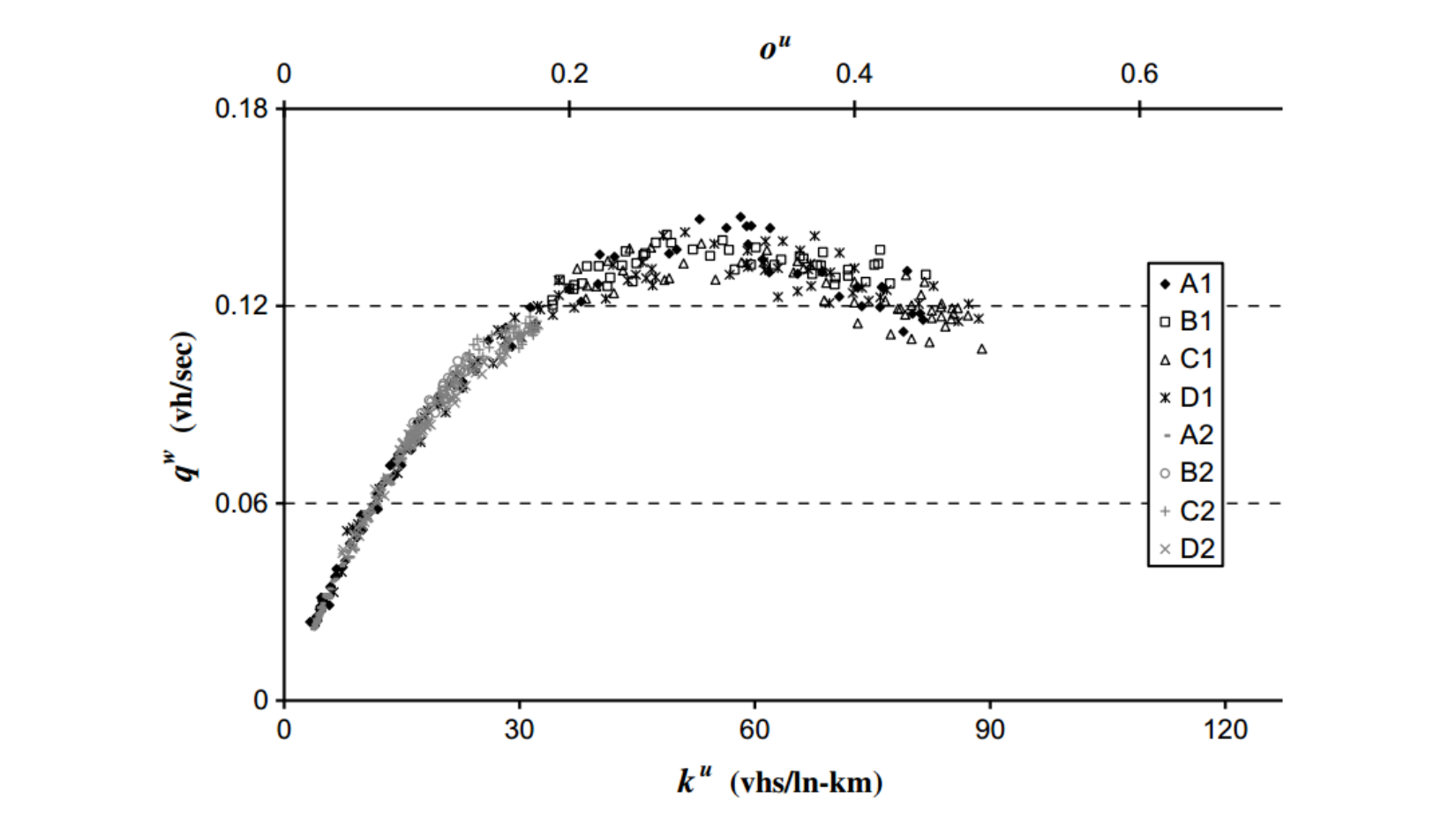}
	\caption{Empirical MFD, showing the relationship between average network flow and density (Reprinted from \citet{geroliminis_existence_2008} with permission from Elsevier)}
	\label{fig:mfd}
\end{figure}

Classical MP control is not specifically designed to regulate regional traffic accumulation under oversaturated conditions. When external demand is excessive, limited downstream storage, queue spillback, and ultimately gridlock can substantially degrade network performance, while locally determined MP decisions may continue directing vehicles toward already congested regions. The Macroscopic Fundamental Diagram (MFD) \citep{geroliminis_existence_2008, DAGANZO2007} provides a network-level representation of traffic conditions by relating network accumulation to network production. A well-defined MFD generally exhibits a unimodal relationship in which network productivity reaches its maximum at a critical accumulation. Beyond this critical level, increasing accumulation is associated with declining network productivity and may eventually lead to queue spillover and gridlock \citep{DAGANZO2007,keyvan-ekbatani_exploiting_2012}. An empirical example of an urban MFD is shown in \autoref{fig:mfd}.

Network perimeter control (PC) is a widely recognized regional congestion mitigation strategy based on MFD. PC regulates traffic entering a protected region at perimeter intersections, typically by adjusting signal timings and restricting green time for inbound movements, with the objective of maintaining regional vehicle accumulation near its critical level. Inspired by this concept, several studies have developed MP-based control methods that integrate or combine MP with PC to tackle regional congestion.

\citet{kouvelas_two-layer_2018} introduced a two-layer control framework which uses an upper layer to regulate transfer flows between regions through perimeter control and a lower layer to control interior intersections using MP. However, the PC and MP algorithms are applied to different intersections independently. The interregional transfer flows at perimeter intersections are regulated by a multivariable Proportional–Integral (PI) feedback controller:
\begin{equation}
\mathbf{u}(t)
=
\mathbf{u}(t-1)
-
\mathbf{K}_{P}\left[\mathbf{n}(t)-\mathbf{n}(t-1)\right]
-
\mathbf{K}_{I}\left[\mathbf{n}(t)-\hat{\mathbf{n}}\right],
\end{equation}
where $\mathbf{u}(t)$ is the vector of perimeter-control inputs at control interval $t$, $\mathbf{n}(t)$ is the vector of regional vehicle accumulations, $\hat{\mathbf{n}}$ is the vector of desired regional accumulation set points, and $\mathbf{K}_{P}$ and $\mathbf{K}_{I}$ are the proportional and integral gain matrices, respectively. In the lower layer, a cyclic MP algorithm without considering downstream traffic states is applied only to selected critical interior intersections, where green times are distributed proportionally to their pressures. 
Following a similar two-layer control concept, \citet{putra_design_2025} combined PI-based perimeter control with MP control at internal intersections and evaluated the framework on a real-world-derived network in SUMO.
\citet{tsitsokas_two-layer_2023} developed a similar two-layer control framework, integrating multi-region PC to regulate transfer flows between adjacent regions and external inflows, and MP control based on normalized queue lengths for signal control at all or a selected subset of eligible interior intersections.

Unlike MFD-based perimeter control, \citet{cui_throughput-delay_2026} proposed a distributed joint admission and MP signal control framework that regulates network entry flows to limit oversaturation. Vehicles are held in external admission queues when estimated entry-link queues exceed a tunable threshold; otherwise, they are admitted subject to a maximum input rate. The framework estimates queue lengths and turning ratios using connected-vehicle data and applies conventional MP phase selection based on these estimates. Using Lyapunov drift optimization, the authors established a lower bound on weighted throughput and an upper bound on average internal queue lengths, explicitly accounting for queue-length estimation errors. Simulations demonstrated improved throughput under high demand and greater resilience to demand surges and incidents.

In contrast to hierarchical approaches that independently apply PC and MP at different control levels, \citet{liu_n-mp_2024} embedded perimeter control directly into the MP formulation by modifying the weights of movements entering the protected region. The proposed Network-state-based MP (N-MP) algorithm is built on the D-MP formulation, although the authors note that the same mechanism can be incorporated into MP controllers using other pressure metrics. When a region is congested, for an inbound movement $(l,m)$ sending vehicles into the congested region, N-MP reduces its weight by:

\begin{equation}
w_{l,m}^{N}(t)
=
w_{l,m}^{D}(t)
-
\Psi(t)
\mathds{1}_{(\rho_{\mathrm{cr}}^{p},+\infty)}
\!\left(\rho^{p}(t)\right)
\mathds{1}_{\pazocal{M}_{p}}(l,m)
\end{equation}

where $w_{l,m}^{N}(t)$ is the N-MP weight of movement $(l,m)$ at time $t$, $w_{l,m}^{D}(t)$ is its original D-MP weight, $\rho^{p}(t)$ is the average density of the protected region, $\rho_{\mathrm{cr}}^{p}$ is the critical regional density corresponding to maximum network productivity, $\pazocal{M}_{p}$ is the set of inbound movements entering the protected region, $\mathds{1}_{(\rho_{\mathrm{cr}}^{p},+\infty)}(\rho^{p}(t))$ indicates whether the protected-region density exceeds its critical value, $\mathds{1}_{\pazocal{M}_{p}}(l,m)$ indicates whether movement $(l,m)$ is an inbound movement, and $\Psi(t)$ is a positive penalty function reflecting both local and regional traffic conditions. Thus, when $\rho^{p}(t)>\rho_{\mathrm{cr}}^{p}$, the weights of inbound movements are reduced, lowering the pressure of phases that serve them and consequently restricting inflow into the protected region.
Therefore, perimeter metering emerges directly from the MP phase-selection process rather than from a separate perimeter controller. Although regional density is required as an input, the authors characterize N-MP as retaining the decentralized MP control architecture because no separate optimization or green-time allocation model is needed. They further proposed a clustered N-MP which can enhance the control performance for heterogeneous vehicle distributions. The authors also prove that both formulations preserve the maximum-stability property provided that the penalty function $\Psi$ is bounded. This is the first framework to incorporate regional perimeter control directly within the MP architecture while retaining the maximum-stability property.

\subsection{Selective Deployment of MP in Networks}
Similar to other adaptive signal control methods, full deployment of MP across an entire network is often impractical due to installation costs, sensing requirements, and infrastructure constraints. Consequently, researchers have also investigated whether MP can retain its effectiveness when deployed at only a subset of intersections and, if so, how these intersections should be selected.

\citet{barman_throughput_2023} proposed a strategic MP deployment method for networks where practical constraints limit its implementation to a subset of intersections. The study used Origin-Destination (OD) demand data to characterize a baseline demand pattern and assumed that the additional network capacity enabled by adaptive signal control could accommodate a uniform proportional increase in demand across all OD pairs. In other words, although the baseline demands may differ across OD pairs, they are assumed to increase by the same percentage. The objective was therefore to determine which intersections should be equipped with MP control, subject to a limited deployment budget, to maximize this common demand growth percentage while maintaining network stability. The resulting deployment problem was formulated as a mixed-integer linear program (MILP) that accounts for network flow conservation, movement capacities, feasible signal allocations, and the unchanged control policies at intersections without MP deployment. The authors also developed a greedy algorithm to solve the optimization model. Simulations on the downtown Austin network showed that strategically deploying MP at only a few intersections could substantially increase the maximum stabilizable demand, while larger deployments further reduced queue lengths, link densities, and travel times.

\citet{tsitsokas_two-layer_2023} also proposed a method to select intersections for MP deployment according to traffic conditions observed under baseline fixed-time control. For each candidate intersection $i$, three indicators were evaluated over a predefined peak period: the average congestion level of the intersection, $m_1^i$,  calculated as the time-average of the normalized queues of its incoming links, the average spatial imbalance among the incoming links, $m_2^i$, calculated as the time-average of the variance of their normalized queues, and the fraction of the peak period during which the intersection is considered congested, $N_c^i$, where a control cycle is classified as congested if the cycle-average queue of at least one incoming link exceeds a specified fraction $p$ of its storage capacity. These indicators are combined into a node-ranking measure,
\begin{equation}
R_i
=
a_1 m_1^i
+
a_2 m_2^i
+
a_3 N_c^i,
\end{equation}
where $a_1$, $a_2$, and $a_3$ weight the relative contribution of the three criteria. Their values are calibrated through a grid search in which candidate parameter combinations are used to generate MP node sets and the combination yielding the lowest total travel time for a reference penetration rate is retained. For a prescribed MP penetration rate, the eligible intersections are then ranked according to $R_i$ and the required number of intersections is selected for MP control. The selection is performed offline during the controller-design stage using baseline traffic conditions, and the resulting MP-controlled node set remains fixed during operation rather than being updated dynamically in real time.

In contrast, \citet{AHMED2024optimal} proposed a bi-level optimization framework to identify the number and locations of intersections for selective MP deployment. At the upper level, a Population-Based Incremental Learning (PBIL) algorithm generated candidate deployment configurations, while the lower level evaluated their total network travel time through AIMSUN microsimulation, using delay-based cyclic MP at the selected intersections and fixed-time control at the remaining intersections. The framework sought the configuration with the lowest total network travel time. To account for deployment limitations, the authors modified PBIL to impose a maximum allowable number of MP-controlled intersections. If a generated configuration exceeded this limit, excess locations were removed using either exploitation, which discarded the selected intersections with the lowest selection probabilities, or exploration, which discarded selected intersections randomly.  They also introduced an informed initialization strategy in which candidate intersections were ranked by their average delays under fixed-time control. Initial selection probabilities decreased linearly with rank, from $0.75$ for the highest-delay intersection to $0.25$ for the lowest-delay intersection. The simulation results on a real network with 33 intersections show that network-wide delay-based MP reduced total travel time by $8.2\%$, whereas PBIL achieved a $10.39\%$ reduction with only eight intersections and a maximum reduction of $11.35\%$ with 20 intersections. These results show that limited, strategically selected deployment can outperform full deployment and that the highest-delay or longest-queue intersections are not necessarily the best locations for adaptive traffic signal control.

\citet{xue_partially_2025} examined selective MP deployment by identifying intersections where MP implementation provides the greatest network-level benefit. They developed a regression model using intersection traffic volume, reciprocal minimum reserve capacity, and a major-intersection indicator to estimate the potential delay reduction from installing MP at an intersection and rank candidate deployment locations. Simulation results indicate that for multi-intersection deployment, selecting the eight intersections with the highest individual Delay Improvement (DI) from the network’s 18 intersections achieved greater delay improvement than deploying MP at the eight intersections located along the main arterial. Unlike \citet{barman_throughput_2023}, however, the study evaluated selective deployment through simulation and regression rather than theoretical analysis, and no stability guarantee was provided.

\subsection{Research Gaps and Future Directions for Network Perimeter Control}
An important issue that remains insufficiently addressed in MP-based perimeter control is how the protected region is defined. In N-MP, the protected region and its perimeter intersections are predefined, and although clustered N-MP accounts for spatial heterogeneity within the protected region, the regional boundary itself does not adapt to the spatial evolution of congestion. Future research could therefore incorporate dynamic boundary identification into MP-based perimeter control, allowing the protected region and its perimeter to adapt to real-time congestion patterns. Another extension is the consideration of multiple interacting protected regions. Although multi-region perimeter control has been combined with MP through hierarchical frameworks, further research is needed to directly incorporate the traffic states of multiple interacting regions into the MP weight formulation while preserving its decentralized architecture and maximum-stability property. Finally, CAV technologies provide opportunities to integrate vehicle-level route guidance with MP-based perimeter control. By jointly influencing vehicle routing and regulating traffic entering congested regions, such a framework could proactively redistribute demand while controlling regional accumulation, particularly under partial CAV penetration.

\section{MP in the Era of Emerging Transportation Technologies}\label{sec: emerging}

Emerging transportation technologies provide new opportunities for extending the MP framework. Connected vehicle (CV) technologies enable vehicle-level traffic information to be directly communicated to signal controllers, while automated vehicles (AVs) further introduce the possibility of controlling certain vehicle movements in coordination with traffic signals. Artificial intelligence (AI), particularly reinforcement learning (RL), has also been increasingly combined with MP to improve traffic representation and controller adaptation. This section reviews studies that explicitly integrate these emerging technologies with MP control.

\subsection{CAV-Enhanced MP Control}
\label{sec:cav_enhanced_mp}

CAV technologies can contribute to MP control in two different ways. CVs provide vehicle-level observations that can be used for pressure calculation, whereas AVs additionally allow vehicle movements to be actively controlled.

\subsubsection{MP in Partially Connected Environments}

The investigation of MP control in partially connected environments is not new. Several studies incorporated CV information into MP control and evaluated their methods under different CV penetration rates \citep{liu_novel_2022, ahmed_occ-mp_2024, ahmed_c-mp_2025}. These studies demonstrated that CV observations can support pressure calculation when complete traffic-state measurements are unavailable. However, the observed CV traffic states were generally used directly as surrogates for the complete traffic state, without explicitly characterizing how missing or heterogeneously distributed CV observations affect the stability of MP control. At low penetration rates, the observed CV sample may therefore provide an incomplete representation of the actual traffic state, particularly when individual movements contain few or no CVs for extended periods.

\citet{tan_cv-mp_2026} explicitly addressed this issue by developing CV-MP, in which pressure is constructed from the real-time link travel times of CVs. The cumulative travel-time state jointly reflects the spatial distribution of observed vehicles and the delays accumulated after entering a link. More importantly, the authors established sufficient conditions for network stability under partially connected and heterogeneously distributed CV environments and derived a necessary condition on CV observations for maintaining stability. The analysis therefore shows that full traffic-state observation is not required for MP stability, provided that the required observations are available with sufficient regularity. It also implies that a persistent lack of CV observations on a congested movement can violate the observation condition and degrade the stability of a controller that relies only on real-time CV data. \citet{tan_transit-mp_2027} then extended CV-MP to multi-modal traffic by incorporating occupancy-based transit priority and explicitly accounting for station dwell in pressure calculation.

Beyond traffic state estimation, another important research direction is how to leverage limited CV data to estimate model parameters in real time, particularly turning ratios. \citet{li_automated_2021} proposed an automated-vehicle-assisted MP controller that utilizes AV trajectory-planning data to estimate real-time turning ratios. \citet{xiao_throughput_2015} extended BP to accommodate bounded queue measurement errors and estimate turning ratios and saturation flow rates using fading-memory estimators. \citet{cao_max_2020,zhang_joint_2020} estimated queue lengths from connected-vehicle positions and speeds, applying a probabilistic correction to account for unobserved vehicles. Similarly, \citet{li_backpressure_2021} estimated vehicle counts from sparse connected-vehicle data, but instead reconstructed spatial speed profiles and converted them into densities using a speed-density relationship. In a joint control framework, \citet{cui_throughput-delay_2026} incorporated connected-vehicle-based queue and turning-ratio estimation into a joint network-entry flow regulation and MP signal control framework and analytically quantified the effects of queue estimation errors on network throughput and average queue lengths.

\subsubsection{Integration of AV Control with MP}

The studies above primarily use connected vehicles as sources of traffic information. Automated vehicles provide an additional control dimension because their movements can be coordinated with traffic signal decisions. Existing studies have considered such integration at both the local vehicle-motion level and the network routing level.

At the vehicle-motion level, several studies have directly incorporated AV controllability into pressure-based network control. \citet{rey_blue_2019} proposed the Blue Phase framework for mixed traffic consisting of legacy and autonomous vehicles. In addition to conventional green phases, the controller introduces AV-restricted blue phases and selects among the feasible traffic states using a decentralized pressure-based policy. The authors proved that the resulting hybrid controller retains network stability under the adopted traffic model. \citet{levin_max-pressure_2019_AIM} subsequently developed a decentralized MP policy for autonomous intersection management and dynamic lane reversal and showed that the resulting controller is throughput optimal under a stochastic queueing model. Similarly, \citet{chen_stability-based_2020} incorporated MP into autonomous intersection management with pedestrians and combined the pressure-based network controller with a trajectory optimization algorithm in the microscopic implementation. A simpler form of vehicle-motion coordination under conventional signalized operation was considered by \citet{tan_dwelling_speed_2026}, who integrated vehicle-specific speed and dwell-time advisory with MP-based transit signal priority. In this framework, MP determines the signal service, while speed and dwell-time advisory adjust the arrival and departure of transit vehicles. Collectively, these studies demonstrate that pressure-based network control can be coordinated with vehicle-level motion control, although existing work has concentrated primarily on autonomous intersection management.

Routing represents a broader, network-level form of vehicle control. Routing and service scheduling are jointly considered in the original multi-commodity backpressure framework \citep{tassiulas1990stability}, whereas most traffic-signal MP formulations subsequently treat vehicle routes or turning proportions as exogenous. \citet{taale_integrated_2015} developed an early framework that applied backpressure principles to both signal control and route guidance and showed that the combined strategy could redistribute traffic according to network conditions, although the appropriate representation of route pressure remained an important design issue. \citet{zaidi_traffic-adaptive_2015} applied multi-commodity backpressure to simultaneously control traffic signals and adaptively reroute vehicles according to their destinations. \citet{gregoire_back-pressure_2016} proposed a joint backpressure signal control and partial routing framework that directs a fraction of controllable vehicles toward eligible downstream links with the shortest queues and selects signal phases based on the resulting pressures. They also established maximum stability under the assumption that the proportion of controllable vehicles at each intersection is independent of the controller's routing decisions.  \citet{le_utility_2017} further formulated desired turning fractions as decision variables in a distributed backpressure framework, where route guidance is used to influence actual route choices, and proved network stability for feasible demand.  \citet{liu_back-pressure_2018} extended joint BP signal control and vehicle routing by introducing shadow buffers and queues that distinguish vehicles traveling along a road from those approaching an intersection, thereby accounting for vehicle travel times in pressure calculations. They incorporated a shortest-travel-time bias into the pressure formulation to encourage vehicles to follow shorter routes. At a broader network-control level, \citet{peng_a_milti_scale_2026} considered a multi-scale architecture in which route guidance and perimeter control operate at an upper level while BP is used for lower-level signal control. These studies show a progression from BP-based route guidance to frameworks in which routing and signal control interact as coordinated control components.

Overall, existing studies demonstrate that vehicle controllability can be integrated with MP at both the local vehicle-motion level and the network routing level. However, the literature remains relatively limited. Existing work has focused mainly on autonomous intersection management and routing, while vehicle-motion coordination under conventional signal operation has received much less attention. A general framework that coordinates vehicle and signal decisions in conventional signalized networks has therefore not yet been extensively investigated.

\subsection{AI-Enhanced MP Control}\label{sec:mp_ai}

AI methods have been integrated into MP control primarily by using pressure to guide reinforcement learning (RL). PressLight introduced pressure-based reward and a corresponding traffic-state representation to ensure that the learned controller is guided by the principle of MP \citep{wei_presslight_2019}, while MPLight extended this approach to large-scale networks through decentralized learning and parameter sharing \citep{MPLight2020_towards_thousands}. Subsequent studies enhanced cooperative learning by incorporating downstream traffic conditions and neighboring-intersection interactions into pressure-based states and rewards, using techniques such as graph attention, temporal convolution, and traffic-state prediction \citep{peng_learning_2021, li_cooperative_2022, boukerche_novel_2022}. Beyond state and reward design, \citet{sun_multi-agent_2024} introduced MP-guided action exploration, while \citet{cheng_reinforcement_2024} developed an MP-based reward with passenger-occupancy weighting for transit signal priority. In these approaches, pressure guides policy learning, while signal actions during deployment are generally selected by the learned policies rather than directly by the analytical MP rule.

Other studies focused on extending pressure representations to better capture traffic conditions for learning. Efficient Pressure simplified the pressure formulation for both analytical and RL-based controllers \citep{EfficientPressure2021}, while \citet{zhang_expression_2022} extended pressure to account for both queued and moving vehicles. Similar extensions incorporated moving- and waiting-vehicle pressures with MP-guided pretraining \citep{tang_duallight_2024}, approaching platoon characteristics \citep{wei_reinforcement_2023}, and lane capacity and vehicle speed \citep{yan_deep_2023}. Pressure representations have also been extended to phase-level traffic states \citep{GeneralizedPhasePressure2025l} and multiple upstream links for anticipatory control \citep{li2025MultiHopupstreamAnticipatory}. Collectively, these studies demonstrate how pressure-based representations can provide physically interpretable traffic information for learning-based signal control.

A different line of research retains MP as the analytical control backbone and uses learning to optimize specific components of the controller. \citet{wang_learning_2022} first developed a switching-curve-based MP controller to account for phase-switching losses and established its throughput optimality under the theoretical formulation. Policy-gradient RL was then used to optimize parameters associated with the switching curve and pressure weighting in the practical controller. \citet{agarwal_dynamic_2024} proposed another hybrid structure in which MP determines the phase order based on traffic pressure, while RL is used to adapt phase durations under heterogeneous traffic conditions. More recently, \citet{zhu2026stability} developed a stability-preserving RL framework for heterogeneity-aware composite MP control, where learning adapts the combination of multiple pressure components while retaining the MP structure and its stability guarantees. In contrast to MP-inspired learned policies, these approaches use learning to enhance particular parts of an MP controller while retaining pressure-based decision logic as an explicit component of the control architecture.

Existing studies therefore combine AI and MP in two fundamentally different ways. Some retain MP as the analytical control backbone and use learning to improve selected components of the controller. Others use pressure to inform a learned policy that directly determines the signal action. In the latter case, using an MP-inspired representation does not make the learned policy equivalent to analytical MP, and the maximum-stability guarantee of MP does not automatically apply to the learned controller.

\subsection{Anticipated Challenges and New Opportunities}
\label{sec:emerging_gaps}

Although several studies have integrated pressure-based control with autonomous intersection management and routing, the use of AV controllability within conventional MP-controlled signalized networks remains relatively underexplored. Conventional MP mainly reacts to the traffic state observed at a decision instant and allocates signal service according to the resulting pressure. AV control provides the possibility of modifying that state before signal service is allocated. Vehicle-motion and routing decisions can therefore reshape future traffic distributions and alter the pressures used by the signal controller. This creates an opportunity to extend MP from reactive service allocation toward more proactive traffic-state shaping within a coordinated vehicle--signal framework. It also changes the theoretical problem because traffic evolution may become partly control-dependent rather than being governed only by exogenous routing and arrivals. The stability region would consequently depend on both vehicle and signal policies, especially under mixed traffic where only part of the traffic stream is controllable.

For the integration of AI and MP, one opportunity is to use learning to enhance specific components of an analytical MP controller. Existing studies have begun to optimize switching behavior or phase duration using RL \citep{wang_learning_2022, agarwal_dynamic_2024}. The same principle can be extended to state estimation, short-term traffic prediction, and parameter adaptation. Under this approach, learning improves selected inputs or components of MP while the analytical pressure-based control structure is retained. This provides a relatively direct path for improving adaptability without completely abandoning the theoretical structure on which MP stability analysis is based.
A second direction is to allow AI to replace the analytical MP decision rule while explicitly retaining a network-stability property comparable to that of MP. Most existing MP-inspired RL controllers use pressure as a reward or state representation, but the learned policy is not required to select the maximum-pressure phase at each decision epoch \citep{wei_presslight_2019, zhang_expression_2022, GeneralizedPhasePressure2025l, li2025MultiHopupstreamAnticipatory}. Consequently, their use of pressure does not by itself provide the maximum-stability guarantee of analytical MP. This creates an opportunity to investigate learning frameworks in which signal actions are selected directly by an AI policy while stability is enforced through Lyapunov-based constraints or an analytical MP-based safety mechanism. The objective is not necessarily to force a learned controller to reproduce the MP decision, but to provide greater decision flexibility while retaining a provable network-stability property.

\section{Conclusion and Future Research Direction}\label{sec: conclusion}

This paper provides a comprehensive and structured review of MP-based traffic signal control. Early studies primarily focused on improving the basic MP formulation and its operational performance. These efforts addressed limitations of the original queue-based formulation through alternative traffic-state variables, capacity- and position-aware pressure definitions, cyclic and semi-cyclic control structures, explicit treatment of phase-switching losses, lane-based formulations, and improved specification of parameters such as turning ratios and saturation flows. More recent research has increasingly shifted toward improving the practical applicability and broader functionality of MP. This includes extensions for multimodal transportation, network coordination and perimeter control, selective deployment, connected and automated vehicle environments, and integration with artificial intelligence. These developments demonstrate how MP has evolved from a relatively simple queue-based signal control algorithm into a broader framework for decentralized traffic management.

Despite extensive research efforts, several gaps remain in the fundamental design and practical implementation of MP control. Existing formulations still provide limited treatment of interactions among movements, particularly for shared lanes, permitted turning movements, and other situations in which the service rate of one movement depends on conflicting traffic. Practical implementation also requires decisions regarding negative pressures, minimum green times, phase transitions, switching losses, work conservation, sensing requirements, and signal-update intervals, but these issues have been treated differently across studies. In particular, most time-step-based MP controllers rely on a fixed decision interval, even though the appropriate update frequency may vary with traffic demand and network conditions. Further research is therefore needed to establish more general implementation principles and to investigate adaptive signal-update mechanisms. Another important question is how much downstream traffic information is actually required, since recent studies suggest that incorporating downstream states does not necessarily improve performance under all traffic conditions.

Research on multimodal and system-level MP has considerably broadened the scope of the framework, but existing studies remain fragmented. Pedestrians, public transit, and non-motorized vehicles have generally been considered separately, while other modes such as emergency vehicles, freight traffic, and emerging micromobility have received comparatively limited attention. A unified multimodal MP framework that represents heterogeneous users consistently while preserving desirable stability properties remains an important research direction. At the network level, existing studies have demonstrated that MP can be extended to provide explicit signal coordination, regional perimeter control, and selective deployment. However, further work is needed to understand how these extensions interact, how they scale to large and heterogeneous networks, and under what operating conditions their additional information and coordination requirements provide meaningful benefits. For MP-based perimeter control, existing approaches generally rely on predefined protected regions and perimeter intersections, leaving the identification of dynamically evolving congestion boundaries insufficiently explored. Future research could therefore investigate dynamic protected-region identification, interactions among multiple congested regions, and integration of perimeter control with CAV-based route guidance to jointly regulate regional accumulation and redistribute traffic demand.

Connected and automated transportation technologies create additional opportunities for MP control. Connected vehicles can provide detailed traffic-state information, but reliable MP operation under low or heterogeneous penetration depends not only on the average penetration rate but also on whether observations remain available for individual movements. Automated vehicles introduce a further control dimension because vehicle trajectories, speeds, dwell times, and routes can potentially be coordinated with signal decisions. Existing research has demonstrated such integration in autonomous intersection management and several specialized applications, but joint vehicle--signal control under conventional signalized operation remains relatively underexplored. Future research could therefore move MP from purely reactive signal service toward proactive traffic-state shaping, in which vehicle-level control influences future pressure states before signal decisions are made. AI can also support MP through traffic-state estimation, prediction, parameter adaptation, and signal-control optimization. However, when AI directly replaces the analytical MP decision rule, the maximum-stability property of conventional MP is no longer automatically retained. Developing learning-based controllers that preserve explicit stability, robustness, interpretability, privacy, and cybersecurity properties therefore remains an important research opportunity.

Real-world implementation of MP remains limited. Existing field studies have demonstrated the feasibility of pressure-based signal control using travel-time and delay information. \citet{mercader_max-pressure_2020} implemented a cyclic travel-time-based MP controller at an intersection in Jerusalem and reported an approximately 4\% reduction in normalized travel time compared with historical fixed-time operation. \citet{dixit_simple_2020} deployed an MP-inspired crowdsourced delay-based controller at seven intersections across India and Indonesia, with maximum approach-delay reductions ranging from 12\% to 30\%. More recently, \citet{zoabi_experimental_2025} implemented a lane-level travel-time-delay MP controller at a T-intersection in Nesher, Israel, while preserving the existing phase sequence, minimum green times, and coordination requirements, and reported lower queues than the existing actuated controller. Nevertheless, most MP studies still rely on simulation, while large-scale and long-term field demonstrations remain scarce. Existing experiments also highlight practical challenges related to sensing coverage, measurement delays, data reliability, communication requirements, and compatibility with existing signal-control practices.

Future research should therefore focus not only on developing new MP formulations, but also on integrating the advances reviewed in this paper into robust and practically deployable control frameworks. Important directions include adaptive control structures, improved treatment of finite-capacity and interacting movements, unified multimodal control, scalable coordination and regional control, selective and low-cost deployment, reliable operation under partial connectivity, and coordinated vehicle--signal control. AI-based methods may further improve prediction, estimation, and adaptation, provided that stability and reliability can be maintained. Most importantly, large-scale and long-term field experiments are needed to determine whether the theoretical and simulation-based advantages demonstrated throughout the MP literature can be translated into reliable improvements under real-world traffic conditions.

\section*{CRediT authorship contribution statement}
\noindent \textbf{Abdullah Al Asif:} Conceptualization, Methodology, Formal analysis, Software, Writing -- original draft.
\textbf{Chaopeng Tan:} Conceptualization, Methodology, Writing -- original draft.
\textbf{Vikash V. Gayah:} Conceptualization, Methodology, Writing -- review \& editing.
\textbf{Hao Liu:} Conceptualization, Methodology, Supervision, Funding acquisition, Writing -- original draft.

\section*{Declaration of Generative AI and AI-assisted technologies in the writing process}
\noindent During the preparation of this work, the authors used Grammarly and ChatGPT to check grammar and improve sentence clarity and readability. After using these tools, the authors reviewed and edited the content as needed, and take full responsibility for the content of the publication.

\section*{Declaration of competing interest}
\noindent The authors declare that they have no known competing financial interests or personal relationships that could have appeared to influence the work reported in this article.

\section*{Data availability}
\noindent Data will be made available on request.

\section*{Acknowledgment}
\noindent This research was supported by NSF Grant CMMI-2554235.

\newpage

\bibliographystyle{author_date}
\bibliography{biblio}

@inproceedings{wongpiromsarn2012distributed,
  title={Distributed traffic signal control for maximum network throughput},
  author={Wongpiromsarn, Tichakorn and Uthaicharoenpong, Tawit and Wang, Yu and Frazzoli, Emilio and Wang, Danwei},
  booktitle={2012 15th international IEEE conference on intelligent transportation systems},
  pages={588--595},
  year={2012},
  organization={IEEE},
  doi={10.1109/ITSC.2012.6338817},
}

@article{keyvan-ekbatani_exploiting_2012,
	title = {Exploiting the fundamental diagram of urban networks for feedback-based gating},
	volume = {46},
	issn = {0191-2615},
	url = {https://www.sciencedirect.com/science/article/pii/S0191261512000926},
	doi = {10.1016/j.trb.2012.06.008},
	number = {10},
	urldate = {2026-09-11},
	journal = {Transportation Research Part B: Methodological},
	author = {Keyvan-Ekbatani, Mehdi and Kouvelas, Anastasios and Papamichail, Ioannis and Papageorgiou, Markos},
	month = dec,
	year = {2012},
	pages = {1393--1403},
	
}

@article{DAGANZO2007,
title = {Urban gridlock: Macroscopic modeling and mitigation approaches},
journal = {Transportation Research Part B: Methodological},
volume = {41},
number = {1},
pages = {49-62},
year = {2007},
issn = {0191-2615},
doi = {10.1016/j.trb.2006.03.001},
url = {https://www.sciencedirect.com/science/article/pii/S0191261506000282},
author = {Carlos F. Daganzo},
}

@article{geroliminis_existence_2008,
	title = {Existence of urban-scale macroscopic fundamental diagrams: {Some} experimental findings},
	volume = {42},
	issn = {0191-2615},
	shorttitle = {Existence of urban-scale macroscopic fundamental diagrams},
	url = {https://www.sciencedirect.com/science/article/pii/S0191261508000180},
	doi = {10.1016/j.trb.2008.02.002},
	number = {9},
	urldate = {2026-09-12},
	journal = {Transportation Research Part B: Methodological},
	author = {Geroliminis, Nikolas and Daganzo, Carlos F.},
	month = nov,
	year = {2008},
	pages = {759--770},
	}

@article{wang_research_2008,
	title = {Research on bicycle conversion factors},
	volume = {42},
	issn = {0965-8564},
	url = {https://www.sciencedirect.com/science/article/pii/S0965856408000839},
	doi = {10.1016/j.tra.2008.03.016},
	number = {8},
	urldate = {2026-09-08},
	journal = {Transportation Research Part A: Policy and Practice},
	author = {Wang, Dianhai and Feng, Tianjun and Liang, Chunyan},
	month = oct,
	year = {2008},
	pages = {1129--1139},
	}

@article{dai_maximum_2005,
	title = {Maximum {Pressure} {Policies} in {Stochastic} {Processing} {Networks}},
	volume = {53},
	issn = {0030-364X},
	url = {https://pubsonline.informs.org/doi/10.1287/opre.1040.0170},
	doi = {10.1287/opre.1040.0170},
	number = {2},
	urldate = {2026-10-04},
	journal = {Operations Research},
	publisher = {INFORMS},
	author = {Dai, J. G. and Lin, Wuqin},
	month = apr,
	year = {2005},
	pages = {197--218},
	}

@article{zhu2026stability,
  title={Stability-Preserving Reinforcement Learning for Heterogeneity-Aware Composite Max-Pressure Traffic Signal Control},
  author={Zhu, Jichen and Tan, Chaopeng},
  journal={arXiv preprint},
  year={2026},
  doi={10.48550/arXiv.2610.31037},
}

@article{tassiulas1990stability,
  author={Tassiulas, Leandros and Ephremides, Anthony},
  journal={IEEE Transactions on Automatic Control}, 
  title={Stability properties of constrained queueing systems and scheduling policies for maximum throughput in multihop radio networks}, 
  year={1992},
  volume={37},
  number={12},
  pages={1936-1948},
  doi={10.1109/9.182479}}

@article{mirchandani2001real,
title = {A real-time traffic signal control system: architecture, algorithms, and analysis},
journal = {Transportation Research Part C: Emerging Technologies},
volume = {9},
number = {6},
pages = {415-432},
year = {2001},
doi = {10.1016/S0968-090X(00)00047-4},
author = {Pitu Mirchandani and Larry Head},
}

@article{lowrie1990scats,
  title={Scats, sydney co-ordinated adaptive traffic system: A traffic responsive method of controlling urban traffic},
  author={Lowrie, PR},
  year={1990}
}

@inproceedings{gartner2001implementation,
  title={Implementation of the OPAC adaptive control strategy in a traffic signal network},
  author={Gartner, Nathan H and Pooran, Farhad J and Andrews, Christina M},
  booktitle={ITSC 2001. 2001 IEEE Intelligent Transportation Systems. Proceedings (Cat. No. 01TH8585)},
  pages={195--200},
  year={2001},
  organization={IEEE}
}

@article{hunt1982scoot,
  title={The SCOOT on-line traffic signal optimisation technique},
  author={Hunt, P. B. and Robertson, D. I. and Bretherton, R. D. and Royle, M. C.},
  journal={Traffic Engineering \& Control},
  volume={23},
  number={4},
  year={1982}
}

@inbook{xue_partially_2025,
	title = {A {Partially} {Implemented} {Max} {Pressure} for {Traffic} {Network} {Control}},
	doi = {10.1061/9780784486269.171},
	publisher = {American Society of Civil Engineers},
	author = {Xue, Xiaolong and Lin, Dianchao and Li, Li},
	month = oct,
	year = {2025},
	  booktitle={CICTP 2025},
	pages = {1797--1807},
	
}

@article{gao_arrival-based_2022,
	title = {Arrival-{Based} {Backpressure} {Traffic} {Signal} {Control}},
	volume = {2676},
	doi = {10.1177/03611981221086627},
	language = {en},
	number = {9},
	urldate = {2026-04-28},
	journal = {Transportation Research Record: Journal of the Transportation Research Board},
	author = {Gao, Hang and Zhang, H. Michael},
	month = sep,
	year = {2022},
	pages = {172--186},
}

@article{chen_stability-based_2020,
	title = {Stability-based analysis of autonomous intersection management with pedestrians},
	volume = {114},
	url = {https://www.sciencedirect.com/science/article/pii/S0968090X19309623},
	doi = {10.1016/j.trc.2020.01.016},
	urldate = {2026-03-19},
	journal = {Transportation Research Part C: Emerging Technologies},
	author = {Chen, Rongsheng and Hu, Jeffrey and Levin, Michael W. and Rey, David},
	month = may,
	year = {2020},
	pages = {463--483},
}

@article{tan_cv-mp_2026,
	title = {{CV}-{MP}: {Max}-pressure control in heterogeneously distributed and partially connected vehicle environments},
	volume = {204},
	shorttitle = {{CV}-{MP}},
	doi = {10.1016/j.trb.2025.103387},
	urldate = {2026-03-19},
	journal = {Transportation Research Part B: Methodological},
	author = {Tan, Chaopeng and Sun, Dingshan and Liu, Hao and Rinaldi, Marco and van Lint, Hans},
	month = feb,
	year = {2026},
	pages = {103387},

}

@inproceedings{cao_max_2020,
	title = {A {Max} {Pressure} {Approach} to {Urban} {Network} {Signal} {Control} with {Queue} {Estimation} using {Connected} {Vehicle} {Data}},
	url = {https://ieeexplore.ieee.org/document/9294361},
	doi = {10.1109/ITSC45102.2020.9294361},
	urldate = {2026-03-19},
	booktitle = {2020 {IEEE} 23rd {International} {Conference} on {Intelligent} {Transportation} {Systems} ({ITSC})},
	author = {Cao, Jingnan and Hu, Yonghui and Diamantis, Manolis and Zhang, Siyu and Wang, Yibing and Guo, Jingqiu and Papamichail, Ioannis and Papageorgiou, Markos and Zhang, Lihui and Hu, Jun Simon},
	month = sep,
	year = {2020},
	pages = {1--8},
}

@article{cui_throughput-delay_2026,
	title = {Throughput-{Delay} {Tradeoff} {Management} for {Partially} {Connected} {Networks} via {Lyapunov} {Drift} {Optimization}},
	url = {https://ieeexplore.ieee.org/document/11370545},
	doi = {10.1109/TITS.2026.3659040},
	urldate = {2026-03-25},
	journal = {IEEE Transactions on Intelligent Transportation Systems},
	author = {Cui, Shaohua and Wang, Lichao and Xue, Yongjie and Yang, Kaidi and Gao, Kun},
	year = {2026},
	pages = {1--17},
}

@inproceedings{li_automated_2021,
	title = {Automated {Vehicle} {Planning} {Data}-assisted {Method} for {Estimating} {Turning} {Ratios} and {Optimizing} {Signals} at {Connected} {Arterials}},
	url = {https://ieeexplore.ieee.org/document/9564583},
	doi = {10.1109/ITSC48978.2021.9564583},
	urldate = {2026-04-29},
	booktitle = {2021 {IEEE} {International} {Intelligent} {Transportation} {Systems} {Conference} ({ITSC})},
	author = {Li, Jiangchen and Peng, Liqun},
	month = sep,
	year = {2021},
	pages = {1057--1062},
	
}

@article{hoang_optimal_2024,
	title = {Optimal decentralized signal control for platooning in connected vehicle networks},
	volume = {167},
	doi = {10.1016/j.trc.2024.104832},
	journal = {Transportation Research Part C: Emerging Technologies},
	author = {Hoang, The Anh and Walton, Neil and Vu, Hai L.},
	month = oct,
	year = {2024},
	pages = {104832},

}

@article{anderson_stability_2018,
	title = {Stability and implementation of a cycle-based max pressure controller for signalized traffic networks},
	volume = {13},
	copyright = {http://creativecommons.org/licenses/by/3.0/},
	url = {https://www.aimsciences.org/en/article/doi/10.3934/nhm.2018011},
	doi = {10.3934/nhm.2018011},
	language = {en},
	number = {2},
	urldate = {2026-03-02},
	journal = {Networks and Heterogeneous Media},
	publisher = {Networks and Heterogeneous Media},
	author = {Anderson, Leah and Pumir, Thomas and Triantafyllos, Dimitrios and Bayen, Alexandre M.},
	month = may,
	year = {2018},
	pages = {241--260},
}

@article{le_decentralized_2015,
	series = {Special {Issue}: {Advanced} {Road} {Traffic} {Control}},
	title = {Decentralized signal control for urban road networks},
	volume = {58},
	url = {https://www.sciencedirect.com/science/article/pii/S0968090X14003325},
	doi = {10.1016/j.trc.2014.11.009},
	urldate = {2026-03-17},
	journal = {Transportation Research Part C: Emerging Technologies},
	author = {Le, Tung and Kovács, Péter and Walton, Neil and Vu, Hai L. and Andrew, Lachlan L. H. and Hoogendoorn, Serge S. P.},
	month = sep,
	year = {2015},
	pages = {431--450},
}

@article{levin_mp_2020_cyclic,
	title = {Max-pressure signal control with cyclical phase structure},
	volume = {120},
	url = {https://www.sciencedirect.com/science/article/pii/S0968090X20307324},
	doi = {10.1016/j.trc.2020.102828},
	urldate = {2026-03-18},
	journal = {Transportation Research Part C: Emerging Technologies},
	author = {Levin, Michael W. and Hu, Jeffrey and Odell, Michael},
	month = nov,
	year = {2020},
	pages = {102828},
}

@inproceedings{pumir_stability_2015,
	title = {Stability of modified max pressure controller with application to signalized traffic networks},
	url = {https://ieeexplore.ieee.org/document/7171007},
	doi = {10.1109/ACC.2015.7171007},
	urldate = {2026-03-25},
	booktitle = {2015 {American} {Control} {Conference} ({ACC})},
	author = {Pumir, Thomas and Anderson, Leah and Triantafyllos, Dimitrios and Bayen, Alexandre M.},
	month = jul,
	year = {2015},
	note = {ISSN: 2378-5861},
	pages = {1879--1886},
}

@article{robbennolt_microsimulation_2022,
	title = {Microsimulation {Study} {Evaluating} the {Benefits} of {Cyclic} and {Non}-{Cyclic} {Max}-{Pressure} {Control} of {Signalized} {Intersections}},
	volume = {2676},
	url = {https://journals.sagepub.com/doi/10.1177/03611981221095520},
	doi = {10.1177/03611981221095520},
	language = {en},
	number = {12},
	urldate = {2026-04-29},
	journal = {Transportation Research Record: Journal of the Transportation Research Board},
	author = {Robbennolt, Jake and Chen, Rongsheng and Levin, Michael},
	month = dec,
	year = {2022},
	pages = {303--317},
}

@article{barman_performance_2022,
	title = {Performance {Evaluation} of {Modified} {Cyclic} {Max}-{Pressure} {Controlled} {Intersections} in {Realistic} {Corridors}},
	volume = {2676},
	url = {https://journals.sagepub.com/doi/10.1177/03611981211072807},
	doi = {10.1177/03611981211072807},
	language = {en},
	number = {6},
	urldate = {2026-04-29},
	journal = {Transportation Research Record: Journal of the Transportation Research Board},
	author = {Barman, Simanta and Levin, Michael W.},
	month = jun,
	year = {2022},
	pages = {110--128},
}

@article{sun_simulation_2018,
	title = {A {Simulation} {Study} on {Max} {Pressure} {Control} of {Signalized} {Intersections}},
	volume = {2672},
	url = {https://journals.sagepub.com/doi/10.1177/0361198118786840},
	doi = {10.1177/0361198118786840},
	language = {en},
	number = {18},
	urldate = {2026-04-29},
	journal = {Transportation Research Record: Journal of the Transportation Research Board},
	author = {Sun, Xiaotong and Yin, Yafeng},
	month = dec,
	year = {2018},
	pages = {117--127},
}

@article{wu_delay-based_2018,
	title = {Delay-{Based} {Traffic} {Signal} {Control} for {Throughput} {Optimality} and {Fairness} at an {Isolated} {Intersection}},
	volume = {67},
	issn = {1939-9359},
	url = {https://ieeexplore.ieee.org/abstract/document/8063346},
	doi = {10.1109/TVT.2017.2760820},
	number = {2},
	urldate = {2026-03-18},
	journal = {IEEE Transactions on Vehicular Technology},
	author = {Wu, Jian and Ghosal, Dipak and Zhang, Michael and Chuah, Chen-Nee},
	month = feb,
	year = {2018},
	pages = {896--909},

}

@article{dixit_simple_2020,
	title = {A simple crowdsourced delay-based traffic signal control},
	volume = {15},
	issn = {1932-6203},
	url = {https://journals.plos.org/plosone/article?id=10.1371/journal.pone.0230598},
	doi = {10.1371/journal.pone.0230598},
	language = {en},
	number = {4},
	urldate = {2026-03-18},
	journal = {PLOS ONE},
	publisher = {Public Library of Science},
	author = {Dixit, Vinayak and Nair, Divya Jayakumar and Chand, Sai and Levin, Michael W.},
	month = apr,
	year = {2020},
	pages = {e0230598},
}

@article{liu_total-delay-based_2023,
	title = {Total-delay-based {Max} {Pressure}: {A} {Max} {Pressure} {Algorithm} {Considering} {Delay} {Equity}},
	volume = {2677},
	issn = {0361-1981},
	shorttitle = {Total-delay-based {Max} {Pressure}},
	doi = {10.1177/03611981221147051},
	language = {EN},
	number = {6},
	urldate = {2026-03-19},
	journal = {Transportation Research Record},
	publisher = {SAGE Publications Inc},
	author = {Liu, Hao and Gayah, Vikash V.},
	month = jun,
	year = {2023},
	pages = {324--339},
}

@article{zoabi_experimental_2025,
	title = {An experimental study on {Max}-{Pressure} traffic controller based on travel time delays},
	volume = {175},
	issn = {0968-090X},
	url = {https://www.sciencedirect.com/science/article/pii/S0968090X25000749},
	doi = {10.1016/j.trc.2025.105070},
	urldate = {2026-03-25},
	journal = {Transportation Research Part C: Emerging Technologies},
	author = {Zoabi, Razi and Haddad, Jack},
	month = jun,
	year = {2025},
	pages = {105070},

}

@INPROCEEDINGS{Zoabi_dynamic_time_step,
  author={Zoabi, Razi and Kulcsar, Balazs and Haddad, Jack},
  booktitle={2025 IEEE 64th Conference on Decision and Control (CDC)}, 
  title={Dynamic Time-Step Max-Pressure Controller Considering Phase-Switching Gaps for Signalized Networks}, 
  year={2025},
  volume={},
  number={},
  pages={218-223},
  doi={10.1109/CDC57313.2025.11312328}}

@article{zoabi_max-pressure_2024,
	series = {17th {IFAC} {Symposium} on {Control} of {Transportation} {Systems} {CTS} 2024},
	title = {Max-{Pressure} {Control} utilizing {Total} {Movement} {Delay}},
	volume = {58},
	doi = {10.1016/j.ifacol.2024.07.341},
	number = {10},
	urldate = {2026-04-28},
	journal = {IFAC-PapersOnLine},
	author = {Zoabi, Razi and Haddad, Jack},
	month = jan,
	year = {2024},
	pages = {206--212},
	
}

@article{smith_backpressure_2024,
	title = {Backpressure or no backpressure? {Two} simple examples},
	volume = {161},
	issn = {0968-090X},
	shorttitle = {Backpressure or no backpressure?},
	url = {https://www.sciencedirect.com/science/article/pii/S0968090X24000366},
	doi = {10.1016/j.trc.2024.104515},
	urldate = {2026-04-30},
	journal = {Transportation Research Part C: Emerging Technologies},
	author = {Smith, Michael J. and Mounce, Richard},
	month = apr,
	year = {2024},
	pages = {104515},
}

@inproceedings{cheng_lane-based_2024,
	title = {Lane-{Based} {Max}-{Pressure} {Traffic} {Signal} {Control}},
	url = {https://ieeexplore.ieee.org/document/10703896},
	doi = {10.1109/ICTLE62418.2024.10703896},
	urldate = {2026-04-28},
	booktitle = {2024 12th {International} {Conference} on {Traffic} and {Logistic} {Engineering} ({ICTLE})},
	author = {Cheng, H. K. and Kou, K. P. and Wong, K. I.},
	month = aug,
	year = {2024},
	pages = {67--71},
}

@article{levin_max-pressure_2019_AIM,
	title = {Max-pressure control of dynamic lane reversal and autonomous intersection management},
	volume = {7},
	issn = {2168-0566},
	url = {https://doi.org/10.1080/21680566.2019.1691675},
	doi = {10.1080/21680566.2019.1691675},
	number = {1},
	urldate = {2026-05-15},
	journal = {Transportmetrica B: Transport Dynamics},
	publisher = {Taylor \& Francis},
	author = {Levin, Michael W. and Rey, David and Schwartz, Adam},
	month = dec,
	year = {2019},
	pages = {1693--1718},

}

@inproceedings{zhang_joint_2020,
	title = {Joint {Queue} {Estimation} and {Max} {Pressure} {Control} for {Signalized} {Urban} {Networks} with {Connected} {Vehicles}},
	url = {https://ieeexplore.ieee.org/document/9264889},
	doi = {10.1109/FISTS46898.2020.9264889},
	urldate = {2026-03-19},
	booktitle = {2020 {Forum} on {Integrated} and {Sustainable} {Transportation} {Systems} ({FISTS})},
	author = {Zhang, Siyu and Diamantis, Manolis and Wang, Yibing and Guo, Jingqiu and Papamichail, Ioannis and Papageorgiou, Markos and Zhang, Lihui and Hu, Jun Simon},
	month = nov,
	year = {2020},
	pages = {211--217},
}

@article{li_backpressure_2021,
	title = {Backpressure control with estimated queue lengths for urban network traffic},
	volume = {15},
	copyright = {© 2021 The Authors. IET Intelligent Transport Systems published by John Wiley \& Sons Ltd on behalf of The Institution of Engineering and Technology},
	issn = {1751-9578},
	url = {https://onlinelibrary.wiley.com/doi/abs/10.1049/itr2.12027},
	doi = {10.1049/itr2.12027},
	language = {en},
	number = {2},
	urldate = {2026-03-19},
	journal = {IET Intelligent Transport Systems},
	author = {Li, Li and Okoth, Victor and Jabari, Saif Eddin},
	year = {2021},
	pages = {320--330},
}

@article{liu_novel_2022,
	title = {A novel {Max} {Pressure} algorithm based on traffic delay},
	volume = {143},
	issn = {0968-090X},
	url = {https://www.sciencedirect.com/science/article/pii/S0968090X22002303},
	doi = {10.1016/j.trc.2022.103803},
	urldate = {2026-03-18},
	journal = {Transportation Research Part C: Emerging Technologies},
	author = {Liu, Hao and Gayah, Vikash V.},
	month = oct,
	year = {2022},
	pages = {103803},
}

@inproceedings{xiao_throughput_2015,
	title = {Throughput optimality of extended back-pressure traffic signal control algorithm},
	url = {https://ieeexplore.ieee.org/document/7158897},
	doi = {10.1109/MED.2015.7158897},
	urldate = {2026-05-14},
	booktitle = {2015 23rd {Mediterranean} {Conference} on {Control} and {Automation} ({MED})},
	author = {Xiao, Nan and Frazzoli, Emilio and Luo, Yiwen and Li, Yitong and Wang, Yu and Wang, Danwei},
	month = jun,
	year = {2015},
	pages = {1059--1064},
}

@inproceedings{gregoire_back-pressure_2016,
	title = {Back-pressure traffic signal control with partial routing control},
	url = {https://ieeexplore.ieee.org/document/7799309},
	doi = {10.1109/CDC.2016.7799309},
	urldate = {2026-04-29},
	booktitle = {2016 {IEEE} 55th {Conference} on {Decision} and {Control} ({CDC})},
	author = {Gregoire, Jean and Samaranayake, Samitha and Frazzoli, Emilio},
	month = dec,
	year = {2016},
	pages = {6753--6758},
}

@inproceedings{zaidi_traffic-adaptive_2015,
	title = {Traffic-adaptive signal control and vehicle routing using a decentralized back-pressure method},
	url = {https://ieeexplore.ieee.org/document/7330998},
	doi = {10.1109/ECC.2015.7330998},
	urldate = {2026-05-15},
	booktitle = {2015 {European} {Control} {Conference} ({ECC})},
	author = {Zaidi, Ali A. and Kulcsár, Balázs and Wymeersch, Henk},
	month = jul,
	year = {2015},
	pages = {3029--3034},
}

@inproceedings{liu_back-pressure_2018,
	title = {Back-{Pressure} {Based} {Adaptive} {Traffic} {Signal} {Control} and {Vehicle} {Routing} with {Real}-{Time} {Control} {Information} {Update}},
	url = {https://ieeexplore.ieee.org/document/8519601},
	doi = {10.1109/ICVES.2018.8519601},
	urldate = {2026-05-15},
	booktitle = {2018 {IEEE} {International} {Conference} on {Vehicular} {Electronics} and {Safety} ({ICVES})},
	author = {Liu, Ying and Gao, Juntao and Ito, Minoru},
	month = sep,
	year = {2018},
	pages = {1--6},
	
}

@article{ahmed_c-mp_2025,
	title = {C-{MP}: {A} decentralized adaptive-coordinated traffic signal control using the {Max} {Pressure} framework},
	volume = {200},
	issn = {0191-2615},
	shorttitle = {C-{MP}},
	url = {https://www.sciencedirect.com/science/article/pii/S0191261525001572},
	doi = {10.1016/j.trb.2025.103308},
	urldate = {2026-03-19},
	journal = {Transportation Research Part B: Methodological},
	author = {Ahmed, Tanveer and Liu, Hao and Gayah, Vikash V.},
	month = oct,
	year = {2025},
	pages = {103308},
}

@inproceedings{tutsch_distributed_2024,
	title = {Distributed {Traffic} {Signal} {Control} via {Coordinated} {Maximum} {Pressure}-{Plus}-{Penalty}},
	issn = {2153-0017},
	url = {https://ieeexplore.ieee.org/document/10920011},
	doi = {10.1109/ITSC58415.2024.10920011},
	urldate = {2026-04-28},
	booktitle = {2024 {IEEE} 27th {International} {Conference} on {Intelligent} {Transportation} {Systems} ({ITSC})},
	author = {Tütsch, Vinzenz and He, Zhiyu and Dörfler, Florian and Zhang, Kenan},
	month = sep,
	year = {2024},
	note = {ISSN: 2153-0017},
	pages = {311--316},
	
}

@article{xu_ped-mp_2024,
	title = {Ped-MP: A Pedestrian-Friendly Max-Pressure Signal Control Policy for City Networks},
	volume = {150},
	shorttitle = {Ped-{MP}},
	url = {https://ascelibrary.org/doi/10.1061/JTEPBS.TEENG-7956},
	doi = {10.1061/JTEPBS.TEENG-7956},
	language = {en},
	number = {7},
	urldate = {2026-03-19},
	journal = {Journal of Transportation Engineering, Part A: Systems},
	publisher = {American Society of Civil Engineers},
	author = {Xu, Te and Bika, Yashveer and Levin, Michael W.},
	month = jul,
	year = {2024a},
	pages = {04024028},
    
	
}

@article{xu_smoothing-mp_2024,
	title = {Smoothing-MP: A novel max-pressure signal control considering signal coordination to smooth traffic in urban networks},
	volume = {166},
	issn = {0968-090X},
	shorttitle = {Smoothing-{MP}},
	doi = {10.1016/j.trc.2024.104760},
	urldate = {2026-04-28},
	journal = {Transportation Research Part C: Emerging Technologies},
	author = {Xu, Te and Barman, Simanta and Levin, Michael W.},
	month = sep,
	year = {2024b},
	pages = {104760},
   
	
}

@article{li_max_2025,
	title = {Max {Pressure} {Signal} {Control} {Method} for {Collaborative} {Optimization} of {Motor} and {Nonmotor} {Vehicles}},
	volume = {151},
	url = {https://ascelibrary.org/doi/10.1061/JTEPBS.TEENG-8889},
	doi = {10.1061/JTEPBS.TEENG-8889},
	language = {en},
	number = {7},
	urldate = {2026-03-25},
	journal = {Journal of Transportation Engineering, Part A: Systems},
	publisher = {American Society of Civil Engineers},
	author = {Li, Bing and Li, Jielong and Yin, Juyuan and Wang, Zhenghui and Gao, Jiandong},
	month = jul,
	year = {2025a},
	pages = {04025051},
}

@article{varaiya_max_2013,
	title = {Max pressure control of a network of signalized intersections},
	volume = {36},
	issn = {0968-090X},
	url = {https://www.sciencedirect.com/science/article/pii/S0968090X13001782},
	doi = {10.1016/j.trc.2013.08.014},
	urldate = {2026-03-03},
	journal = {Transportation Research Part C: Emerging Technologies},
	author = {Varaiya, Pravin},
	month = nov,
	year = {2013},
	pages = {177--195},
}

@article{kouvelas_maximum_2014,
	title = {Maximum {Pressure} {Controller} for {Stabilizing} {Queues} in {Signalized} {Arterial} {Networks}},
	volume = {2421},
	issn = {0361-1981},
	url = {https://doi.org/10.3141/2421-15},
	doi = {10.3141/2421-15},
	language = {EN},
	number = {1},
	urldate = {2026-08-23},
	journal = {Transportation Research Record},
	publisher = {SAGE Publications Inc},
	author = {Kouvelas, Anastasios and Lioris, Jennie and Fayazi, S. Alireza and Varaiya, Pravin},
	month = jan,
	year = {2014},
	pages = {133--141},
	
}

@inproceedings{wunderlich_stable_2007,
	title = {A stable longest queue first signal scheduling algorithm for an isolated intersection},
	url = {https://ieeexplore.ieee.org/document/4456393},
	doi = {10.1109/ICVES.2007.4456393},
	urldate = {2026-03-17},
	booktitle = {2007 {IEEE} {International} {Conference} on {Vehicular} {Electronics} and {Safety}},
	author = {Wunderlich, R. and Elhanany, I. and Urbanik, T.},
	month = dec,
	year = {2007},
	pages = {1--6},
}

@article{liu_max_2024,
	title = {A max pressure algorithm for traffic signals considering pedestrian queues},
	volume = {169},
	issn = {0968-090X},
	url = {https://www.sciencedirect.com/science/article/pii/S0968090X24003863},
	doi = {10.1016/j.trc.2024.104865},
	urldate = {2026-03-19},
	journal = {Transportation Research Part C: Emerging Technologies},
	author = {Liu, Hao and Gayah, Vikash V. and Levin, Michael W.},
	month = dec,
	year = {2024},
	pages = {104865},
}

@inproceedings{kouvelas_two-layer_2018,
	title = {Two-layer hierarchical control for large-scale urban traffic networks},
	url = {https://ieeexplore.ieee.org/document/8550499},
	doi = {10.23919/ECC.2018.8550499},
	urldate = {2026-04-28},
	booktitle = {2018 {European} {Control} {Conference} ({ECC})},
	author = {Kouvelas, Anastasios and Triantafyllos, Dimitris and Geroliminis, Nikolas},
	month = jun,
	year = {2018},
	pages = {1295--1300},
}

@inproceedings{putra_design_2025,
	title = {Design of {Perimeter} {Control} for {Urban} {Traffic} {Network} {Combining} {Proportional} {Integral} {Control} with the {Max} {Pressure} {Algorithm}},
	issn = {2835-3382},
	url = {https://ieeexplore.ieee.org/document/11183900},
	doi = {10.1109/ICEVT67191.2025.11183900},
	urldate = {2026-04-28},
	booktitle = {2025 8th {International} {Conference} on {Electric} {Vehicular} {Technology} ({ICEVT})},
	author = {Putra, I Putu Agus Ananda Giri and Sergio, Mesa Anugerah Kenan and Joelianto, Endra and Burohman, Azka Muji and Sutarto, Herman Y. and Mukhlish, Faqihza},
	month = aug,
	year = {2025},
	note = {ISSN: 2835-3382},
	pages = {172--178},
}

@article{liu_n-mp_2024,
	series = {Collected {Papers} from the 25th {International} {Symposium} on {Transportation} and {Traffic} {Theory} ({ISTTT25})},
	title = {N-{MP}: {A} network-state-based {Max} {Pressure} algorithm incorporating regional perimeter control},
	volume = {168},
	issn = {0968-090X},
	shorttitle = {N-{MP}},
	url = {https://www.sciencedirect.com/science/article/pii/S0968090X24002468},
	doi = {10.1016/j.trc.2024.104725},
	urldate = {2026-04-28},
	journal = {Transportation Research Part C: Emerging Technologies},
	author = {Liu, Hao and Gayah, Vikash V.},
	month = nov,
	year = {2024},
	pages = {104725},
}

@article{tsitsokas_two-layer_2023,
	title = {Two-layer adaptive signal control framework for large-scale dynamically-congested networks: {Combining} efficient {Max} {Pressure} with {Perimeter} {Control}},
	volume = {152},
	issn = {0968-090X},
	shorttitle = {Two-layer adaptive signal control framework for large-scale dynamically-congested networks},
	url = {https://www.sciencedirect.com/science/article/pii/S0968090X23001171},
	doi = {10.1016/j.trc.2023.104128},
	urldate = {2026-04-28},
	journal = {Transportation Research Part C: Emerging Technologies},
	author = {Tsitsokas, Dimitrios and Kouvelas, Anastasios and Geroliminis, Nikolas},
	month = jul,
	year = {2023},
	pages = {104128},
}

@article{sun_max-pressure_2025,
	title = {Max-{Pressure} {Controller} for {Traffic} {Networks} {Considering} the {Phase} {Switching} {Loss}},
	volume = {17},
	copyright = {http://creativecommons.org/licenses/by/3.0/},
	issn = {2071-1050},
	url = {https://www.mdpi.com/2071-1050/17/10/4492},
	doi = {10.3390/su17104492},
	language = {en},
	number = {10},
	urldate = {2026-03-25},
	journal = {Sustainability},
	publisher = {Multidisciplinary Digital Publishing Institute},
	author = {Sun, Jiayu and Wang, Yibing and Yang, Hang and Zhang, Zhao and Papageorgiou, Markos and Liu, Guiyun and Zheng, Pengjun},
	month = may,
	year = {2025},
}

@article{wang_learning_2022,
	title = {Learning the max pressure control for urban traffic networks considering the phase switching loss},
	volume = {140},
	issn = {0968-090X},
	url = {https://www.sciencedirect.com/science/article/pii/S0968090X22001139},
	doi = {10.1016/j.trc.2022.103670},
	urldate = {2026-04-28},
	journal = {Transportation Research Part C: Emerging Technologies},
	author = {Wang, Xingmin and Yin, Yafeng and Feng, Yiheng and Liu, Henry X.},
	month = jul,
	year = {2022},
	pages = {103670},
}

@incollection{hsieh_throughput-optimal_2017,
	title = {Throughput-{Optimal} {Scheduling} for {Multi}-{Hop} {Networked} {Transportation} {Systems} {With} {Switch}-{Over} {Delay}},
	publisher = {Association for Computing Machinery},
	url = {https://dl.acm.org/doi/10.1145/3084041.3084065},
	doi = {10.1145/3084041.3084065},
	urldate = {2026-04-28},
	booktitle = {Proceedings of the 18th {ACM} {International} {Symposium} on {Mobile} {Ad} {Hoc} {Networking} and {Computing}},
	author={Hsieh, Ping-Chun and Liu, Xi and Jiao, Jian and Hou, I-Hong and Zhang, Yunlong and Kumar, PR},
	month = jul,
	year = {2017},
	pages = {1--10},

}

@article{li_position_2019,
	title = {Position weighted backpressure intersection control for urban networks},
	volume = {128},
	issn = {0191-2615},
	url = {https://www.sciencedirect.com/science/article/pii/S0191261518307896},
	doi = {10.1016/j.trb.2019.08.005},
	urldate = {2026-03-17},
	journal = {Transportation Research Part B: Methodological},
	author = {Li, Li and Jabari, Saif Eddin},
	month = oct,
	year = {2019},
	pages = {435--461},
}

@article{gregoire_capacity-aware_2015,
	title = {Capacity-{Aware} {Backpressure} {Traffic} {Signal} {Control}},
	volume = {2},
	issn = {2325-5870},
	url = {https://ieeexplore.ieee.org/document/6979216},
	doi = {10.1109/TCNS.2014.2378871},
	number = {2},
	urldate = {2026-03-18},
	journal = {IEEE Transactions on Control of Network Systems},
	author = {Gregoire, Jean and Qian, Xiangjun and Frazzoli, Emilio and de La Fortelle, Arnaud and Wongpiromsarn, Tichakorn},
	month = jun,
	year = {2015},
	pages = {164--173},
}

@inproceedings{riehl_green-pressure_2025,
	title = {Green-{Pressure} – {A} {Weighted} {Queue}-{Length} {Approach} {Towards} {Sustainable} {Intersection} {Management}},
	issn = {2576-2370},
	url = {https://ieeexplore.ieee.org/document/11312349},
	doi = {10.1109/CDC57313.2025.11312349},
	urldate = {2026-03-18},
	booktitle = {2025 {IEEE} 64th {Conference} on {Decision} and {Control} ({CDC})},
	author = {Riehl, Kevin and Kouvelas, Anastasios and Makridis, Michail A.},
	month = dec,
	year = {2025},
	note = {ISSN: 2576-2370},
	pages = {4964--4969},
}

@article{zoabi_enhanced_2026,
	title = {Enhanced queue-based {Max}-{Pressure} traffic signal control},
	volume = {186},
	issn = {0968-090X},
	url = {https://www.sciencedirect.com/science/article/pii/S0968090X26000264},
	doi = {10.1016/j.trc.2026.105538},
	urldate = {2026-03-25},
	journal = {Transportation Research Part C: Emerging Technologies},
	author = {Zoabi, Razi and Haddad, Jack},
	month = may,
	year = {2026},
	pages = {105538},
}

@article{xu_approximate_2025,
	title = {An approximate position-weighted back-pressure traffic signal control policy for traffic networks},
	doi = {10.1080/23249935.2025.2509089},
	urldate = {2026-04-28},
	journal = {Transportmetrica A: Transport Science},
	publisher = {Taylor \& Francis},
	author = {Xu, Te and Bika, Yashveer and Levin, Michael W.},
	month = may,
	year = {2025},
	pages = {1--34},

}

@article{hao_backpressure_2020,
	title = {Backpressure based traffic signal control considering capacity of downstream links},
	volume = {35},
	copyright = {Copyright (c) 2020 The Author(s). Published by Vilnius Gediminas Technical University.},
	issn = {1648-3480},
	url = {https://journals.vilniustech.lt/index.php/Transport/article/view/13288},
	doi = {10.3846/transport.2020.13288},
	language = {en},
	number = {4},
	urldate = {2026-04-28},
	journal = {Transport},
	author = {Hao, Shenxue and Yang, Licai and Shi, Yunfeng and Guo, Yajuan},
	month = sep,
	year = {2020},
	pages = {347--356},
}

@article{maripini_queue_2025,
	title = {Queue {Dissipation}-{Based} {Max} {Pressure} {Signal} {Control} {Using} {Vehicle} {Re}-{Identification} {Data}},
	volume = {13},
	issn = {2169-3536},
	url = {https://ieeexplore.ieee.org/document/11173433},
	doi = {10.1109/ACCESS.2025.3612125},
	urldate = {2026-04-28},
	journal = {IEEE Access},
	author = {Maripini, Himabindu and Ghosh, Tamojit and Vanajakshi, Lelitha},
	year = {2025},
	pages = {166530--166541},
}

@article{levin_max-pressure_2023_review,
	title = {Max-{Pressure} {Traffic} {Signal} {Timing}: {A} {Summary} of {Methodological} and {Experimental} {Results}},
	volume = {149},
	shorttitle = {Max-{Pressure} {Traffic} {Signal} {Timing}},
	url = {https://ascelibrary.org/doi/10.1061/JTEPBS.TEENG-7578},
	doi = {10.1061/JTEPBS.TEENG-7578},
	language = {en},
	number = {4},
	urldate = {2026-03-17},
	journal = {Journal of Transportation Engineering, Part A: Systems},
	publisher = {American Society of Civil Engineers},
	author = {Levin, Michael W.},
	month = apr,
	year = {2023},
	pages = {03123001},
}

@article{arel_reinforcement_2010,
	title = {Reinforcement learning-based multi-agent system for network traffic signal control},
	volume = {4},
	url = {https://digital-library.theiet.org/doi/10.1049/iet-its.2009.0070},
	doi = {10.1049/iet-its.2009.0070},
	number = {2},
	urldate = {2026-03-02},
	journal = {IET Intelligent Transport Systems},
	publisher = {The Institution of Engineering and Technology},
	author = {Arel, I. and Liu, C. and Urbanik, T. and Kohls, A.G.},
	month = jun,
	year = {2010},
	pages = {128--135},
}

@inproceedings{wei_presslight_2019,
	address = {New York, NY, USA},
	series = {{KDD} '19},
	title = {{PressLight}: {Learning} {Max} {Pressure} {Control} to {Coordinate} {Traffic} {Signals} in {Arterial} {Network}},
	isbn = {978-1-4503-6201-6},
	shorttitle = {{PressLight}},
	url = {https://dl.acm.org/doi/10.1145/3292500.3330949},
	doi = {10.1145/3292500.3330949},
	urldate = {2026-03-19},
	booktitle = {Proceedings of the 25th {ACM} {SIGKDD} {International} {Conference} on {Knowledge} {Discovery} \& {Data} {Mining}},
	publisher = {Association for Computing Machinery},
	author = {Wei, Hua and Chen, Chacha and Zheng, Guanjie and Wu, Kan and Gayah, Vikash and Xu, Kai and Li, Zhenhui},
	month = jul,
	year = {2019},
	pages = {1290--1298},
}

@article{agarwal_dynamic_2024,
	title = {Dynamic traffic signal control for heterogeneous traffic conditions using {Max} {Pressure} and {Reinforcement} {Learning}},
	volume = {254},
	issn = {0957-4174},
	url = {https://www.sciencedirect.com/science/article/pii/S095741742401282X},
	doi = {10.1016/j.eswa.2024.124416},
	urldate = {2026-04-28},
	journal = {Expert Systems with Applications},
	author = {Agarwal, Amit and Sahu, Deorishabh and Mohata, Rishabh and Jeengar, Kuldeep and Nautiyal, Anuj and Saxena, Dhish Kumar},
	month = nov,
	year = {2024},
	pages = {124416},
}

@article{cheng_reinforcement_2024,
	title = {Reinforcement {Learning} for {Transit} {Signal} {Priority} with {Priority} {Factor}},
	volume = {7},
	copyright = {http://creativecommons.org/licenses/by/3.0/},
	issn = {2624-6511},
	url = {https://www.mdpi.com/2624-6511/7/5/111},
	doi = {10.3390/smartcities7050111},
	language = {en},
	number = {5},
	urldate = {2026-04-28},
	journal = {Smart Cities},
	publisher = {Multidisciplinary Digital Publishing Institute},
	author = {Cheng, Hoi-Kin and Kou, Kun-Pang and Wong, Ka-Io},
	month = oct,
	year = {2024},
	pages = {2861--2886},
}

@inproceedings{sun_multi-agent_2024,
	title = {Multi-{Agent} {Q}-{Learning} and {Max}-{Pressure} {Control} for {Large}-{Scale} {Traffic} {Signal} {Control}},
	issn = {2766-8665},
	url = {https://ieeexplore.ieee.org/document/10760228},
	doi = {10.1109/ICNSC62968.2024.10760228},
	urldate = {2026-04-28},
	booktitle = {2024 {International} {Conference} on {Networking}, {Sensing} and {Control} ({ICNSC})},
	author = {Sun, Yuanzhen and Qi, Liang and Li, Mengqi and Luan, Wenjing and Zhang, Fuxin},
	month = oct,
	year = {2024},
	note = {ISSN: 2766-8665},
	pages = {1--6},
}

@inproceedings{yan_deep_2023,
	title = {A {Deep} {Reinforcement} {Learning}-{Based} {Cooperative} {Traffic} {Signal} {System} {Through} {Dual}-{Sensing} {Max} {Pressure} {Control}},
	url = {https://ieeexplore.ieee.org/document/10525376},
	doi = {10.1109/ICMEE59781.2023.10525376},
	urldate = {2026-04-28},
	booktitle = {2023 9th {International} {Conference} on {Mechanical} and {Electronics} {Engineering} ({ICMEE})},
	author = {Yan, Tianwen and Zuo, Lei and Yan, Maode and Zhang, Jinqi},
	month = nov,
	year = {2023},
	pages = {258--264},
}

@inproceedings{wei_reinforcement_2023,
	title = {A {Reinforcement} {Learning} {Traffic} {Signal} {Control} {Method} {Based} on {Traffic} {Intensity} {Analysis}},
	issn = {1934-1768},
	url = {https://ieeexplore.ieee.org/document/10240019},
	doi = {10.23919/CCC58697.2023.10240019},
	urldate = {2026-04-28},
	booktitle = {2023 42nd {Chinese} {Control} {Conference} ({CCC})},
	author = {Wei, Lu and Gao, Lei and Yang, Jian and Li, Jinhong},
	month = jul,
	year = {2023},
	note = {ISSN: 1934-1768},
	pages = {6719--6724},
}

@inproceedings{li_cooperative_2022,
	title = {Cooperative {Max}-{Pressure} {Enhanced} {Traffic} {Signal} {Control}},
	isbn = {978-1-4503-9236-5},
	
	doi = {10.1145/3511808.3557569},
	booktitle = {Proceedings of the 31st {ACM} {International} {Conference} on {Information} \& {Knowledge} {Management}},
    author = {Li, Lin and Li, Renbo and Peng, Yuquan and Huang, Chuanming and Yuan, Jingling},
	month = oct,
	year = {2022},
	pages = {4173--4177},

}

@article{boukerche_novel_2022,
	title = {A {Novel} {Reinforcement} {Learning}-{Based} {Cooperative} {Traffic} {Signal} {System} {Through} {Max}-{Pressure} {Control}},
	volume = {71},
	issn = {1939-9359},
	url = {https://ieeexplore.ieee.org/document/9392372},
	doi = {10.1109/TVT.2021.3069921},
	number = {2},
	urldate = {2026-04-28},
	journal = {IEEE Transactions on Vehicular Technology},
	author = {Boukerche, Azzedine and Zhong, Dunhao and Sun, Peng},
	month = feb,
	year = {2022},
	pages = {1187--1198},
}

@inproceedings{zhang_expression_2022,
	title = {Expression might be enough: representing pressure and demand for reinforcement learning based traffic signal control},
	issn = {2640-3498},
	shorttitle = {Expression might be enough},
	url = {https://proceedings.mlr.press/v162/zhang22ah.html},
	language = {en},
	urldate = {2026-08-29},
	booktitle = {Proceedings of the 39th {International} {Conference} on {Machine} {Learning}},
	publisher = {PMLR},
	author = {Zhang, Liang and Wu, Qiang and Shen, Jun and Lü, Linyuan and Du, Bo and Wu, Jianqing},
	month = jun,
	year = {2022},
	pages = {26645--26654},
}

@inproceedings{peng_learning_2021,
	address = {Cham},
	title = {Learning {Cooperative} {Max}-{Pressure} {Control} by {Leveraging} {Downstream} {Intersections} {Information} for {Traffic} {Signal} {Control}},
	isbn = {978-3-030-85899-5},
	doi = {10.1007/978-3-030-85899-5_29},
	language = {en},
	booktitle = {Web and {Big} {Data}},
	publisher = {Springer International Publishing},
	author = {Peng, Yuquan and Li, Lin and Xie, Qing and Tao, Xiaohui},
	editor = {U, Leong Hou and Spaniol, Marc and Sakurai, Yasushi and Chen, Junying},
	year = {2021},
	pages = {399--413},
}

@inproceedings{tang_duallight_2024,
	title = {{DualLight}: {Intelligent} {Traffic} {Signal} {Control} {Method} {Based} on {Deep} {Reinforcement} {Learning}},
	issn = {2158-9208},
	shorttitle = {{DualLight}},
	url = {https://ieeexplore.ieee.org/document/10885217},
	doi = {10.1109/ISPA63168.2024.00040},
	urldate = {2026-04-28},
	booktitle = {2024 {IEEE} {International} {Symposium} on {Parallel} and {Distributed} {Processing} with {Applications} ({ISPA})},
	author = {Tang, Huanqiang and He, Fan and Cheng, Yanfen and Shao, Xun},
	month = oct,
	year = {2024},
	note = {ISSN: 2158-9208},
	pages = {254--261},
}

@article{barman_throughput_2023,
	title = {Throughput properties and optimal locations for limited deployment of {Max}-pressure controls},
	volume = {150},
	issn = {0968-090X},
	url = {https://www.sciencedirect.com/science/article/pii/S0968090X23000943},
	doi = {10.1016/j.trc.2023.104105},
	urldate = {2026-05-14},
	journal = {Transportation Research Part C: Emerging Technologies},
	author = {Barman, Simanta and Levin, Michael W.},
	month = may,
	year = {2023},
	pages = {104105},
}

@article{ahmed_occ-mp_2024,
	title = {{OCC}-{MP}: {A} {Max}-{Pressure} framework to prioritize transit and high occupancy vehicles},
	volume = {166},
	issn = {0968-090X},
	shorttitle = {{OCC}-{MP}},
	doi = {10.1016/j.trc.2024.104795},
	urldate = {2026-03-18},
	journal = {Transportation Research Part C: Emerging Technologies},
	author = {Ahmed, Tanveer and Liu, Hao and Gayah, Vikash V.},
	month = sep,
	year = {2024},

}

@article{tan_transit-mp_2027,
title = {Transit-MP: Transit-prioritized max-pressure control in sparse connected vehicle environments},
journal = {Transportation Research Part B: Methodological},
author = {Chaopeng Tan and Hao Liu and Dingshan Sun and Marco Rinaldi and Hans {van Lint}},
volume = {215},
pages = {103622},
year = {2027},
doi = {10.1016/j.trb.2026.103622},
}

@article{xu_integrating_2022,
	title = {Integrating public transit signal priority into max-pressure signal control: {Methodology} and simulation study on a downtown network},
	volume = {138},
	issn = {0968-090X},
	shorttitle = {Integrating public transit signal priority into max-pressure signal control},
	url = {https://www.sciencedirect.com/science/article/pii/S0968090X22000602},
	doi = {10.1016/j.trc.2022.103614},
	urldate = {2026-03-19},
	journal = {Transportation Research Part C: Emerging Technologies},
	author = {Xu, Te and Barman, Simanta and Levin, Michael W. and Chen, Rongsheng and Li, Tianyi},
	month = may,
	year = {2022},
	pages = {103614},
}

@article{mercader_max-pressure_2020,
	title = {Max-pressure traffic controller based on travel times: {An} experimental analysis},
	volume = {110},
	issn = {0968-090X},
	shorttitle = {Max-pressure traffic controller based on travel times},
	url = {https://www.sciencedirect.com/science/article/pii/S0968090X19307442},
	doi = {10.1016/j.trc.2019.10.002},
	urldate = {2026-03-18},
	journal = {Transportation Research Part C: Emerging Technologies},
	author = {Mercader, Pedro and Uwayid, Wasim and Haddad, Jack},
	month = jan,
	year = {2020},
	pages = {275--290},
}

@article{gregoire_back-pressure_2014,
	series = {19th {IFAC} {World} {Congress}},
	title = {Back-pressure traffic signal control with unknown routing rates},
	volume = {47},
	issn = {1474-6670},
	url = {https://www.sciencedirect.com/science/article/pii/S1474667016434178},
	doi = {10.3182/20140824-6-ZA-1003.01585},
	number = {3},
	urldate = {2026-03-17},
	journal = {IFAC Proceedings Volumes},
	author = {Gregoire, Jean and Frazzoli, Emilio and Fortelle, Arnaud de La and Wongpiromsarn, Tichakorn},
	month = jan,
	year = {2014},
	pages = {11332--11337},
}

@article{MPLight2020_towards_thousands, 
title={Toward A Thousand Lights: Decentralized Deep Reinforcement Learning for Large-Scale Traffic Signal Control}, volume={34}, url={https://ojs.aaai.org/index.php/AAAI/article/view/5744}, DOI={10.1609/aaai.v34i04.5744}, 
number={04}, 
journal={Proceedings of the AAAI Conference on Artificial Intelligence}, author={Chen, Chacha and Wei, Hua and Xu, Nan and Zheng, Guanjie and Yang, Ming and Xiong, Yuanhao and Xu, Kai and Li, Zhenhui}, 
year={2020}, 
month={Apr.}, 
pages={3414–3421} }

@article{EfficientPressure2021,
      title={Efficient Pressure: Improving efficiency for signalized intersections}, 
      author={Qiang Wu and Liang Zhang and Jun Shen and Linyuan Lü and Bo Du and Jianqing Wu},
      year={2021},
      archivePrefix={arXiv},
      primaryClass={cs.LG},
      DOI={10.48550/arXiv.2112.02336},
      journal={arXiv preprint},
}

@article{GeneralizedPhasePressure2025l,
      title={Generalized Phase Pressure Control Enhanced Reinforcement Learning for Traffic Signal Control}, 
      author={Xiao-Cheng Liao and Yi Mei and Mengjie Zhang and Xiang-Ling Chen},
      year={2025},
      archivePrefix={arXiv},
      primaryClass={cs.LG},
      DOI={10.48550/arXiv.2503.20205},
      journal={arXiv preprint},
}

@ARTICLE{li2025MultiHopupstreamAnticipatory,
  author={Li, Xiaocan and Wang, Xiaoyu and Smirnov, Ilia and Sanner, Scott and Abdulhai, Baher},
  journal={IEEE Open Journal of Intelligent Transportation Systems}, 
  title={Multi-Hop Upstream Anticipatory Traffic Signal Control With Deep Reinforcement Learning}, 
  year={2025b},
  volume={6},
  number={},
  pages={554-567},
  doi={10.1109/OJITS.2025.3562757}}

@article{tan_dwelling_speed_2026,
  author={Tan, Chaopeng and Laskaris, Georgios and Sun, Dingshan and Abohariri, Robin and Rinaldi, Marco and van Lint, Hans},
  journal={IEEE Transactions on Intelligent Transportation Systems}, 
  title={Dwelling and Speed Advisory Enhanced Max-Pressure Control With Transit Signal Priority}, 
  year={2026},
  volume={},
  number={},
  pages={1-14},
  doi={10.1109/TITS.2026.3695849}}

@ARTICLE{peng_a_milti_scale_2026,
  author={Peng, Xianyue and Wang, Hao and Chen, Shenyang and Zhang, H. Michael},
  journal={IEEE Transactions on Intelligent Transportation Systems}, 
  title={A Multi-Scale Perimeter Control and Route Guidance System for Large-Scale Road Networks}, 
  year={2026},
  volume={27},
  number={6},
  pages={6257-6272},
  doi={10.1109/TITS.2026.3678163}}

@article{AHMED2024optimal,
title = {Identification of optimal locations of adaptive traffic signal control using heuristic methods},
journal = {International Journal of Transportation Science and Technology},
volume = {13},
pages = {122-136},
year = {2024},
issn = {2046-0430},
doi = {10.1016/j.ijtst.2023.12.003},
author = {Tanveer Ahmed and Hao Liu and Vikash V. Gayah},
}

@article{lin_unveiling_2025a,
title = {Unveiling network capacity potential with imminent supply information part {I}: {Theoretical} derivation},
journal = {Transportation Research Part B: Methodological},
volume = {192},
pages = {103150},
year = {2025},
issn = {0191-2615},
doi = {10.1016/j.trb.2024.103150},
url = {https://www.sciencedirect.com/science/article/pii/S0191261524002741},
author = {Dianchao Lin and Li Li},
}

@INPROCEEDINGS{sumo2018,
  author={Lopez, Pablo Alvarez and Behrisch, Michael and Bieker-Walz, Laura and Erdmann, Jakob and Flötteröd, Yun-Pang and Hilbrich, Robert and Lücken, Leonhard and Rummel, Johannes and Wagner, Peter and Wiessner, Evamarie},
  booktitle={2018 21st International Conference on Intelligent Transportation Systems (ITSC)}, 
  title={Microscopic Traffic Simulation using SUMO}, 
  year={2018},
  volume={},
  number={},
  pages={2575-2582},
  doi={10.1109/ITSC.2018.8569938}
  }

@article{agarwal_fusing_2024,
	title = {Fusing crowdsourced data to an adaptive wireless traffic signal control system architecture},
	volume = {26},
	issn = {2542-6605},
	url = {https://www.sciencedirect.com/science/article/pii/S2542660524001100},
	doi = {10.1016/j.iot.2024.101169},
	urldate = {2026-09-02},
	journal = {Internet of Things},
	author = {Agarwal, Amit and Sahu, Deorishabh and Nautiyal, Anuj and Gupta, Manan and Agarwal, Pratham},
	month = jul,
	year = {2024},
	pages = {101169},
	}

@article{ghosh_pascal_2026,
	title = {{PASCAL}: {Pressure} {Adaptive} {Signal} {Control} {ALgorithm}},
	volume = {191},
	issn = {0968-090X},
	shorttitle = {{PASCAL}},
	url = {https://www.sciencedirect.com/science/article/pii/S0968090X26002925},
	doi = {10.1016/j.trc.2026.105804},
	urldate = {2026-08-29},
	journal = {Transportation Research Part C: Emerging Technologies},
	author = {Ghosh, Tamojit and Vanajakshi, Lelitha Devi and Mazharul (Shimul) Haque, Md. and Bhaskar, Ashish},
	month = oct,
	year = {2026},
	pages = {105804},
	}

@article{lin_unveiling_2025b,
title = {Unveiling network capacity potential with imminent supply information part {II}: {Backpressure-based} validation},
journal = {Transportation Research Part B: Methodological},
volume = {192},
pages = {103153},
year = {2025},
issn = {0191-2615},
doi = {10.1016/j.trb.2025.103153},
url = {https://www.sciencedirect.com/science/article/pii/S0191261525000025},
author = {Dianchao Lin and Li Li},

}

@article{rey_blue_2019,
  author  = {Rey, David and Levin, Michael W.},
  title   = {Blue phase: Optimal network traffic control for legacy and autonomous vehicles},
  journal = {Transportation Research Part B: Methodological},
  year    = {2019},
  volume  = {130},
  pages   = {105--129},
  doi     = {10.1016/j.trb.2019.11.001}
}

@article{taale_integrated_2015,
  author  = {Taale, Henk and van Kampen, Joost and Hoogendoorn, Serge},
  title   = {Integrated Signal Control and Route Guidance based on Back-pressure Principles},
  journal = {Transportation Research Procedia},
  year    = {2015},
  volume  = {10},
  pages   = {226--235},
  doi     = {10.1016/j.trpro.2015.09.072}
}

@article{le_utility_2017,
  author  = {Le, Tung and Vu, Hai L. and Walton, Neil and Hoogendoorn, Serge P. and Kov{\'a}cs, P{\'e}ter and Queija, Rudesindo N.},
  title   = {Utility optimization framework for a distributed traffic control of urban road networks},
  journal = {Transportation Research Part B: Methodological},
  year    = {2017},
  volume  = {105},
  pages   = {539--558},
  doi     = {10.1016/j.trb.2017.10.004}
}
\end{document}